\documentclass[aps,prb,twocolumn,amsmath,amssymb,showpacs,superscriptaddress]{revtex4-2}

\usepackage{graphicx}
\usepackage{epstopdf}

\usepackage{amsmath,amssymb,amsfonts}
\usepackage{textcomp}
\usepackage{hyperref}
\usepackage{gensymb}
\usepackage{verbatim}
\usepackage{amsmath}
\hypersetup{breaklinks=true,colorlinks=true,urlcolor=black}
\usepackage{color}
\usepackage{soul}
\usepackage{float}
\usepackage{gensymb}
\usepackage{caption}
\usepackage{subcaption}
\usepackage[english]{babel}
\usepackage[T1]{fontenc}
\usepackage[utf8]{inputenc}
\usepackage{booktabs}
\DeclareGraphicsExtensions{.jpg,.pdf,.png,.eps}

\begin{document}
\title{Superconductivity and phase diagrams of  CaK(Fe$_{1-x}$Mn$_{x}$)$_{4}$As$_{4}$ single crystals}
\author{M. Xu}
\author{J. Schmidt}
\affiliation{Ames Laboratory, Iowa State University, Ames, Iowa 50011, USA}
\affiliation{Department of Physics and Astronomy, Iowa State University, Ames, Iowa 50011, USA}

\author{E. Gati}
\affiliation{Ames Laboratory, Iowa State University, Ames, Iowa 50011, USA}
\affiliation{Department of Physics and Astronomy, Iowa State University, Ames, Iowa 50011, USA}
\affiliation{Max Planck Institute for Chemical Physics of Solids, 01187 Dresden, Germany}

\author{L. Xiang}
\affiliation{Ames Laboratory, Iowa State University, Ames, Iowa 50011, USA}
\affiliation{Department of Physics and Astronomy, Iowa State University, Ames, Iowa 50011, USA}
\affiliation{National High Magnetic Field Laboratory, Florida State University, Tallahassee, Florida 32310, USA}

\author{\\W. R. Meier}
\affiliation{Ames Laboratory, Iowa State University, Ames, Iowa 50011, USA}
\affiliation{Department of Physics and Astronomy, Iowa State University, Ames, Iowa 50011, USA}
\affiliation{Department of Materials Science and Engineering, University of Tennessee Knoxville, Knoxville, Tennessee, 37996, USA}
\author{V. G. Kogan}
\affiliation{Ames Laboratory, Iowa State University, Ames, Iowa 50011, USA}
\author{S. L. Bud'ko}
\affiliation{Ames Laboratory, Iowa State University, Ames, Iowa 50011, USA}
\affiliation{Department of Physics and Astronomy, Iowa State University, Ames, Iowa 50011, USA}

\author{P. C. Canfield}
\affiliation{Ames Laboratory, Iowa State University, Ames, Iowa 50011, USA}
\affiliation{Department of Physics and Astronomy, Iowa State University, Ames, Iowa 50011, USA}
\email[]{canfield@ameslab.gov}

\date{\today}

\begin{abstract}
In order to study the effects of Mn substitution on the superconducting and magnetic ground state of CaKFe$_4$As$_4$ ($T_c$ = 35 K), members of the CaK(Fe$_{1-x}$Mn$_{x}$)$_{4}$As$_{4}$ series have been synthesized by high-temperature solution growth in single crystalline form and characterized by elemental analysis, thermodynamic and transport measurements. These measurements show that the superconducting transition temperature decreases monotonically and is finally suppressed below 1.8 K as $x$ is increased from 0 to 0.036. For $x$-values greater than 0.016, signatures of a magnetic transition can be detected in both thermodynamic and transport measurements in which kink-like features allow for the determination of the transition temperature, $T^*$, that increases as Mn substitution increases. A temperature-composition (\textit{T}-\textit{x}) phase diagram is constructed, revealing a half-dome of superconductivity with the magnetic transition temperature, $T^*$, appearing near 26~K for $x$ $\sim$ 0.017 and rising slowly up to 33~K for $x$ $\sim$ 0.036. In addition to the creation of the $T$-$x$ phase diagram for CaK(Fe$_{1-x}$Mn$_{x}$)$_{4}$As$_{4}$, specific heat data are used to track the jump in specific heat at $T_c$; The CaK(Fe$_{1-x}$Mn$_x$)$_4$As$_4$ data does not follow the scaling of $\Delta$$C_{p}$ with $T_{c}^3$ as many of the other Fe-based superconducting systems do. These data suggest that, as magnetic pair-breaking is present, the jump in $C_{p}$ for a given $T_{c}$ is reduced. Elastoresistivity coefficients, $2m_{66}$ and $m_{11}-m_{12}$, as a function of temperature are also measured. $2m_{66}$ and $m_{11}-m_{12}$ are qualitatively similar to CaK(Fe$_{1-x}$Ni$_x$)$_4$As$_4$. This may indicate that the magnetic order in Mn substituted system may be still the same as CaK(Fe$_{1-x}$Ni$_x$)$_4$As$_4$. Superconductivity of CaK(Fe$_{1-x}$Mn$_{x}$)$_{4}$As$_{4}$ is also studied as a function of magnetic field. A clear change in $H^\prime_{c2}$($T$)/$T_c$, where $H^\prime_{c2}$($T$) is d$H_{c2}$($T$)/d$T$, at $x$ $\sim$ 0.015 is observed and probably is related to change of the Fermi surface due to magnetic order. Coherence lengths and the London penetration depths are also calculated based on $H_{c1}$ and $H_{c2}$ data. Coherence lengths as the function of $x$ also shows the changes near $x$ = 0.015, again consistent with Fermi surfaces changes associated with the magnetic ordering seen for higher $x$-values.
\end{abstract}

\maketitle

\section{introduction}

Since their discovery in 2008 {\color{blue}\cite{Kamihara2008}}, the study of Fe-based superconductors has lead to extensive experimental interest and their variety offers the opportunity of understanding unconventional superconductivity in a broader sense. Though diverse, the several families of Fe-based superconductors share similar crystal structures {\color{blue}\cite{Johnston2010,Paglione2010,Hosono2015}} all featuring edge-shared tetrahedral Fe-As or Fe-Se layers, and phase diagrams 
{\color{blue}\cite{Canfield2009a,Ni2009,Ni20101,Johnston2010,Paglione2010,Canfield2010f,Stewart2011,Hosono2015,Yoshida2016,Meier2016,Meier2017,Meier2018,Hsu14262,Bohmer2017a,Kawashima2016,Bao2018,Gati2020}} that suggest a relationship between, or proximity of, superconducting and magnetic and/or nematic ordering or fluctuations. The role of magnetic as well as structural transitions and fluctuations in unconventional superconductivity is still an open topic, attracting both theoretical and experimental research. The relationship between these states is believed to be key to understanding unconventional superconductivity {\color{blue}\cite{Mazin2010}}.

Within the Fe-based superconductors, three structural classes, $Ae$Fe$_2$As$_2$ ($Ae$=Alkaline Earth) (122) families {\color{blue}\cite{Canfield2009a,Ni2009,Ni20101,Johnston2010,Paglione2010,Canfield2010f,Stewart2011,Hosono2015}}, $AeA$Fe$_4$As$_4$ (A=Alkaline Metal) (1144) family {\color{blue}\cite{Yoshida2016,Meier2016,Meier2017,Meier2018,Kawashima2016,Bao2018}} and FeSe {\color{blue}\cite{Hsu14262,Bohmer2017a}} provide a microcosm of many of the key questions at hand. At ambient pressure, doped 122-systems manifest the most common interplay between stripe-like antiferromagnetic order, nematicity, and superconductivity {\color{blue}\cite{Canfield2010f,Paglione2010}}. On the other hand the Ni-substituted 1144-system has hedgehog-spin-vortex-crystal (h-SVC) type antiferromagnetic (AFM) order and superconductivity interacting with each other without any structural phase transition{\color{blue}\cite{Meier2018,Meier2019T}}. In contrast, at ambient pressure, FeSe has a nematic-like transition and superconductivity, but no magnetic order {\color{blue}\cite{Hsu14262,Bohmer2017a}}. The comparison of these three classes of Fe-based superconductors allows us to examine the effects of the presence or absence of nematicity and magnetic order on the superconducting state. 

CaKFe$_4$As$_4$ is a structurally ordered, quaternary compound that intrinsically superconducts with $T_c$ around 35~K {\color{blue}\cite{Yoshida2016,Meier2016}}. As such it offers the possibility of studying the effects of Mn substitution on the superconducting state {\color{blue}\cite{Li2012a,Leboeuf2014}}. Given that it was not possible to induce  superconductivity in the Ba(Fe$_{1-x}$Mn$_x$)$_2$As$_2$ system {\color{blue}\cite{Thaler2011,Pandey2011}}, one of the motivations for this study was to gain insight into how Mn affects superconductivity in the CaK(Fe$_{1-x}$Mn$_{x}$)$_{4}$As$_{4}$ system. This is of particular interest given the fact that Mn, of all the 3-d transition metals, is most likely to manifest local-moment-like properties akin to what can be found for rare earths such as Gd. 

A second motivation for this study of the CaK(Fe$_{1-x}$Mn$_{x}$)$_{4}$As$_{4}$ system is based on the similarities between CaKFe$_4$As$_4$ and Ba$_{0.5}$K$_{0.5}$Fe$_2$As$_2$, both having the same nominal electron count and presenting only a superconducting transition upon cooling. This analogy was explored {\color{blue}\cite{Meier2018}} in the study of the similarities between the underdoped ($x$ $<$ 0.5) Ba$_{1-x}$K$_x$Fe$_2$As$_2$ and Ni/Co doped CaKFe$_4$As$_4$ phase diagrams, both showing a 1/2 dome of superconductivity with increasing AFM ordering ($T_{N}$) coinciding with decreasing superconducting $T_{c}$. Study of CaK(Fe$_{1-x}$Mn$_{x}$)$_{4}$As$_{4}$ allows us to investigate to what extent this analogy between the K-doped-122 and transition-metal-doped 1144 phase diagrams holds. In particular, we compare nominally hole doped CaK(Fe$_{1-x}$Mn$_x$)$_4$As$_4$ with overdoped ($x$ $>$ 0.5) Ba$_{1-x}$K$_x$Fe$_2$As$_2$. Even though superconductivity is suppressed in both cases, we found a fundamental difference between the two phase diagrams: a magnetic phase transition line appears in the CaK(Fe$_{1-x}$Mn$_{x}$)$_{4}$As$_{4}$ compounds upon doping, which is not present for overdoped Ba$_{1-x}$K$_x$Fe$_2$As$_2$. In addition, at a quantitative level, Mn substitution suppresses $T_{c}$ much more rapidly than what was found for Co or Ni substitution.

In this paper, we detail the synthesis and characterization of CaK(Fe$_{1-x}$Mn$_x$)$_4$As$_4$ single crystals. A temperature-composition ($T$-$x$) phase diagram is constructed by elemental analysis, thermodynamic and transport measurements. In addition to creating the $T$-$x$ phase diagram, specific heat and elastoresistivity is also measured to track the jump in specific heat at $T_c$ and investigate the evolution of nematic fluctuations. Coherence lengths and the London penetration depths are also calculated based on $H_{c1}$ and $H_{c2}$ data obtained from measurements. Finally, temperature vs change of electron count, |$\Delta e^-$|, phase diagram of CaK(Fe$_{1-x}TM_{x}$)$_{4}$As$_{4}$ single crystals, $TM$ = Mn, Ni and Co, is also presented and discussed.

\section{Crystal Growth and Experimental Method}

Single crystalline CaK(Fe$_{1-x}$Mn$_{x}$)$_{4}$As$_{4}$ samples were grown by high-temperature solution growth {\color{blue}\cite{Canfield2020}} out of FeAs flux in the manner similar to pure CaKFe$_{4}$As$_{4}$ {\color{blue}\cite{Meier2017,Meier2016}}. Lumps of potassium metal (Alfa Aesar 99.95\%), distilled calcium metal pieces (Ames Laboratory, Materials Preparation Center (MPC 99.9\%)) and Fe$_{0.512}$As$_{0.488}$ and Mn$_{0.512}$As$_{0.488}$ precursor powders were loaded into a 1.7ml fritted alumina Canfield Crucible Set {\color{blue}\cite{Canfield2016a}} (LSP Industrial Ceramics, Inc.) in an argon filled glove-box. The ratio of K:Ca:Fe$_{0.512}$As$_{0.488}$ and Mn$_{0.512}$As$_{0.488}$ was 1.2:0.8:20. A 1.3~cm outer diameter and 6.4~cm long tantalum tube which was used to protect the silica ampoule from reactive vapors was welded with the crucible set in partial argon atmosphere inside. The sealed Ta tube was then itself sealed into a silica ampoule and the ampoule was placed inside a box furnace. The furnace was held for 2 hours at 650~\textcelsius\ before increasing to 1180~\textcelsius\ and held there for 5 hours to make sure the precursor was fully melted. The furnace was then fast cooled from 1180~\textcelsius\ to 960~\textcelsius\ in 1.5 hours. Crystals were grown during a slow cool-down from 960~\textcelsius\ to 920~\textcelsius\ over 48 hours. After 1-2 hours at 920~\textcelsius\, the ampoule was inverted into a centrifuge and spun to separate the remaining liquid from the grown crystals. Metallic, plate-like, crystals were obtained but, unfortunately, with both average size and thickness diminishing by factor 2-4 as $x$ is increased.

Similar to CaKFe$_{4}$As$_{4}$ {\color{blue}\cite{Meier2017}}, single crystals of CaK(Fe$_{1-x}$Mn$_{x}$)$_{4}$As$_{4}$ are soft and malleable which makes them difficult to grind for powder x-ray diffraction measurements. Diffraction measurements were carried out on single crystal samples, which were cleaved along the (001) plane, using a Rigaku MiniFlex II powder diffactometer in Bragg-Brentano geometry with Cu K$\alpha$ radiation ($\lambda$ = 1.5406 \AA{}) {\color{blue}\cite{Jesche2016}}.

The Mn substitution levels ($x$) of the CaK(Fe$_{1-x}$Mn$_{x}$)$_{4}$As$_{4}$ crystals were determined by energy dispersive spectroscopy (EDS) quantitative chemical analysis using an EDS detector (Thermo NORAN Microanalysis System, model C10001) attached to a JEOL scanning-electron microscope (SEM). The compositions of platelike crystals were measured at three separate positions on each crystal's face (parallel to the crystallographic \textit{ab}-plane) after cleaving them. An acceleration voltage of 16~kV, working distance of 10~mm and take off angle of 35$^{\circ}$ were used for measuring all standards and crystals with unknown composition. Pure CaKFe$_4$As$_4$ was used as a standard for Ca, K, Fe and As quantification. SmMn$_2$Ge$_2$ and GdMn$_2$Ge$_2$ were used as standards for Mn, both leading to consistent results without significant difference within the experimental error ($\sim$ 0.0015). The spectra were fitted using NIST-DTSA II Lorentz 2020-06-26 software\cite{Newbury2014}. Different measurements on the same sample reveal good homogeneity in each crystal and the average compositions and error bars were obtained from these data, accounting for both inhomogeneity and goodness of fit of each spectra.

Temperature and magnetic field-dependent magnetization and resistance measurements as well as temperature dependent specific heat measurements were carried out by using Quantum Design (QD), Magnetic Property Measurement Systems (MPMS) and Physical Property Measurement Systems (PPMS). Field-dependent magnetization data from which $H_{c1}$ is obtained, was collected by Quantum Design MPMS-3 using quartz sample holder with magnetic field parallel to the crystallographic \textit{ab} plane. Temperature dependent magnetization measurements were taken for \textit{H}$\parallel$\textit{ab} by placing the plate-like sample between two collapsed plastic straws with the third, uncollapsed, straw providing support as a sheath on the outside. The single crystal samples of CaK(Fe$_{1-x}$Mn$_{x}$)$_{4}$As$_{4}$ measured in the MPMS have plate-like morphology with length and width from 3 mm to 10 mm and thickness (\textit{c} axis) 50 - 200 $\mu$m. The approximate effective demagnetizing factor $N$ ranges from 0.007 to 0.077 with magnetic field applied parallel to the crystallographic \textit{ab} plane{\color{blue}\cite{Prozorov2018}}. Since the value of $N$ is small, we neglect the demagnetizing field when magnetic field was applied parallel to the crystallographic \textit{ab} plane. AC electrical resistance measurements were performed in a standard four-contact geometry using the ACT option of the PPMS, with a 3 mA excitation and a frequency of 17 Hz. 50$\mu$m diameter Pt wires were bonded to the samples with silver paint (DuPont 4929N) with contact resistance values of about 2-3 Ohms. The magnetic field, up to 90 kOe, was applied along \textit{c} or \textit{ab} directions, with the current flowing in the \textit{ab} plane. Temperature dependent specific heat measurements were carried out by using the relaxation technique as implemented in the Heat Capacity option of the PPMS. Due to small masses of the individual crystals for some concentrations, stacks of several crystals were used when needed, with thin layers of Apiezon N grease in between.

Elastoresistance measurements were performed using a technique similar to that described in Refs. {\color{blue}\cite{Kuo2016,Kuo2013}}. To this end, samples of CaK(Fe$_{1-x}$Mn$_x$)$_4$As$_4$ were cleaved and cut into bar-like shapes of dimensions $\approx\,1.5\,\times\,0.3\,\times\,0.02$\,mm$^3$. The orientation of the long side of the crystals was chosen to be either the [1\,1\,0]$_T$ or the [1\,0\,0]$_T$ direction (the notion []$_T$ refers to the tetragonal unit cell). For each direction, two crystals were glued orthogonally on a piezoelectric actuator from Piezomechanik using Devcon 5-Minute Epoxy. In previous works {\color{blue}\cite{Kuo2016}}, it was demonstrated that this glue provides a strain transmission of $\approx80\,\%$ below 250\,K. Strictly speaking, samples glued on the piezoelectric actuator experience biaxial strain upon application of a voltage. However, the biaxial strain is highly anisotropic, i.e., has opposite signs in the two orthogonal directions. Thus, we denote the strain along the poling directions of the actuator as $\epsilon_{xx}$, and the strain perpendicular to it as $\epsilon_{yy}$. Correspondingly, we denote the resistance of the sample, whose long axis is parallel to $\epsilon_{xx}$ ($\epsilon_{yy}$), as $R_{xx}$ ($R_{yy}$). The resistances $R_{xx}$ and $R_{yy}$ were measured in a standard four-point configuration with LakeShore's resistance bridges LS370 and LS372, with contacts made by spot welding 25$\mu$m diameter Pt wires to the sample resulting in contact resistances of typically 1\,$\Omega$. The strains $\epsilon_{xx}$ and $\epsilon_{yy}$, created by the application of voltages between -30\,V and 120\,V to the actuator, were measured by two orthogonal strain gauges that were placed on the surface of the actuator opposite to the one on which samples were attached. The low-temperature environment was provided by a Janis SHI-950T-X closed cycle refrigerator, with the sample in $^4$He exchange gas.

\section{C\lowercase{a}K(F\lowercase{e}$_{1-x}$M\lowercase{n}$_{x}$)$_{4}$A\lowercase{s}$_{4}$ Structure and Composition}
\begin{figure}[H]
	\includegraphics[width=1.5\columnwidth]{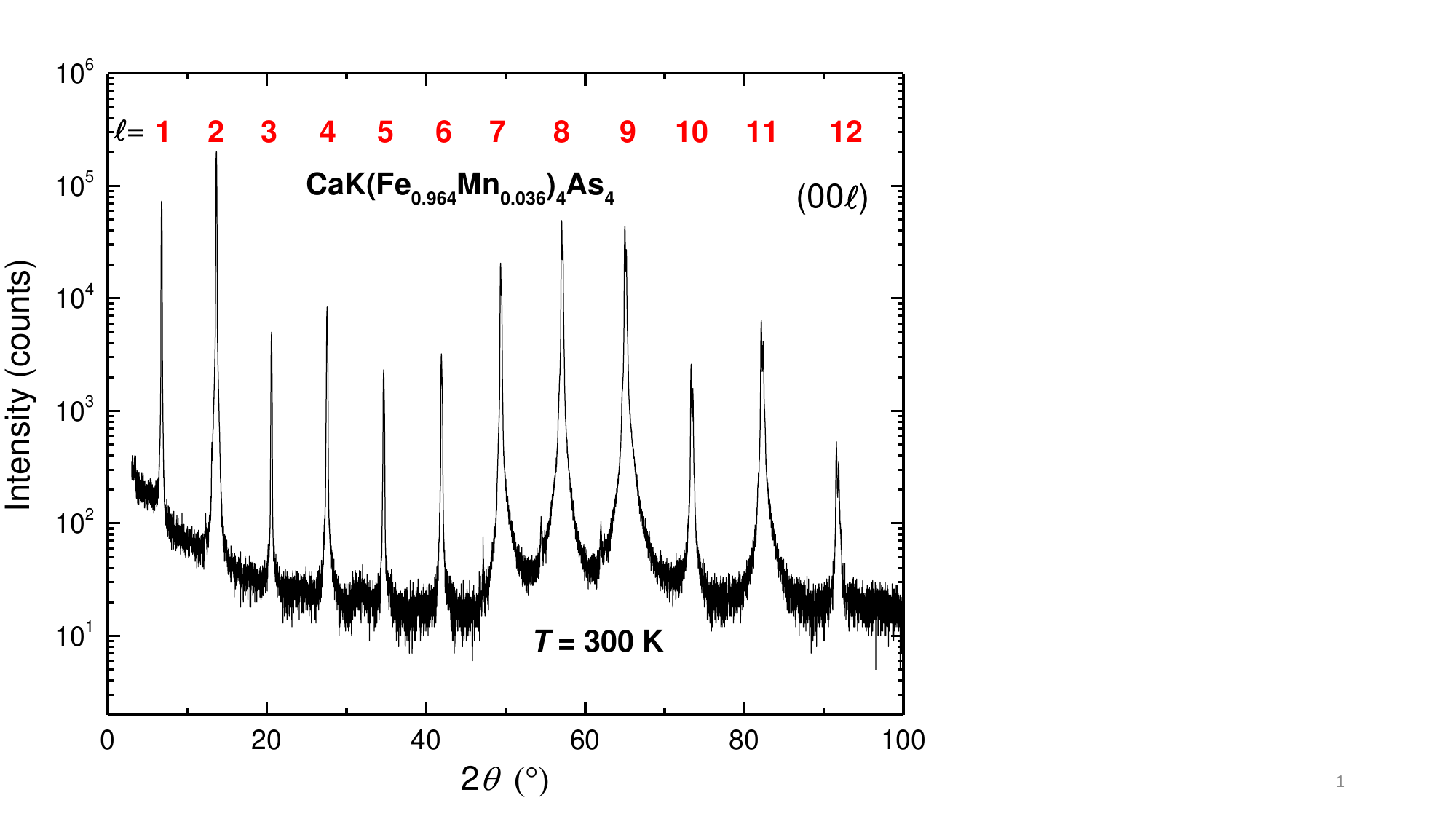}		
	\caption{X-ray diffraction data showing ($00l$) diffraction peaks from in-lab diffraction measurements on a single-crystalline plate plotted on a semi-log scale. The value of $l$ is shown in red above each peak. Note that $l$ = odd ($00l$) lines are evidence of the ordered CaKFe$_{4}$As$_{4}$ structure  {\color{blue}\cite{Yoshida2016,Meier2016}}\label{figure1}}.
\end{figure}
\begin{figure}[H]
	\includegraphics[width=1.5\columnwidth]{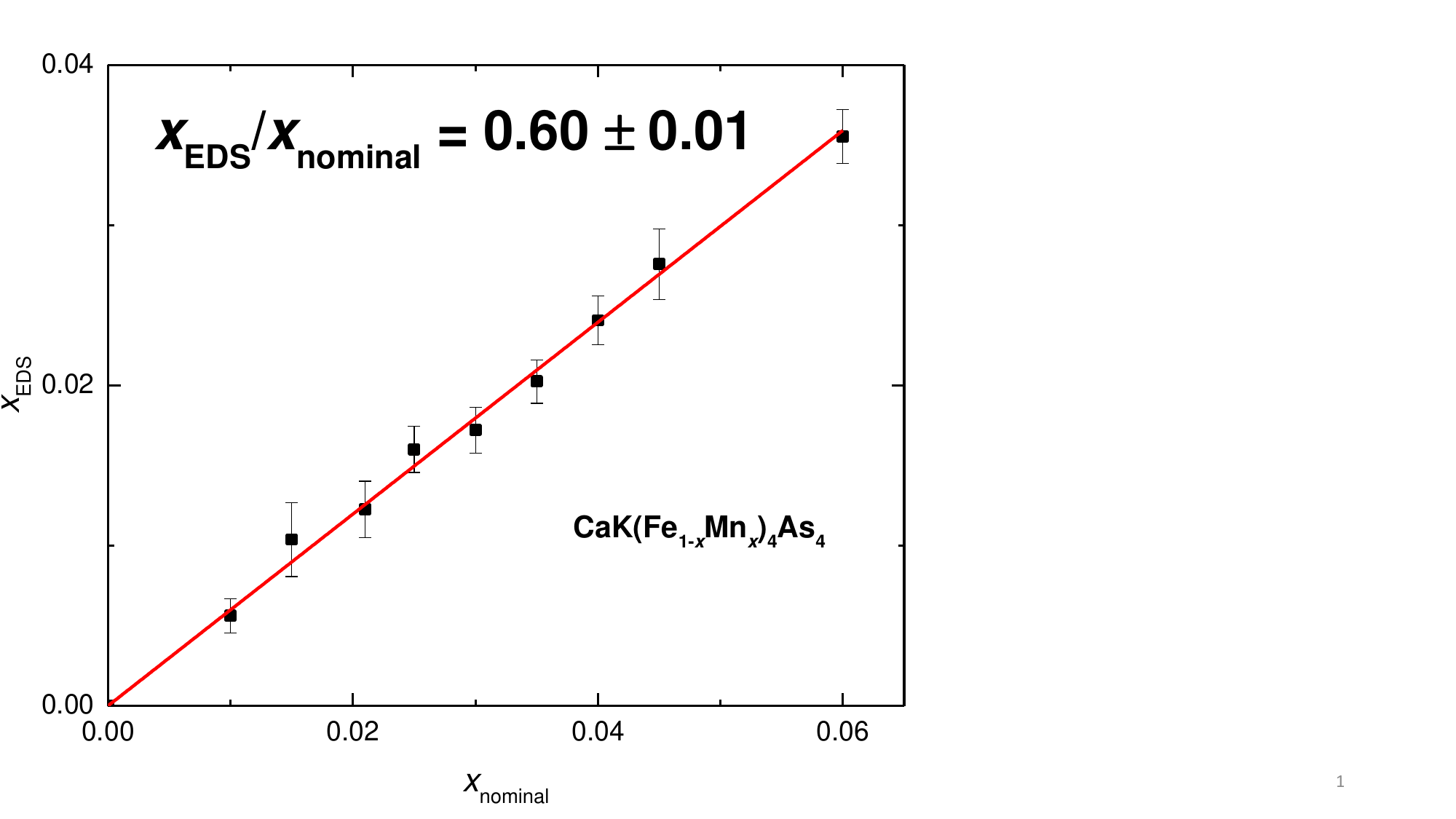}	
	\caption{EDS measured Mn concentration vs nominal Mn concentration for the CaK(Fe$_{1-x}$Mn$_{x}$)$_{4}$As$_{4}$ series. The line is the linear fit of data with fixing intercept to 0. \label{figure5}}
\end{figure}
 Figure \ref{figure1} presents single crystal diffraction data of CaK(Fe$_{1-x}$Mn$_{x}$)$_{4}$As$_{4}$ with $x_{EDS}=0.036$. From the figure, we can see that all ($00l$), $l\leq12$, are detected. The $h+k+l$ = odd peaks which are forbidden for the \textit{I}4/mmm  structure {\color{blue}\cite{Yoshida2016}} can be clearly found. This indicates that sample has the anticipated \textit{P}4/mmm structure associated with the CaKFe$_4$As$_4$ structure {\color{blue}\cite{Yoshida2016,Meier2018,Meier2016}}.

 The Mn substitution, $x$$_{EDS}$, determined by EDS is shown in Fig. \ref{figure5} for different crystals as a function of the nominal Mn fraction $x$$_{nominal}$ that was originally used for the growth. Error bars account for both possible inhomogeneity of substitution and goodness of fit of each EDS spectra. A clear correlation can be seen between the nominal and the measured substitution levels, with a proportionality factor of 0.60 $\pm$ 0.01. From this point onward, $x$$_{EDS}$ will be used when referring to Mn substitution level. For comparison, the ratio of measured to nominal Ni and Co fraction in the corresponding CaK(Fe$_{1-x}$\textit{TM}$_x$)$_{4}$As$_{4}$ are 0.64 and 0.79 respectively {\color{blue}\cite{Meier2018}}.
 
\section{Data analysis and Phase Diagram}
 
\begin{figure}[H]
	\includegraphics[width=1.5\columnwidth]{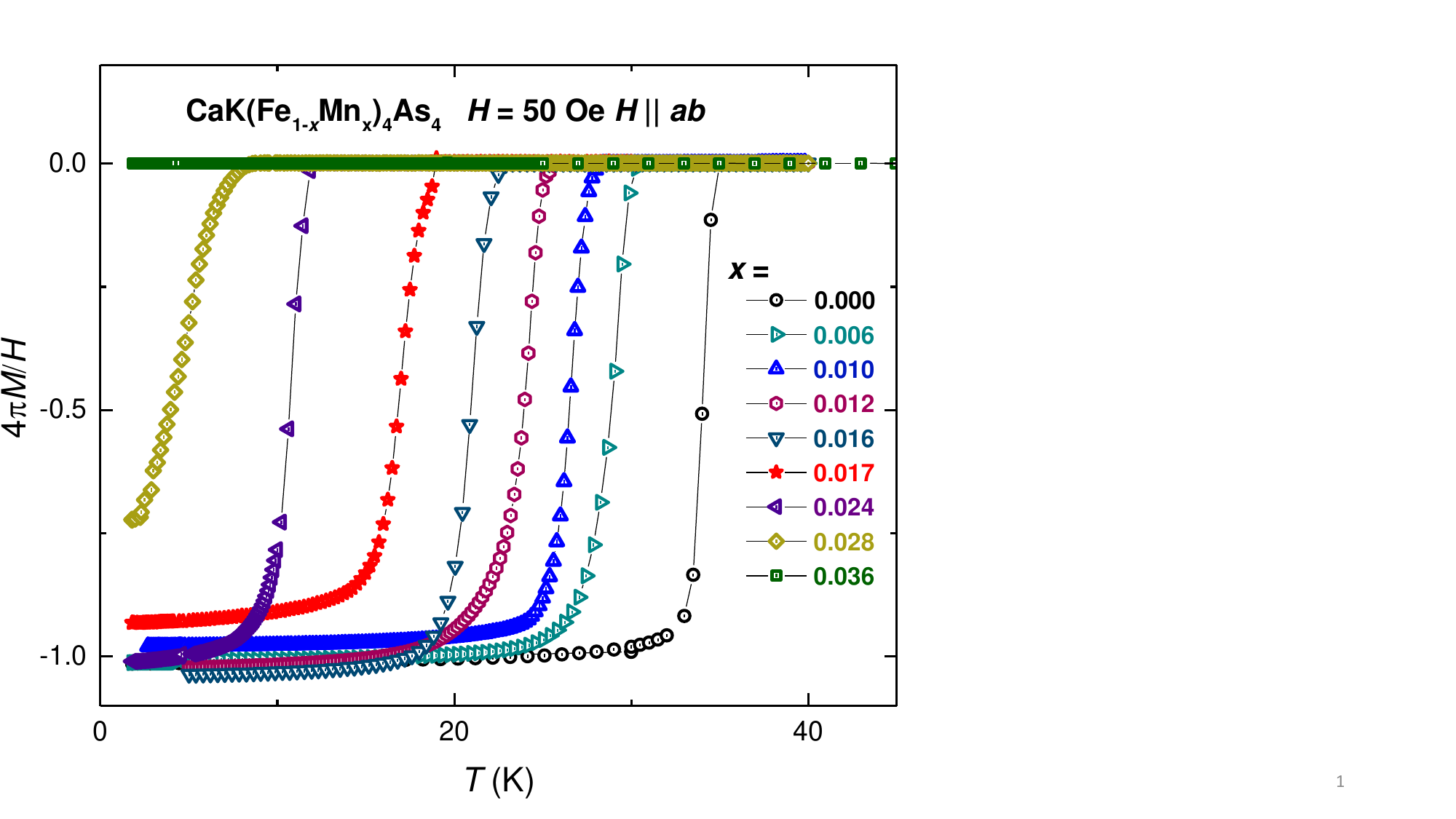}	
	\caption{Zero-field-cooled-warming (ZFCW) low temperature magnetization as a function of temperature for CaK(Fe$_{1-x}$Mn$_{x}$)$_{4}$As$_{4}$ single crystals with a field of 50~Oe applied parallel to the crystallographic \textit{ab} plane. $M$ is the volumetric magnetic moment with cgs unit emu cm$^{-3}$ or Oe. \label{figure2}}
\end{figure}
\begin{figure}[H]
	\includegraphics[width=1.5\columnwidth]{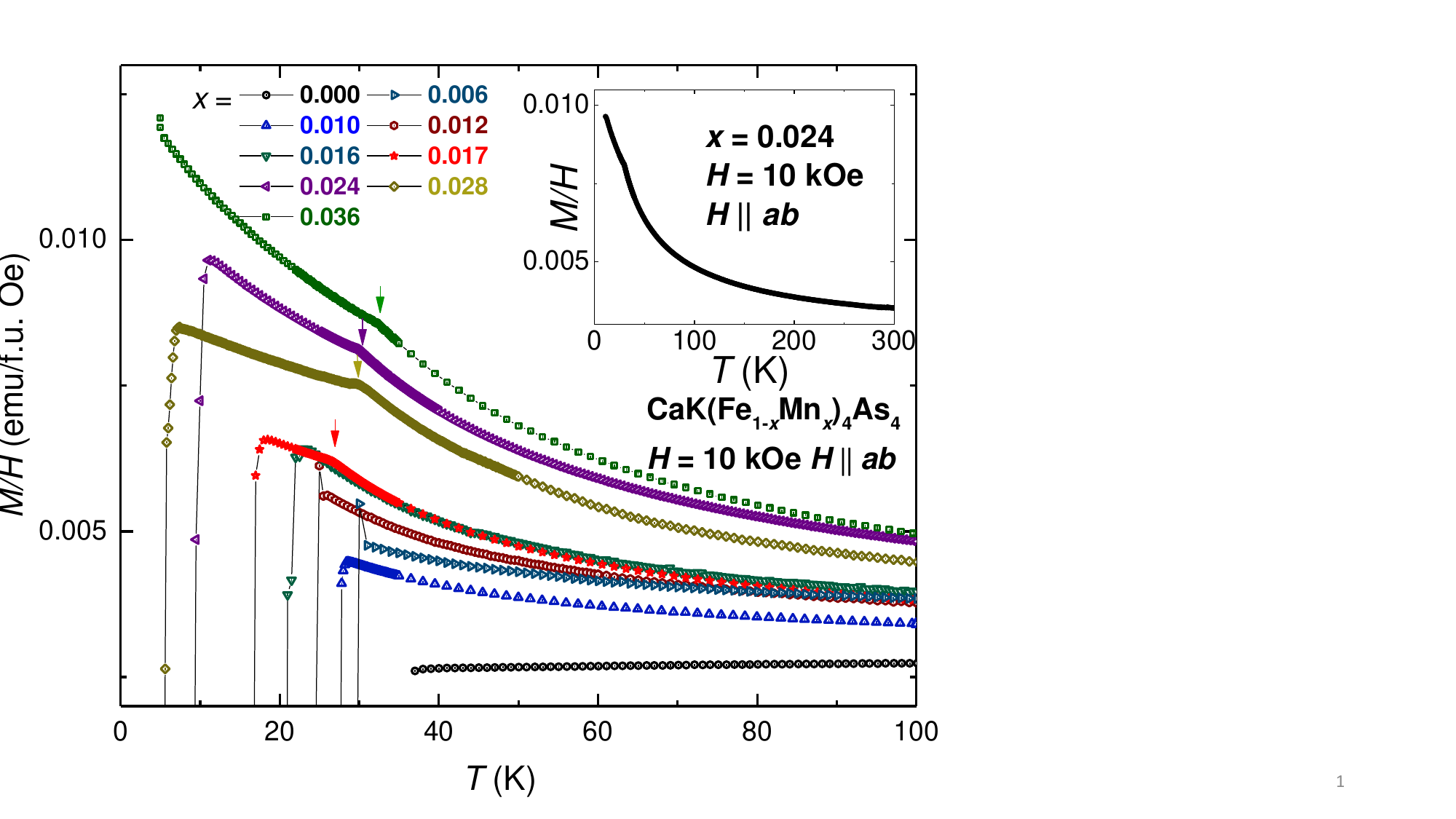}		
	\caption{Low temperature magnetization divided by applied field as a function of temperature for CaK(Fe$_{1-x}$Mn$_{x}$)$_{4}$As$_{4}$ single crystals with a field of 10~kOe applied parallel to the crystallographic \textit{ab} plane. The inset shows the CaK(Fe$_{0.976}$Mn$_{0.024}$)$_{4}$As$_{4}$ single crystal magnetization for 5 K < $T$ < 300 K. Small vertical arrows indicate the location of $T^*$, see Appendix for criterion. \label{figure3}}
\end{figure}
\begin{figure}[H]
	\includegraphics[width=1.5\columnwidth]{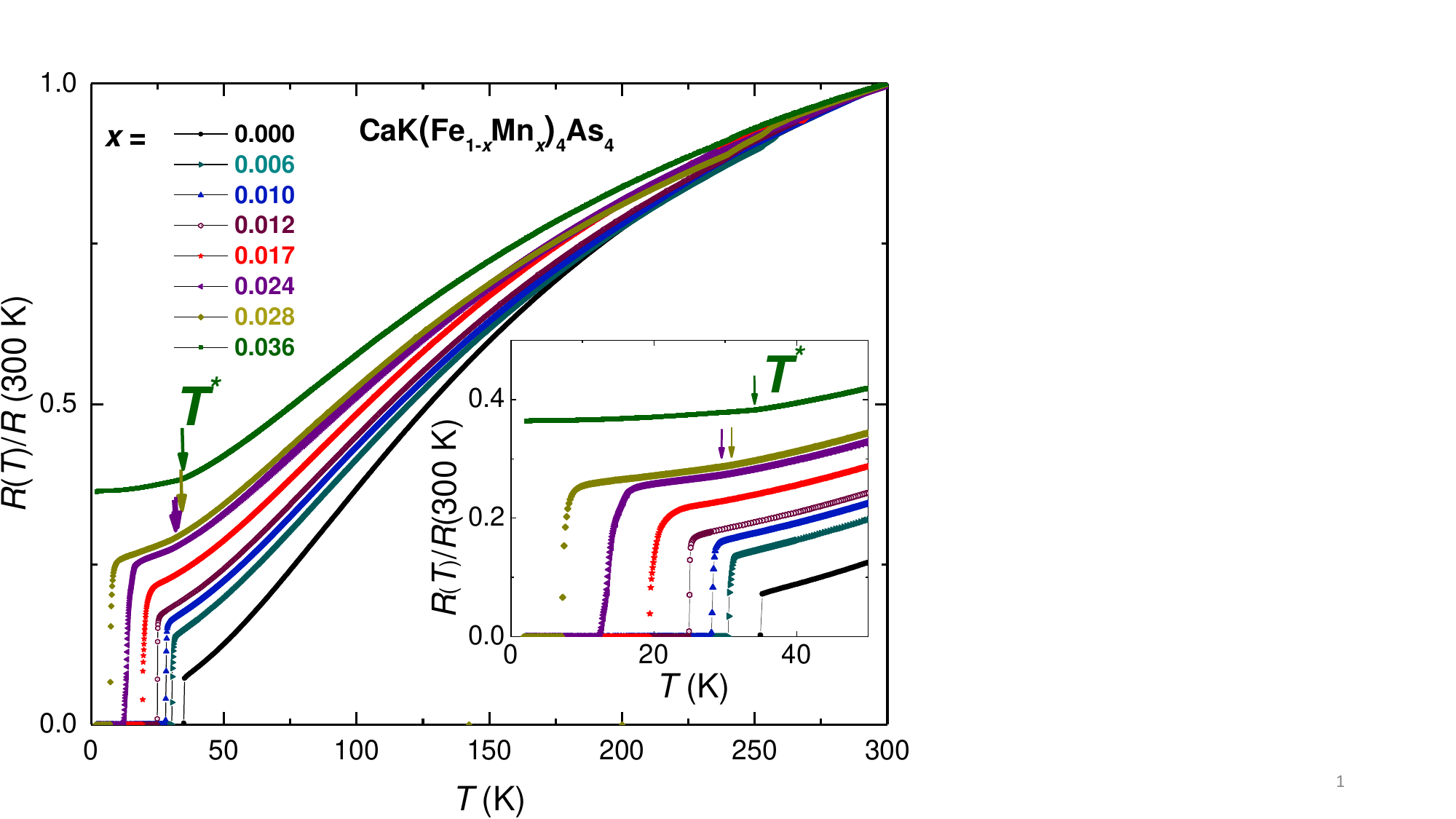}		
	\caption{Temperature dependence of normalized resistance, $\textit{R}(\textit{T})/\textit{R}$(300~K), of CaK(Fe$_{1-x}$Mn$_{x}$)$_{4}$As$_{4}$ single crystals showing the suppression of the superconducting transition $T_c$ and the appearance and evolution of a kink-like feature, marked with arrows. The criterion used to determine $T^*$ from this kink-feature is outlined and discussed in the appendix. \label{figure4}}
\end{figure}

\begin{figure*}
	\includegraphics[width=1.6\linewidth]{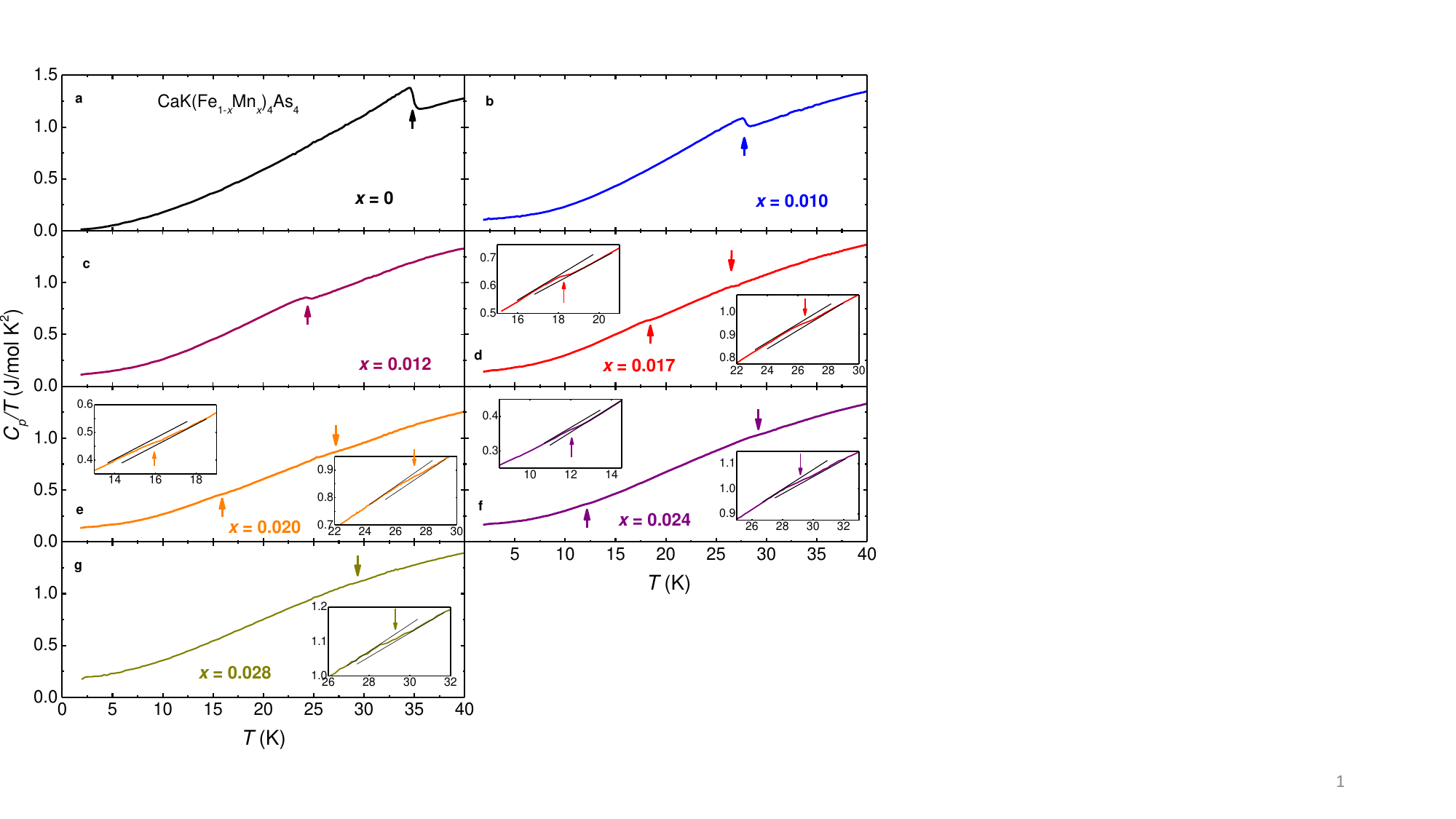}	
	\caption{Heat capacity of CaK(Fe$_{1-x}$Mn$_{x}$)$_{4}$As$_{4}$ single crystals, and criteria used to determine $T^*$ and $T_c$, show transition temperatures of $T_c$ (upward arrows) and $T^*$ (downward arrows) for different substitution levels. \label{figureHC1}}
\end{figure*}
Figure \ref{figure2} shows the low temperature (1.8~K - 45~K), zero-field-cooled-warming (ZFCW) magnetization for CaK(Fe$_{1-x}$Mn$_{x}$)$_{4}$As$_{4}$ single crystals for $H_{|| ab}$ = 50 Oe (Field-cooled-warming (FCW) and ZFCW magnetization data for an $x$ = 0.016 sample can be found in figure \ref{50OeFCZFCz} in the Appendix). $M$ is the volumetric magnetic moment in this figure and is calculated by using the density of CaKFe$_4$As$_4$, which is determined to be 5.22~g/cm$^3$ from lattice parameters at room temperature {\color{blue}\cite{Yoshida2016}}. A magnetic field of 50~Oe was applied parallel to \textit{ab} plane (i.e. parallel to the surface of the plate-like crystal). The superconducting transitions ($T_c$) are clearly seen in this graph except for the highest substitution value \textit{x} = 0.036. As the value of the Mn substitution, $x$, increases, the superconducting transition temperature decreases. For $x=0.028$, a full magnetic shielding is not reached by 1.8 K.

Figure \ref{figure3} shows the low temperature (5~K - 100~K) $M$($T$)/$H$ data for CaK(Fe$_{1-x}$Mn$_{x}$)$_{4}$As$_{4}$ single crystals with 10~kOe field applied parallel to the crystallographic \textit{ab} plane. There is an appearance of a Curie-Weiss tail after adding Mn and kink-like features found around 30~K for $x>0.010$, which may indicate new phase transition ($T^*$). The inset shows $M$($T$)/$H$ of a CaK(Fe$_{0.976}$Mn$_{0.024}$)$_{4}$As$_{4}$ single crystal over a wider temperature range. As Mn is added the Curie-tail-like feature grows. Although the Mn doping levels are low ($x$ < 0.05), the $M$($T$) data can be fit by a C/($T$+$\theta$) + $\chi_{0}$ function when we take $\theta$ to be equal to $T^{*}$, or its extrapolation by linear fitting $T^*$ from figure \ref{figure11} to $x$ = 0.006 for low $x$, where $T^*$ is absent (see table \ref{Table1} in the Appendix). The effective moment calculated per Mn is found to be $\sim$ 5 $\mu$B. This means that the substitution of Mn brings more local-moment-like behavior to this system. For $x$ > 0.016, a kink-like feature can be seen at a temperature $T^*$. As $x$ increases from 0.016 to 0.036, the temperature $T^*$  increases from $\sim$ 30 K to $\sim$ 35 K. The criterion for determining $T^*$ is shown in Fig. \ref{figure92} in the Appendix.

Figure \ref{figure4} presents the temperature dependent, normalized, electrical resistance of CaK(Fe$_{1-x}$Mn$_{x}$)$_{4}$As$_{4}$ single crystals. RRR (the ratio of $R$(300 K) and resistance before $T_c$) decreases as Mn substitution increases, which indicates disorder increasing. The superconducting transition temperatures decrease as Mn is added to the system. when \textit{x} = 0.036, there is no signature of a superconducting transition detectable above 1.8 K. With increasing Mn content, a kink appears for $x$ > 0.017 and rises to about 33~K for $x$ = 0.036. A similar feature also appeared in Ni and Co-substituted CaKFe$_{4}$As$_{4}$ electrical resistance measurements {\color{blue}\cite{Meier2018}}. The criterion for determining the transition temperature, $T^*$, associated with this kink is shown in the appendix in Fig. \ref{figure92}c where $R(T)$ and d$R(T)$/d$T$ are both shown.

Figure \ref{figureHC1} presents the temperature dependent specific heat data divided by temperature for $T$ < 45 K.  Whereas the feature associated with superconductivity (centered around $T_c$) is relatively clear (marked by upward pointing arrows), at least for lower $x$-values, the feature associated with $T^*$ (marked by downward pointing arrows) is quite subtle and is shown more clearly in the insets. Transition temperature is inferred from the mid-point of the feature. As has been already seen in the magnetization as well as resistance data, the addition of Mn (increase of $x$-value) leads to a suppression of $T_c$ as well as an onset and gradual increase of $T^*$ above a threshold value of $x$ $\sim$ 0.017.

Figure \ref{figure11} summarizes the transition temperature results inferred from magnetization, resistance and specific heat measurements, plotting the superconducting and magnetic transitions as a function of substitution, constructing the $T$-$x$ phase diagram for the CaK(Fe$_{1-x}$Mn$_{x}$)$_{4}$As$_{4}$ system. As depicted in this phase diagram, increasing Mn substitution (i) suppresses $T_c$ monotonically with it extrapolating to 0 K by $x$ $\lesssim$ 0.036 and (ii) stabilizes a new transition, presumably an antiferromagnetic one, for $x$ $\gtrsim$ 0.016 with the  transition temperature rising from $\sim$ 26 K for $x$ = 0.017 to $\sim$ 33 K for $x$ = 0.036. Each phase line is made out of data points inferred from \textit{R}(\textit{T}), \textit{M}(\textit{T}) and \textit{$C_{p}$}(\textit{T}) measurements, illustrating the good agreement between our criterion for inferring $T_{c}$ and $T^{*}$. The CaK(Fe$_{1-x}$Co$_{x}$)$_{4}$As$_{4}$ and CaK(Fe$_{1-x}$Ni$_{x}$)$_{4}$As$_{4}$ series have qualitatively similar phase diagrams, with the quantitative differences being associated with the doping levels necessary to induce the magnetic phase and to suppress superconductivity. Further comparisons of the CaK(Fe$_{1-x}$Mn$_{x}$)$_{4}$As$_{4}$ phase diagram to the phase diagrams of the CaK(Fe$_{1-x}$Co$_{x}$)$_{4}$As$_{4}$ and CaK(Fe$_{1-x}$Ni$_{x}$)$_{4}$As$_{4}$ series, as well as doped BaFe$_2$As$_2$ series will be made in the discussion section below.

Given that the $R$($T$) and $C_p$($T$) data were taken in zero applied field whereas the $M/H$($T$) data shown in Fig. \ref{figure3} were taken in 10 kOe, it is prudent to examine the field dependence of transition associated with $T^*$. In figure \ref{figure15} we show the d($MT/H$)/d$T$ data {\color{blue}\cite{MFisher1962}} for the $x$ = 0.028 sample for $H$ || \textit{ab} = 10, 30 and 50 kOe. As is commonly seen for antiferromagnetic phase transition, increasing a magnetic field leads to a monotonic suppression of $T^{*}$. The inset to Fig. \ref{figure15} shows that the extrapolated, $H$ = 0, $T^*$ value would be 30.7 K as compared to the value of 30.4 K for 10 kOe. This further confirms that there should be (and is) good agreement between the $T^{*}$ values inferred from 10 kOe magnetization data and the $T^{*}$ values inferred from the zero field specific heat and resistance data in Fig. \ref{figure11}. In addition these data suggest that magnetic field could be used to fine tune the value of $T^*$, if needed.

\begin{figure}[H]
	\includegraphics[width=1.5\columnwidth]{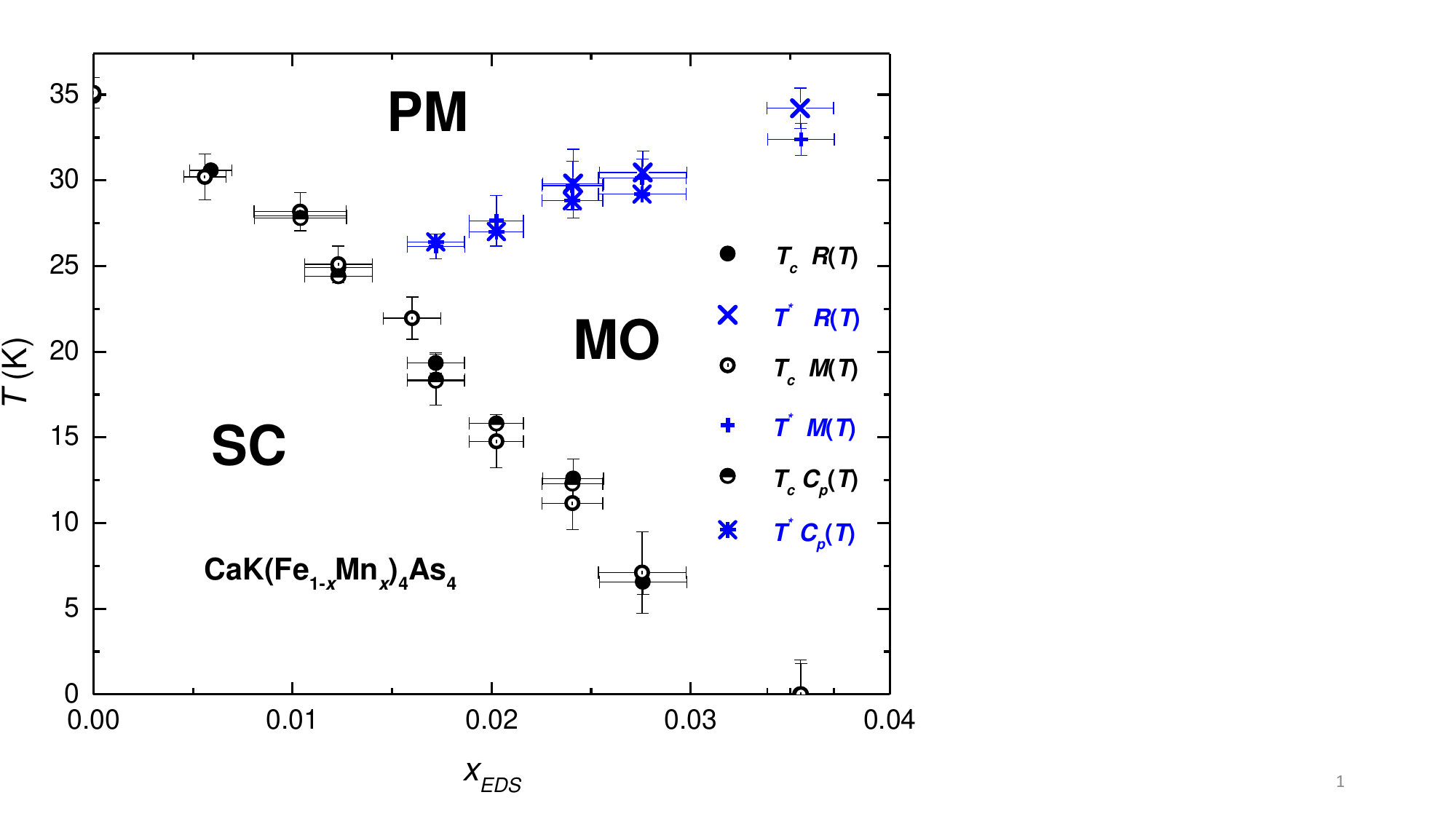}		
	\caption{Temperature - composition phase diagram of CaK(Fe$_{1-x}$Mn$_{x}$)$_{4}$As$_{4}$ single crystals as determined from resistance ($R$($T$)), magnetization ($M$($T$)) and specific heat ($C_p$($T$)) measurements. The circular symbols denote the $T_{c}$ phase line and the cross-like symbols denote the $T^{*}$ phase line, most likely associated with antiferromagnetic order. Superconducting (SC), magnetically ordered (MO) and paramagnetic (PM) regions are denoted. Details of how the MO line extends into the SC state are not addressed in this phase diagram.    \label{figure11}}
\end{figure}

\begin{figure}[H]
	\includegraphics[width=1.5\columnwidth]{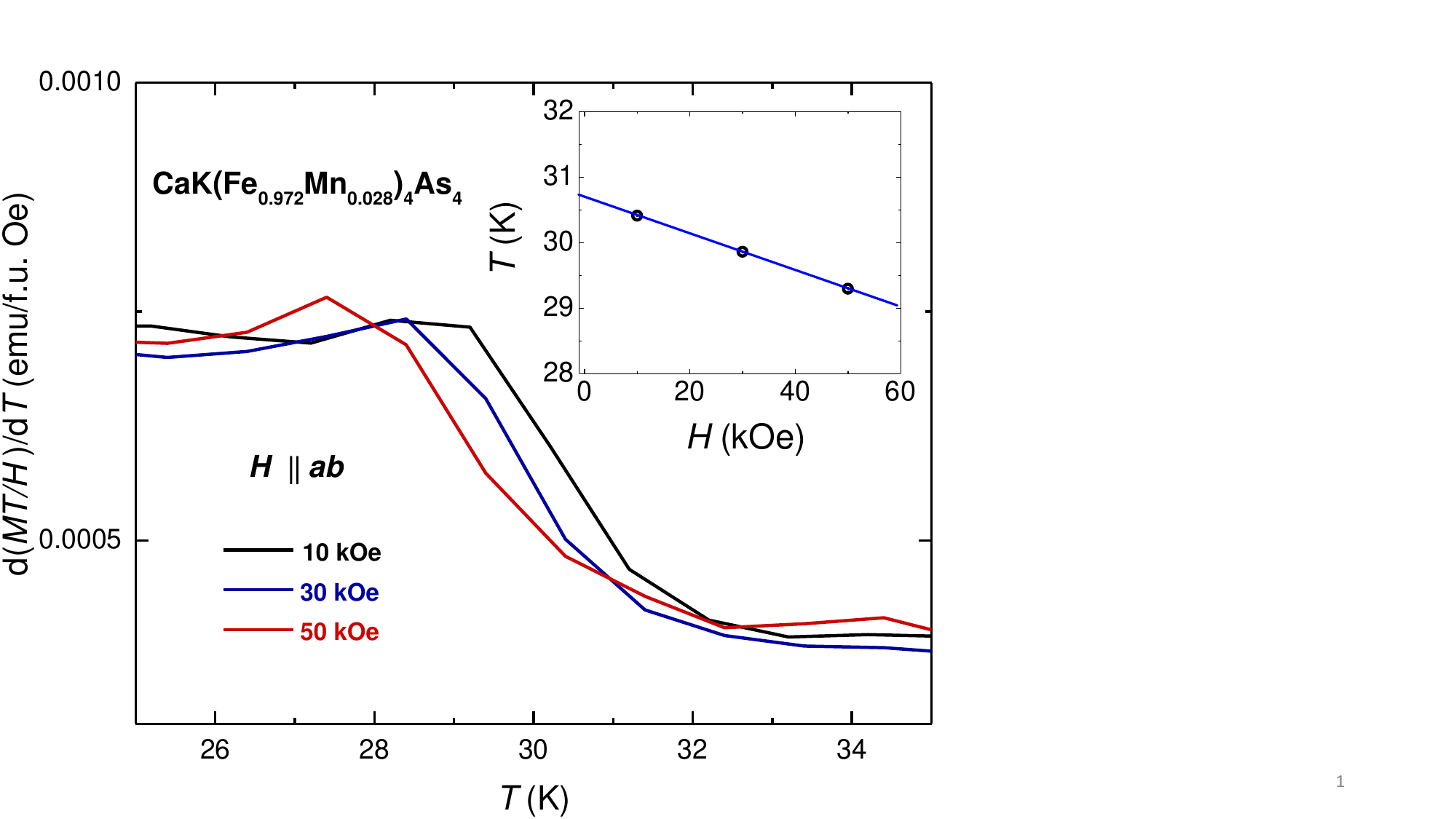}		
	\caption{d($MT/H$)/d$T$ vs. $T$ of CaK(Fe$_{0.972}$Mn$_{0.028}$)$_{4}$As$_{4}$ single crystal with 10 kOe, 30 kOe and 50 kOe applied parallel to the crystallographic \textit{ab} plane. Inset shows transition temperature, $T^*$, inferred for different applied field values using the same criterion shown in appendix. The solid blue line is linear fit to the data points, extrapolating to 30.7 K for $H$ = 0.   \label{figure15}}
\end{figure}

\section{Superconducting critical fields and anisotropy }

 \begin{figure}[H]
 	\includegraphics[width=1.5\columnwidth]{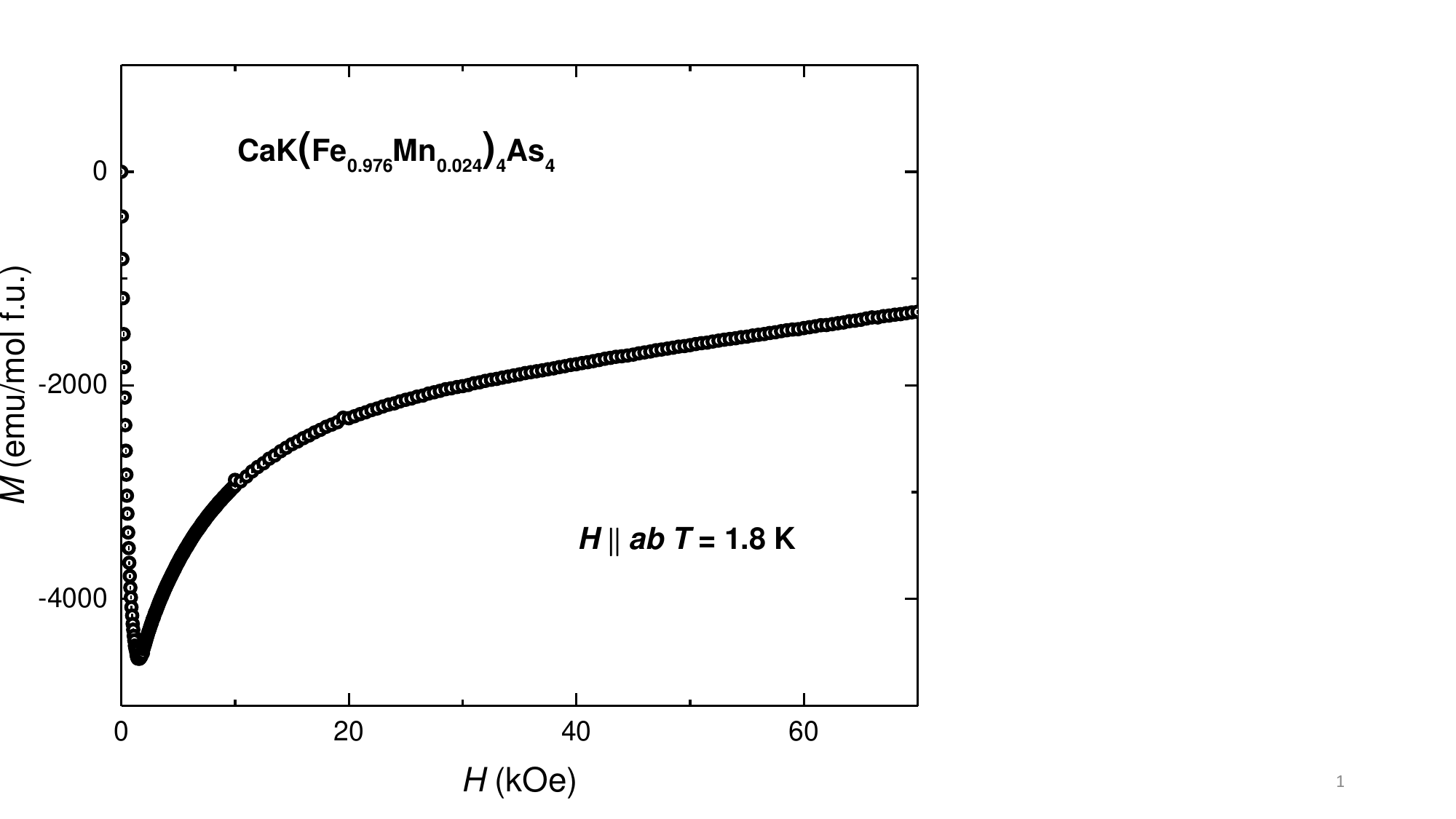}		
 	\caption{Magnetization of a single crystal of CaK(Fe$_{0.976}$Mn$_{0.024}$)$_{4}$As$_{4}$ as a function of magnetic field applied parallel to the crystallographic \textit{ab} plane at 1.8 K. Temperature is ZFC to 1.8 K and demagnetization is done at 60 K before cooling to minimize the remnant magnetic field.     \label{figure6}}
 \end{figure}

Superconductivity can be studied as a function of field (in addition to temperature and doping). Before we present our $H$$_{c2}$($T$) results, based on $R$($T,H$) data, it is useful to check the $M$($H$) data. We start with \textit{M}(\textit{H}) data for $x$ = 0.024, $T^*$ = 29.7 K and $T_c$ = 9.9 K, taken over a wide field range. The 1.8 K \textit{M}(\textit{H}) data for shown in Fig. \ref{figure6} is classically non-linear, showing a local minimum near $H$ $\sim$ 2.3 kOe. For $T$ = 1.8 K < $T_{c}$ the $H_{c2}$ value is clearly higher than the 65 kOe maximum field we applied. $H_{c1}$ can be inferred from the lower field \textit{M}(\textit{H}) data.

\begin{figure}[H]
	\centering
	\begin{minipage}{0.44\textwidth}
		\centering
		\includegraphics[width=1.5\columnwidth]{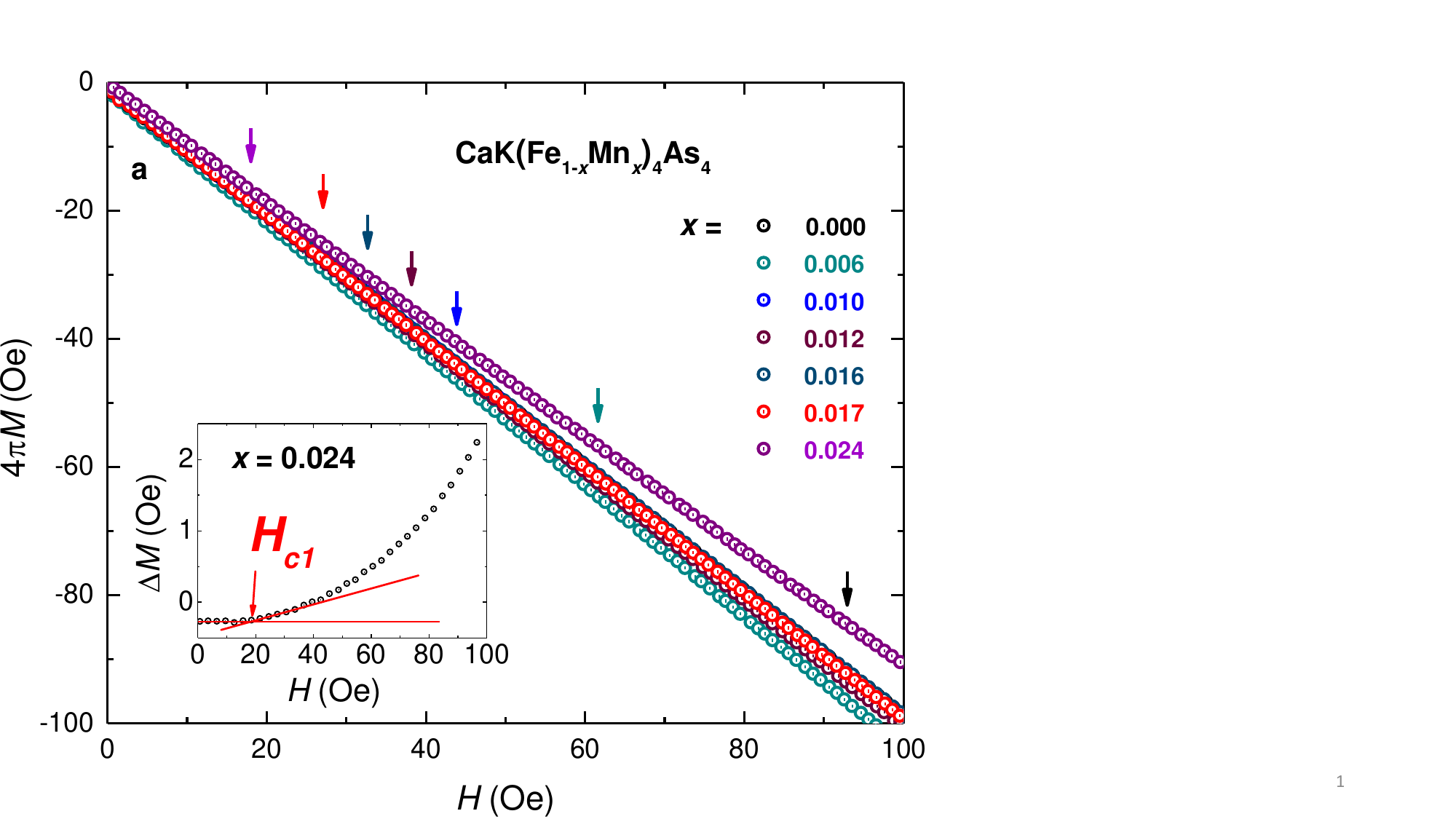}		
	\end{minipage}\hfill
	\centering
	\begin{minipage}{0.44\textwidth}
		\centering
		\includegraphics[width=1.5\columnwidth]{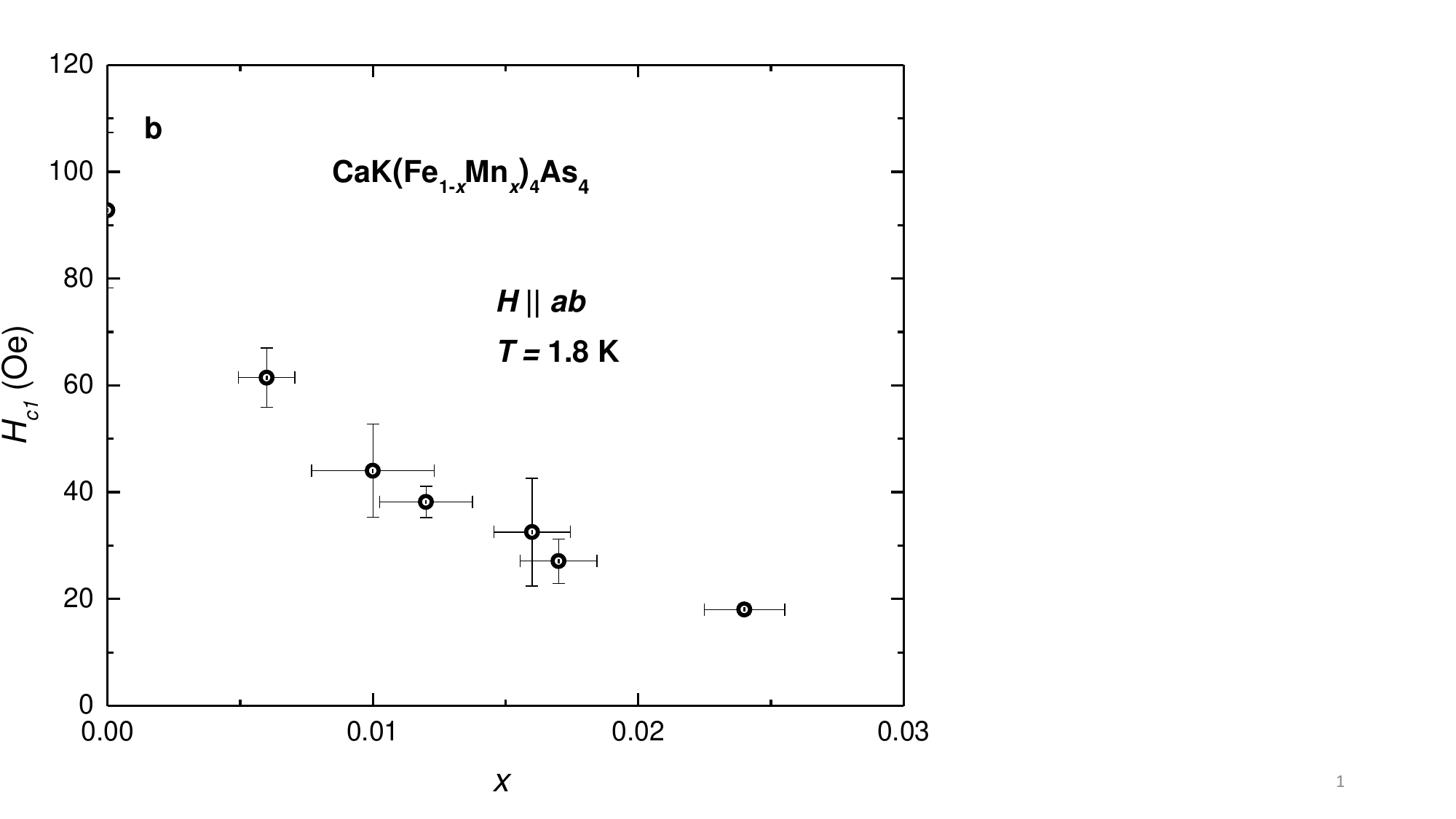}		
	\end{minipage}\hfill
	\caption{(a) magnetization as a function of magnetic field applied parallel to the crystallographic \textit{ab} axis at 1.8 K for different substitution levels. Arrows mark the value of the magnetic field ($H_{c1}$) where $M$(\textit{H}) deviates from linear behavior. Inset shows the criterion we use to determine the $H_{c1}$ values. Remnant field of measurements are smaller than 1 Oe, which is consistent with $M$($H$) plots shown in figure. Temperature is ZFC to 1.8 K and demagnetization is done at 60 K before cooling to minimize the remnant magnetic field (b) $H_{c1}$ value versus $x$.}\label{HC12sum}
\end{figure}

In order to better estimate $H_{c1}$ values we performed low field, $M(H)$ sweeps at base temperature. In Fig. \ref{HC12sum}a we show the \textit{M}(\textit{H}) data for 0 $\leqslant$ $x$ $\leqslant$ 0.024 for $H$ $\leqslant$ 100 Oe. As $x$ increases the deviation from the fully diamagnetic, linear behavior, that occurs at $H_{c1}$, appears at lower and lower fields. As shown in inset of figure \ref{HC12sum}a, $\Delta$$M$ is determined by subtracting the linear, lowest field behavior of $4\pi M$ from $H$. Because of the finite thickness of samples, even magnetic field is apply in $ab$ plane, there is a small demagnetizing factor ($n$ < 0.077). Therefore, $H_c1$ is taken as the vortices start to enter the sample and determined when deviates from the low field, the no-zero value is due to remnant field of MPMS. The standard error of $H_{c1}$ comes from at least 4 different samples' measurements. We should note that using this method with higher data density we were able to identify $H_{c1}$ for the pure CaKFe$_{4}$As$_{4}$ (x = 0) as $\sim$ 100 Oe as opposed to the $\sim$ 1.3 kOe estimated in Ref. {\color{blue}\cite{Meier2016}}. Figure \ref{HC12sum}b shows a monotonic decrease in $H_{c1}$ as Mn substitution increases and $T_{c}$ decreases.

\begin{figure}[H]
	\centering
	\begin{minipage}{0.44\textwidth}
		\centering
		\includegraphics[width=3\columnwidth]{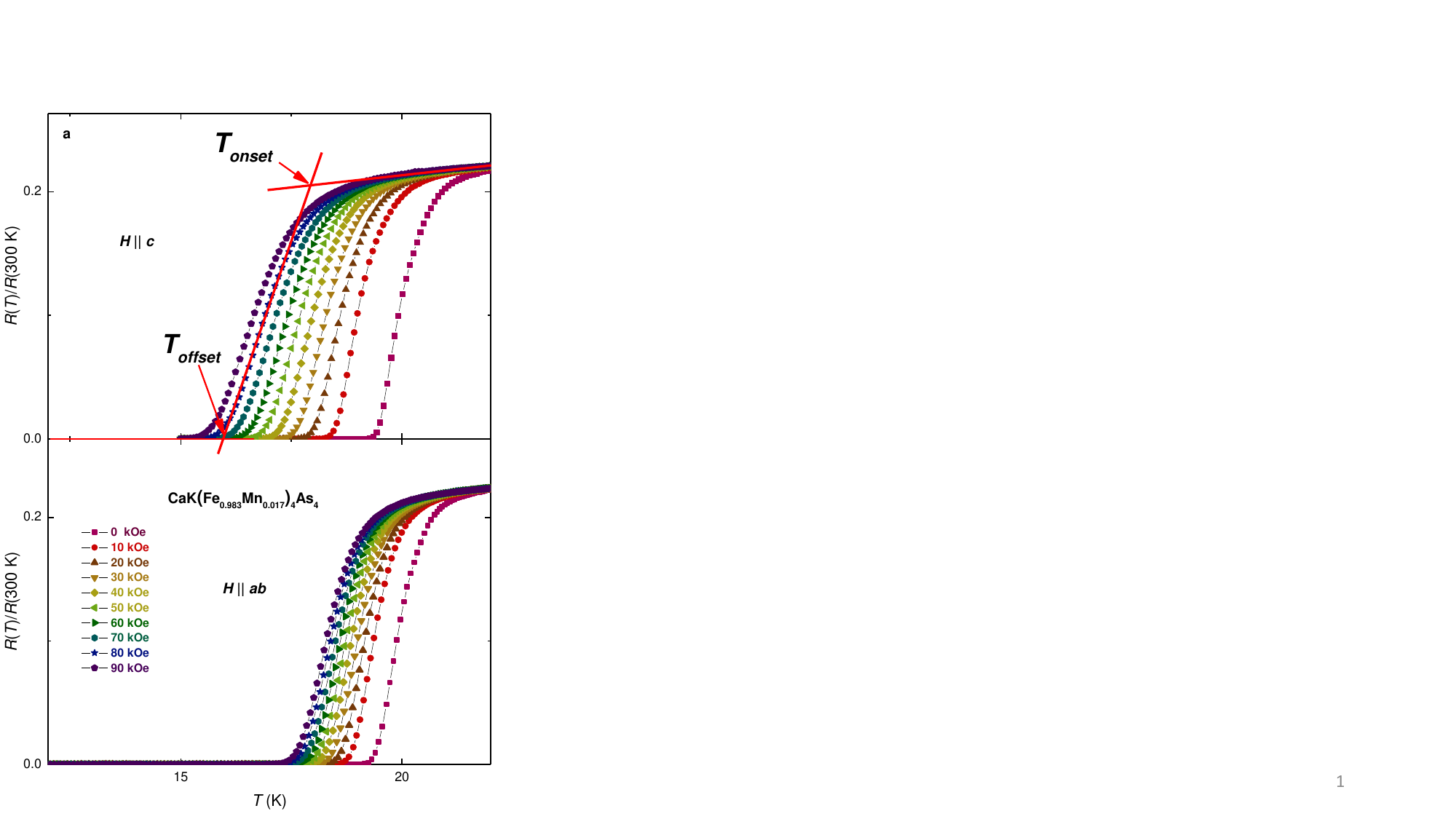}		
	\end{minipage}\hfill
	\centering
	\begin{minipage}{0.44\textwidth}
		\centering
		\includegraphics[width=1.5\columnwidth]{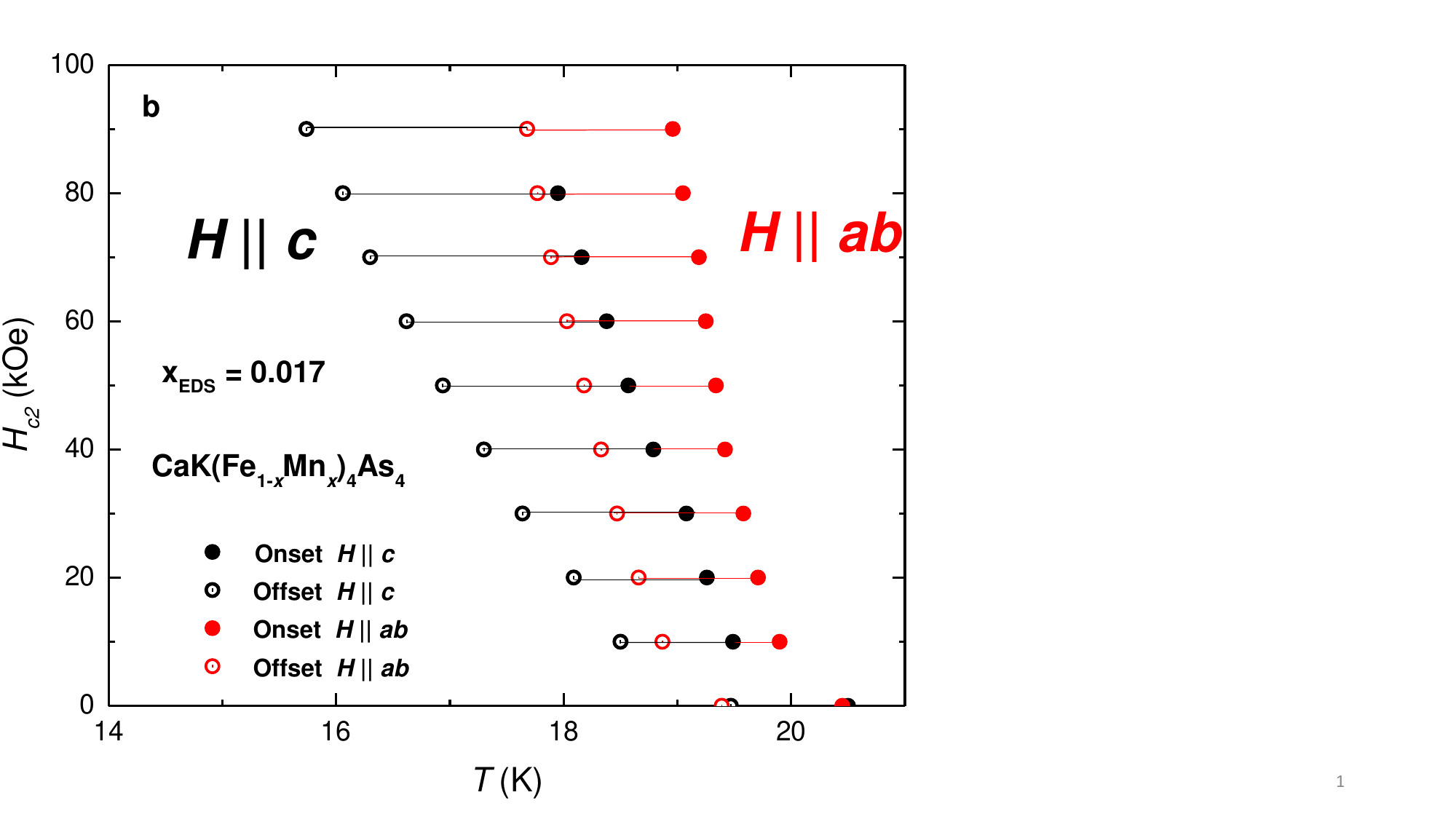}		
	\end{minipage}\hfill
	\caption{(a) Temperature-dependent electrical resistance of CaK(Fe$_{0.983}$Mn$_{0.017}$)$_{4}$As$_{4}$ single crystal for magnetic field parallel to the crystallographic \textit{c} axis (upper panel) and $ab$ plane (lower panel) for representative fields $\textit{H} \leqq 90$ kOe. Onset and offset criteria for $T_c$ are shown by the red solid lines. (b) Anisotropic $H_{c2}(T)$ data determined for two single crystalline samples of CaK(Fe$_{0.983}$Mn$_{0.017}$)$_{4}$As$_{4}$ using onset criterion (solid) and offset criterion (hollow) inferred from the data shown in (a).}\label{figure7}
\end{figure}

In order to further study the effects of Mn substitution on the superconducting state, anisotropic $H_{c2}(T)$ data for temperatures near $T_{c}$ were determined for the substitution levels that have superconductivity. Figure \ref{figure7}a shows a representative set of $R$($T$) data taken for fixed applied magnetic fields, $H$ $\parallel$ $c$ axis and $ab$ plane $\leqq$ 90 kOe for $x$ = 0.017. Figure \ref{figure7}a also shows an example of the onset and offset criteria used for the evaluation of $T_c$. Figure \ref{figure7}b present the $H_{c2}(T)$ curves for CaK(Fe$_{0.983}$Mn$_{0.017}$)$_{4}$As$_{4}$ single crystals with both $H$ || $c$ and $H$ || $ab$, showing both the onset (T$_{onset}$) and offset (T$_{offset}$) temperatures. For $x$ = 0.006, 0.010, 0.012, 0.024 and 0.028, $H_{c2}(T)$ curves are shown in Appendix figs. \ref{figure81} - \ref{figure86}. From $H_{c2}(T)$ plots, we can see that $T_{c}$ is only suppressed by about 3~K when 90~kOe magnetic field is applied, so the complete $H_{c2}(T)$ plots of the CaK(Fe$_{1-x}$Mn$_x$)$_4$As$_4$ compounds cannot be fully determined, however we still can observed several trends in these data.

\begin{figure}
 	\includegraphics[width=1.4\columnwidth]{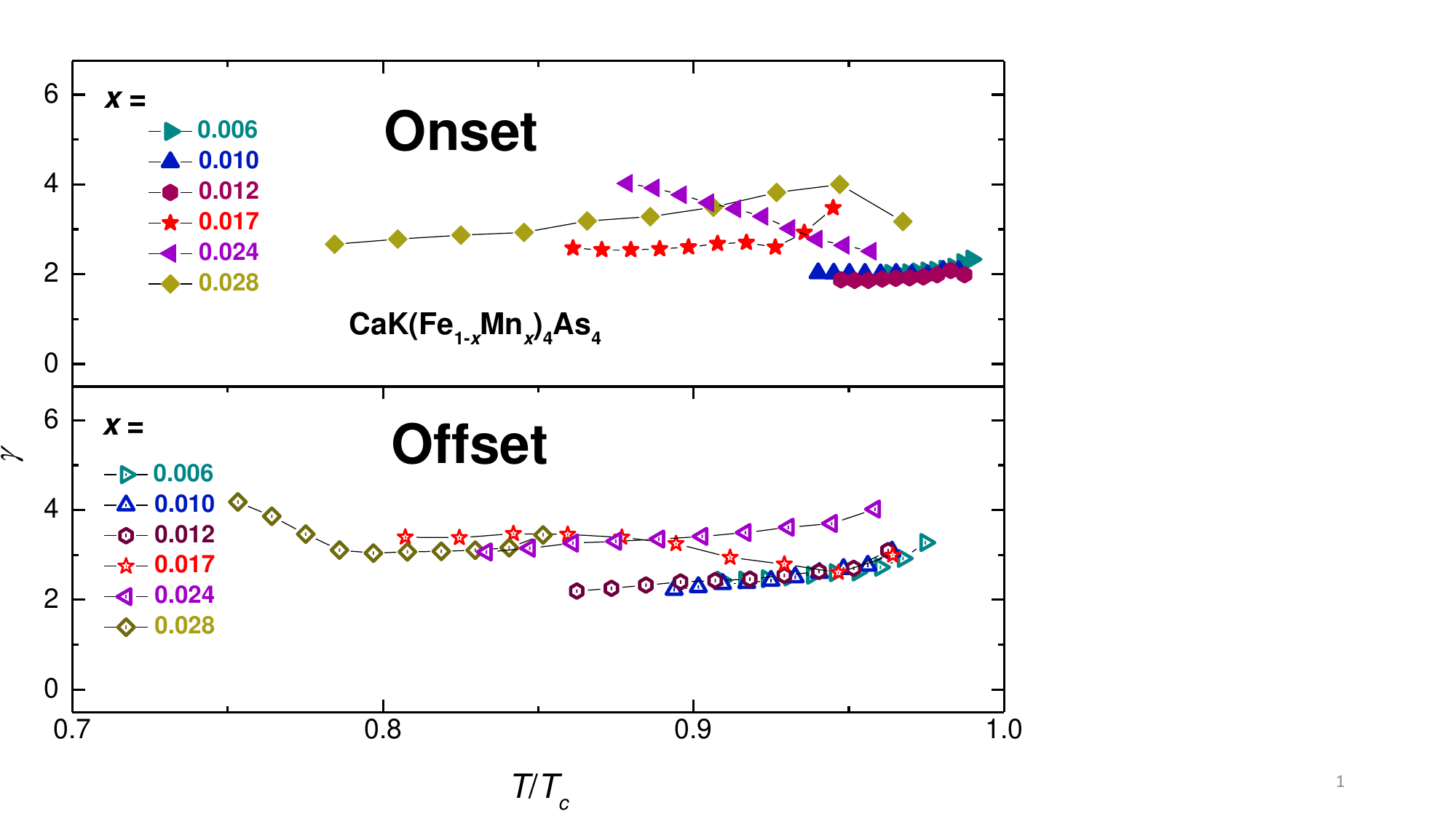}	
 	\caption{Anisotropy of the upper critical field, $\gamma=H_{c2}^{\parallel ab}(T)/H_{c2}^{\parallel c}(T)$, as a function of effective temperature, $T$/$T_{c}$, for CaK(Fe$_{1-x}$Mn$_{x}$)$_{4}$As$_{4}$ single crystals, using onset criterion (upper panel) and offset criterion (lower panel), inferred from the temperature-dependent electrical resistance data. The $T_{c}$ value used to calculate the effective temperature (\textit{T}/$T_{c}$) is the zero-field superconductivity transition temperature for each Mn-substitution levels (see fig. \ref{figure11} above).  \label{figure12}}
 \end{figure}

\begin{figure}[H]
	\includegraphics[width=1.5\columnwidth]{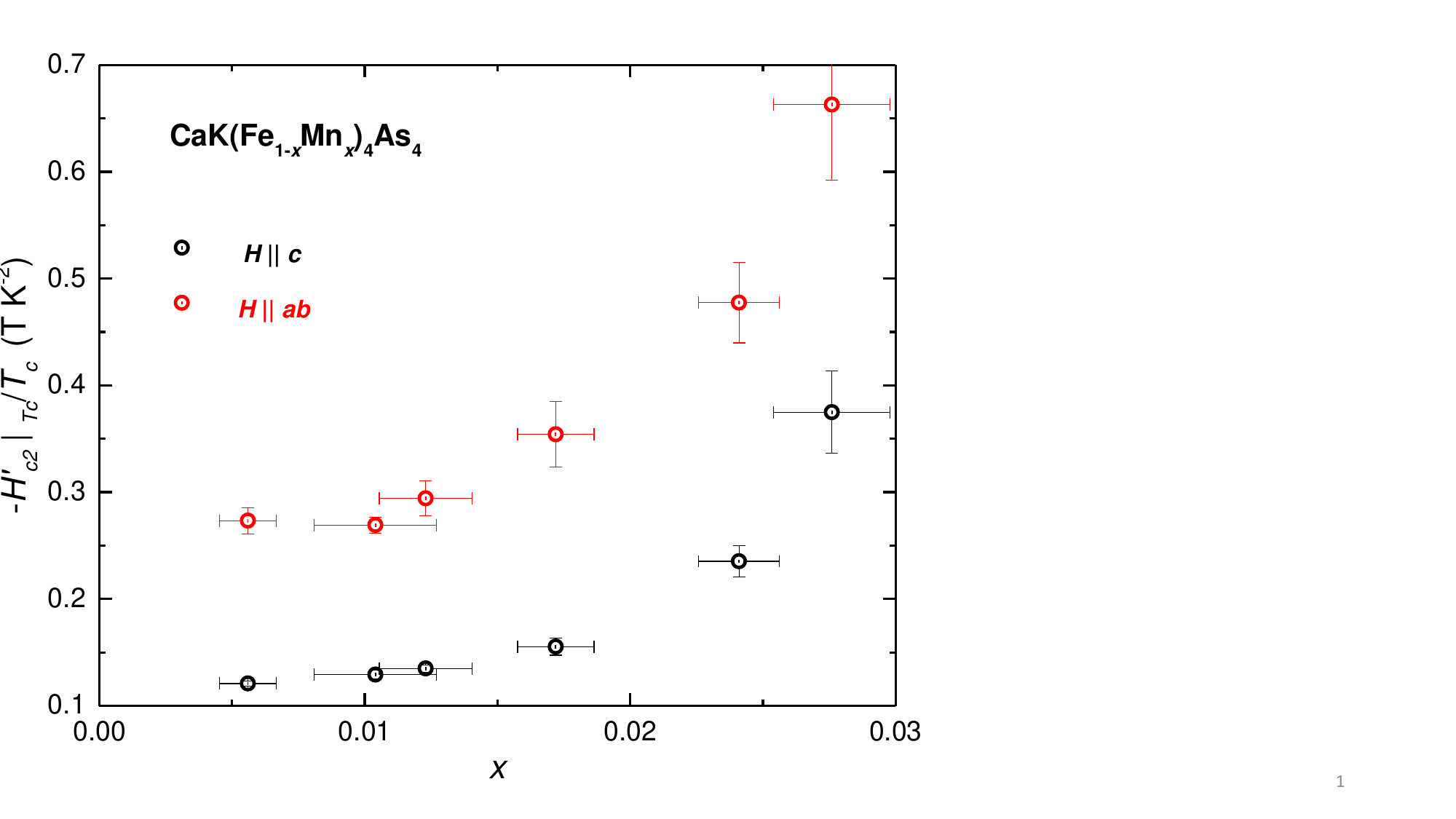}			
	\caption{Substitution dependence of the upper critical field slope -$H^\prime_{c2}$($T$)/$T_c$, where $H^\prime_{c2}$($T$) is d$H_{c2}$($T$)/d$T$. $T_c$ is determined by $T_{offset}$ from criteria shown in figure \ref{figure7}. Qualitatively similar, albeit somewhat weaker, results can be seen using $T_{onset}$ criterion data. \label{figureHc2dt}}
\end{figure}

Figure \ref{figure12} shows the temperature dependent anisotropy ratio, $\gamma=H_{c2}^{\parallel ab}(T)/H_{c2}^{\parallel c}(T)$ $\sim$ 3 for these samples over the 0.8 < $T$/$T_c$ < 1.0 range. This value is  similar to other 122 and 1144 materials {\color{blue}\cite{Ni20101,Meier2016}}.  
Given that we have determined $H_{c2}$($T$) for temperatures close to $T_c$, we can evaluate the $H^\prime_{c2}$($T$)/$T_c$ close to $T_{c}$, where $H^\prime_{c2}$($T$) is d$H_{c2}$($T$)/d$T$, specifically seeing how it changes as $T_c$ drops below $T^*$ with increasing x. Error of $H^\prime_{c2}$($T$)/$T_c$ comes from linear fit of  $H_{c2}$($T$) near the $T_c$. In the case of other Fe-based systems {\color{blue}\cite{Xiang2018,Kaluarachchi2016,Xiang2017,Taufour2014}} clear changes in $H^\prime_{c2}$($T$)/$T_c$ were associated with changes in the magnetic sublattice coexisting with superconductivity (i.e. ordered or disordered). In Fig. \ref{figureHc2dt} we can see that there is a change in the x-dependence of $H^\prime_{c2}$($T$)/$T_c$ for $x$ $>$ 0.015, beyond which substitution level suppresses $T_c$ below $T^*$. Comparison with the slope change of $H_{c2}$ in the pressure-temperature phase diagram of CaK(Fe$_{1-x}$Ni$_{x}$)$_4$As$_4$ {\color{blue}\cite{Xiang2018}}, further suggest that this is probably related to changes in the Fermi surface, caused by the onset of the new periodicity associated with the AFM order.


\section{Elastoresistance }

\begin{figure}[H]
	\includegraphics[width=2.4\columnwidth]{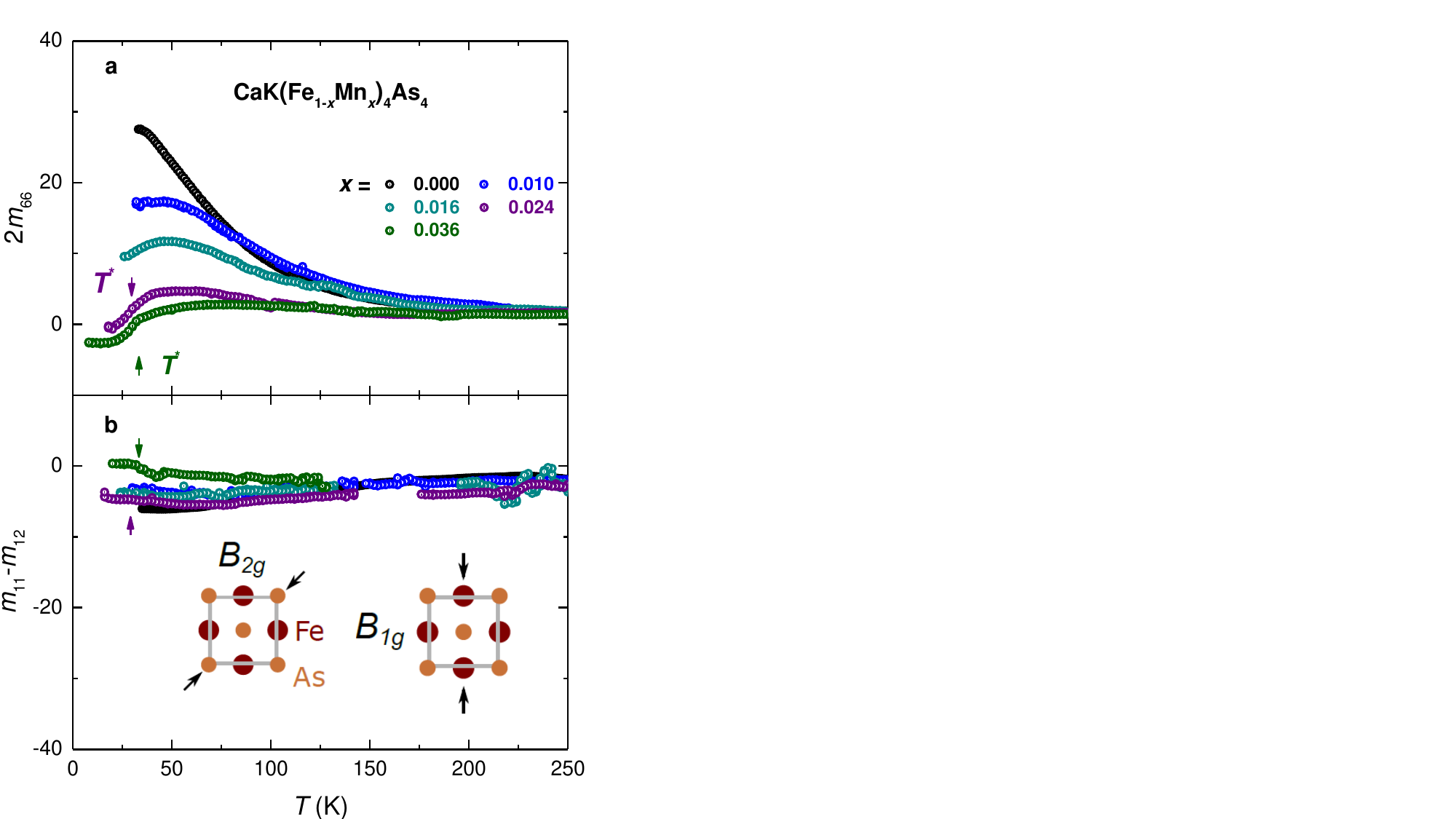} 
	\caption{Elastoresistance coefficients, 2$m_{66}$ (a) and $m_{11}-m_{12}$ (b), of CaK(Fe$_{1-x}$Mn$_x$)$_4$As$_4$ for $0\,\leq\,x\,\leq\,0.036$ as a function of temperature, $T$. Data for $x\,=\,0$ was taken from Ref.\,\cite{Meier2016}. For each $x$, data were taken down to the superconducting transition temperature $T_c$. Arrows mark the position of the magnetic transition, $T^*$, whenever present, taken from Fig. \ref{figure11}. The insets in (b) visualize the $B_{2g}$ and $B_{1g}$ strains that are experienced by the sample with respect to the FeAs plane.}
	\label{fig:elastoresistance}
\end{figure}

In order to investigate the evolution of nematic fluctuations in CaK(Fe$_{1-x}$Mn$_x$)$_4$As$_4$ and compare it with Ni-substituted CaKFe$_4$As$_4$ as well as Co-substituted BaFe$_2$As$_2$, we measured elastoresistance. Figure \ref{fig:elastoresistance} shows the temperature dependence of the elastoresistance coefficients $2m_{66}$ and $m_{11}-m_{12}$ for Mn-1144 samples. These quantities were determined from the strain-induced resistance change $\Delta R/R$ \cite{Kuo2013} via

\begin{align}
	2m_{66} &=& (\Delta R/R)_{xx} - (\Delta R/R)_{yy} \\ &&\textnormal{ for } \epsilon_{xx},\epsilon_{yy}\parallel [110]_T \\
	m_{11}-m_{12} &=& (\Delta R/R)_{xx} - (\Delta R/R)_{yy} \\ &&\textnormal{ for } \epsilon_{xx},\epsilon_{yy}\parallel [100]_T .
\end{align}

Thus, $2m_{66}$ and $m_{11}-m_{12}$ reflect the elastoresistance coefficients associated with $B_{2g}$ and $B_{1g}$ strain, respectively. For each temperature measured, we found that $(\Delta R/R)_{xx}$ has the opposite sign of $(\Delta R/R)_{yy}$, indicating that the $B_{2g}$- and $B_{1g}$-induced resistance changes are much larger than the isotropic $A_{1g}$-induced strain component\cite{Kuo2013}. Given that anisotropic strain is the conjugate field to $B_{2g}$ and $B_{1g}$ nematic order, the respective elastoresistance coefficients are often considered as a measure of the respective nematic susceptibilites {\color{blue}\cite{Chu2012}}. One should note, however, that the proportionality factor, between the elastoresistance coefficients and the nematic susceptibilities, $\chi_{nem,i}\,=\,k\,\cdot\,m_{i}$ with $m_i\,=\,2m_{66}$ or  ($m_{11}-m_{12}$), does not have to be constant. $k$ can, in principle, be temperature- and/or doping-dependent, as it depends on microscopic details of the Fermi surface and scattering mechanisms. Nonetheless, the observation of a diverging elastoresistance with Curie-Weiss-like temperature dependence in several Fe-based superconductors, which is in agreement with the theoretical expectations based on a Ginzburg-Landau ansatz and a constant $k$, can be taken as a good indication that elastoresistance measurements can be used to infer the strength of nematic fluctuations {\color{blue}\cite{Kuo2016,Chu2012}}.

As shown in Fig.\,\ref{fig:elastoresistance}, $2m_{66}$ of CaK(Fe$_{1-x}$Mn$_x$)$_4$As$_4$ is characterized by a marked temperature dependence for low $x$ and also changes significantly with doping, whereas $m_{11}-m_{12}$ is only weakly-temperature dependent and is essentially doping-independent. This suggests that nematic fluctuations, if present, occur only in the $B_{2g}$ channel. A closer look at $2m_{66}$, however, clearly suggests that nematic fluctuations are suppressed with increasing Mn substitution level. In fact, for $x\,=\,0.024$ and 0.036, i.e., those concentrations for which a clear signature of a magnetic transition is observed, the values of $2m_{66}$ across the entire temperature range are comparable to the ones of $m_{11}-m_{12}$. Thus, the magnetic order is very likely not characterized by a nematic component. In other words, a stripe-type magnetic order, as realized in many other pnictides and which would break tetragonal symmetry, can likely be ruled out based on this elastoresistance data. These data suggest that (i) for CaK(Fe$_{1-x}$Mn$_x$)$_4$As$_4$ the magnetic order may well be a spin-vortex crystal type of antiferromagnetism similar to what was found CaK(Fe$_{1-x}$Ni$_x$)$_4$As$_4$ {\color{blue}\cite{Meier2018}}, and (ii) there may be stripe-type magnetic and nematic fluctuations and susceptibility present in the undoped and weakly doped CaK(Fe$_{1-x}$Mn$_x$)$_4$As$_4$ but they are suppressed when the doping levels are increased.

Qualitatively, our elastoresistance results for CaK(Fe$_{1-x}$Mn$_x$)$_4$As$_4$ are similar to the elastoresistance results {\color{blue}\cite{Bohmer2020}} obtained for the Ni-doped CaK(Fe$_{1-x}$Ni$_x$)$_4$As$_4$, where Ni substitution stabilizes a spin-vortex crystal magnetic phase {\color{blue}\cite{Meier2018}}. This statement relates to the decrease of $2m_{66}$ with increasing $x$ as well as its complex temperature dependence which does not simply follow Curie-Weiss behavior. Whereas the former is clearly surprising, given that Ni- and Mn- substitutions lead to a different nominal electron counts, the latter might be related to the proportionality factor $k$ between $2m_{66}$ and $\chi_{nem}$ not being constant. Another possibility that could give rise to deviations from Curie-Weiss type behavior is the close proximity of competing magnetically-ordered phases, as pointed out from calculations of the nematic susceptibility in a phenomenological model in Ref.\,{\color{blue}\cite{Bohmer2020}}. A qualitative difference between the elastoresistance results on Mn- and Ni-substituted CaKFe$_4$As$_4$ is the sign of $2m_{66}$ for those $x$ where magnetic order is present. In our experiment, $2m_{66}$ is positive above $T_N$ across the entire range of $x$ studied, whereas for Ni-substituted samples a sign change of $2m_{66}$ from positive to negative was observed for Ni substitution values that are large enough for the magnetic transition to emerge. It is likely that the differences of the Fermi surface resulting from different electron counts affect the sign of $k$.

\section{discussion and summary}

The $T$-$x$ phase diagram for CaK(Fe$_{1-x}$Mn$_x$)$_4$As$_4$ is qualitatively similar to those found for Co- and Ni-substituted CaKFe$_4$As$_4$, there is a clear suppression of $T_{c}$ with increasing Mn substitution as well as an onset of what is likely to be a AFM ordering for $x$ > 0.015. For CaK(Fe$_{1-x}$Mn$_x$)$_4$As$_4$, elastoresistance measurements shows that $2m_{66}$ is small compare to the Fe-based superconductors showing stripe-type order. This means that the order is not strongly coupling to an orthorhombic distortion and is analogous to Ni-substituted CaKFe$_4$As$_4$, which suggests that AFM structure in Mn-substituted 1144 doesn't break $C_{4}$ symmetry and is still h-SVC. 

In figures \ref{HC12sum} and \ref{figureHc2dt} we presented measurements and analysis of $H_{c1}$ and $H_{c2}$ data. Whereas we do not see any clear effect of the onset of AFM ordering on $H_{c1}$ (Fig. \ref{HC12sum}b), there is a clear effect on $H_{c2}$ (Fig. \ref{figureHc2dt}). Using our $H_{c1}$ and $H_{c2}$ data we can extract information about the superconducting coherence length and London penetration depth as well.

\begin{figure}
	\includegraphics[width=1.5\columnwidth]{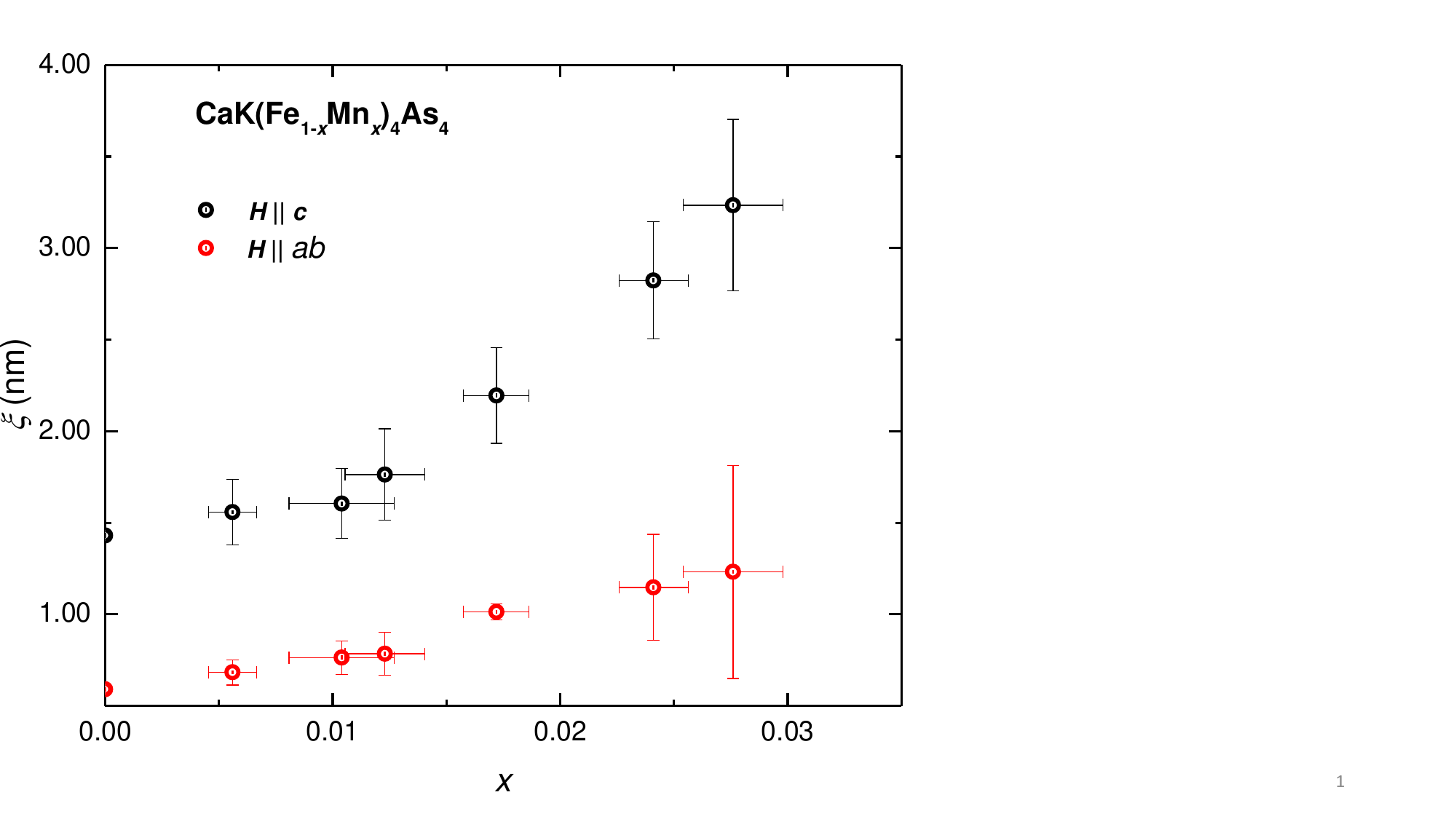}		
	\caption{Coherence length versus $x$ plot of CaK(Fe$_{1-x}$Mn$_x$)$_4$As$_4$ single crystals with applied field in $c$ direction and $ab$ plane.     \label{CoHL}}
\end{figure}

Figure~\ref{CoHL} shows coherence length, $\xi$, of CaK(Fe$_{1-x}$Mn$_x$)$_4$As$_4$ as a function of $x$. $\xi$ is estimated by using the anisotropic scaling relations |d$H_{c2}^{||c}$/d$T$|= $\phi_0$/2$\pi$$\xi^2_{\perp}$$T_c$ and |d$H_{c2}^{\perp c}$/d$T$|= $\phi_0$/2$\pi$$\xi_{||}\xi_{\perp}$$T_c$ {\color{blue}\cite{Meier2016}}. Coherence lengths increase as substitution levels increase, and, given that $\xi$ depends on d$H_{c2}$/d$T$, it is not surprising that a change in behavior of $\xi(x)$ occurs as the AFM state is entered.

\begin{figure}
	\centering
	\begin{minipage}{0.44\textwidth}
		\centering
		\includegraphics[width=1.5\columnwidth]{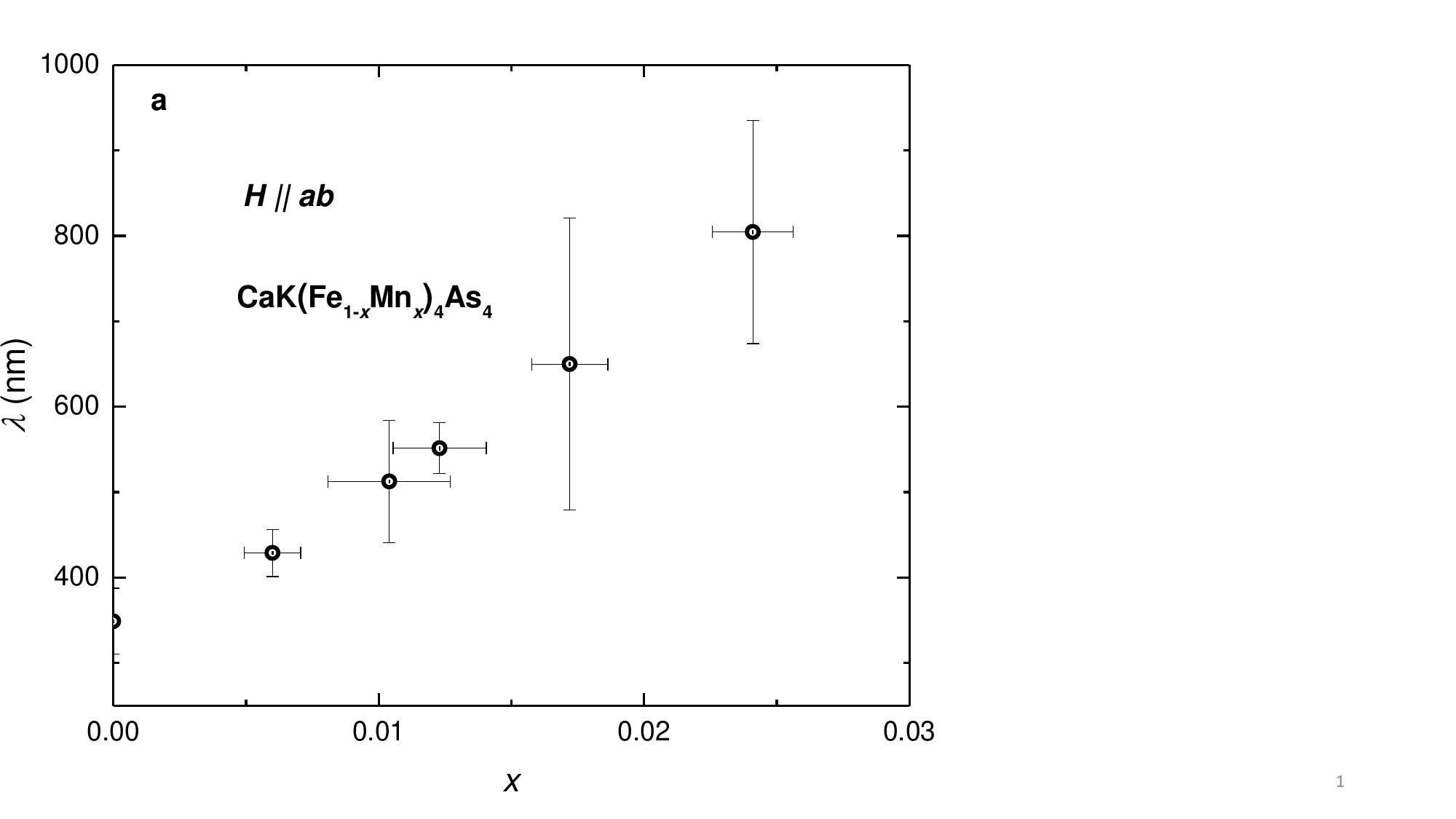}		
	\end{minipage}\hfill
	\centering
	\begin{minipage}{0.44\textwidth}
		\centering
		\includegraphics[width=1.5\columnwidth]{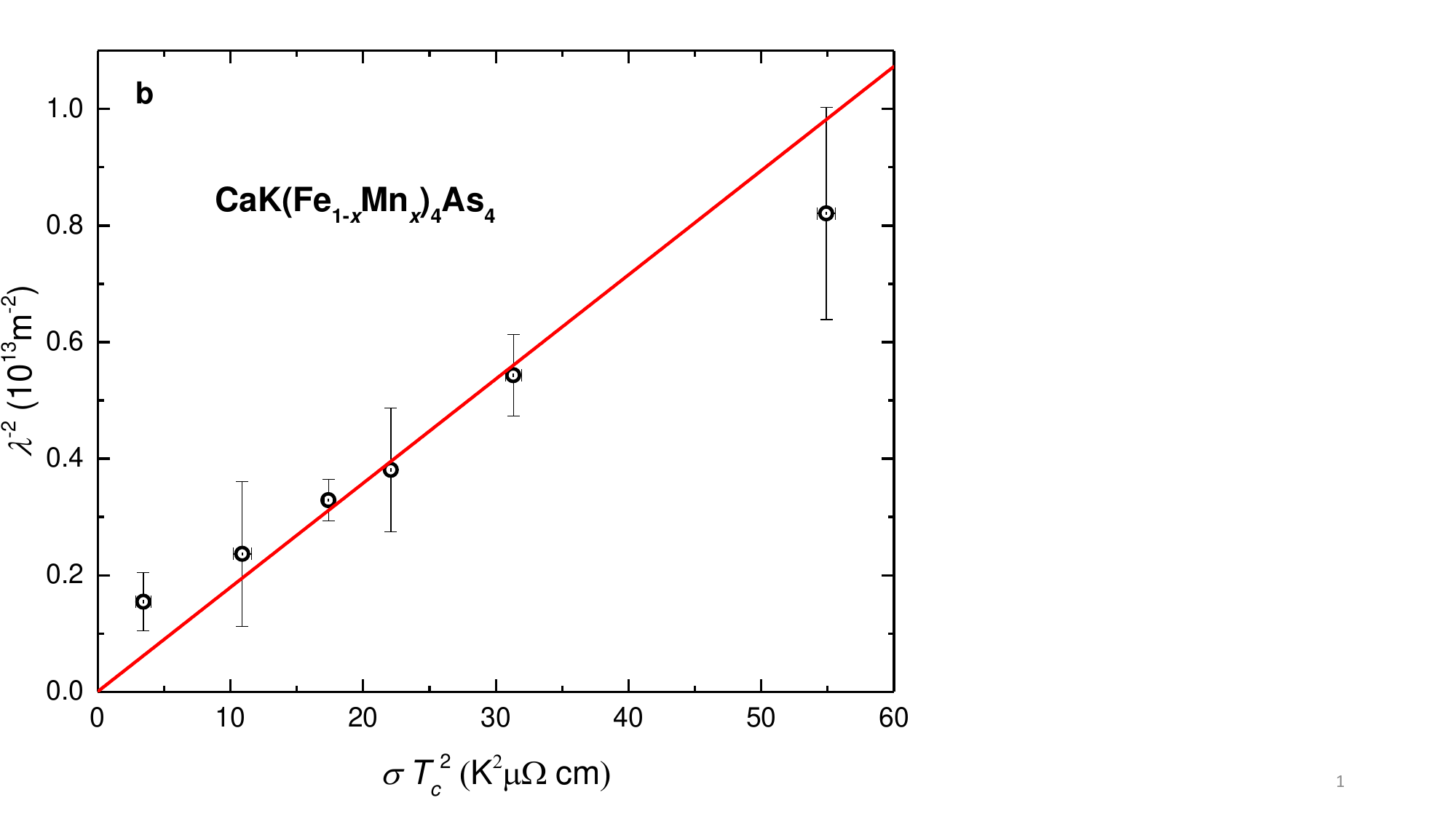}		
	\end{minipage}\hfill
	\caption{(a) shows the London penetration depth as a function of $x$ obtained by magnetic field applied parallel to the crystallographic \textit{ab} axis with different substitution levels. (b) shows $\lambda ^{-2}$ versus $\sigma$ $T_c^2$. Red line is the linear fit of $\lambda ^{-2}$ versus $\sigma$ $T_c^2$ with fixed zero intercept.  \label{Pena}}
\end{figure}

Figure \ref{Pena}a shows the London penetration depth, $\lambda$, as a function of $x$. $\lambda$ is obtained by using: $H_{c1}$ = $\phi_0$(ln $\lambda$/$\xi$ + 0.5)/(4$\pi$$\lambda ^2$) {\color{blue}\cite{Cho2017}}. Figure \ref{Pena}b shows $\lambda^{-2}$ versus $\sigma$ $T_c^2$, where $\sigma$ is normal state conductivity which was measured near the $T_c$. These data roughly follow the behavior associated with Homes type scaling in the presence of pair breaking {\color{blue}\cite{Dordevic2013,Kogan2013a,Kogan2013}}. Given that there is no clear break in behavior in $H_{c1}$ for $x$ $\sim$ 0.015 (Fig. \ref{HC12sum}), it is not surprising that there is no clear feature in Fig. \ref{Pena}a for $x$ $\sim$ 0.015.

\begin{figure}
	\includegraphics[width=\linewidth]{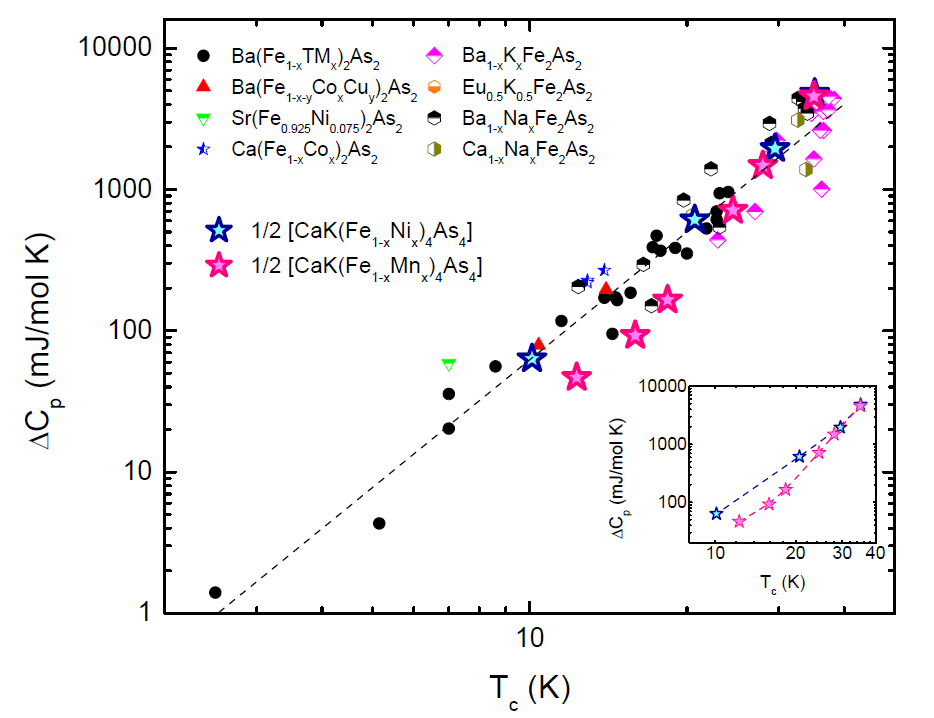}		
	\caption{Scaling of $\Delta$$C_{p}$ and $T_{c}$ for 122 and 1144 systems. Figure modified from {\color{blue}\cite{BudKo2018}}. For consistency, half of the molecular weight of the 1144 samples was taken for this plot.   \label{BNC}}
\end{figure}

Figure \ref{BNC} presents the jump in specific heat at $T_{c}$ versus $T_{c}$ for representative 122 systems, as well as for CaK(Fe$_{1-x}$Ni$_x$)$_4$As$_4$ and CaK(Fe$_{1-x}$Mn$_x$)$_4$As$_4$, on a log-log plot {\color{blue}\cite{BudKo2009}}; the inset shows just the CaK(Fe$_{1-x}$Ni$_x$)$_4$As$_4$ and CaK(Fe$_{1-x}$Mn$_x$)$_4$As$_4$ data on their own for clarity. The CaK(Fe$_{1-x}$Mn$_x$)$_4$As$_4$ data follows the basic scaling that has been seen for many of the Fe-based superconducting systems, but, at a more quantitative level, the CaK(Fe$_{1-x}$Mn$_x$)$_4$As$_4$ data do fall on or below the lower edge of the manifold, especially when compared to the CaK(Fe$_{1-x}$Ni$_x$)$_4$As$_4$. This would be consistent with the jump in $C_{p}$ for a given $T_{c}$ being lower when magnetic pair-breaking is present {\color{blue}\cite{Skalski1964,BudKo2010}}. Given that the Mn ions appear to have a relatively large local moment ($\sim$ 5 $\mu_{B}$) such magnetic pair breaking could be an additional effect that is not present in the Ni- or Co-doped 1144 or Ni- or Co-doped 122 systems. As mentioned in the introduction section, since in Ba(Fe$_{1-x}$Mn$_{x}$)$_{2}$As$_{2}$ could not stabilize superconductivity, the CaK(Fe$_{1-x}$Mn$_x$)$_4$As$_4$ system is important for assessing the role Mn may play in these Fe-based superconductors.

\begin{figure}
	\includegraphics[width=1.5\columnwidth]{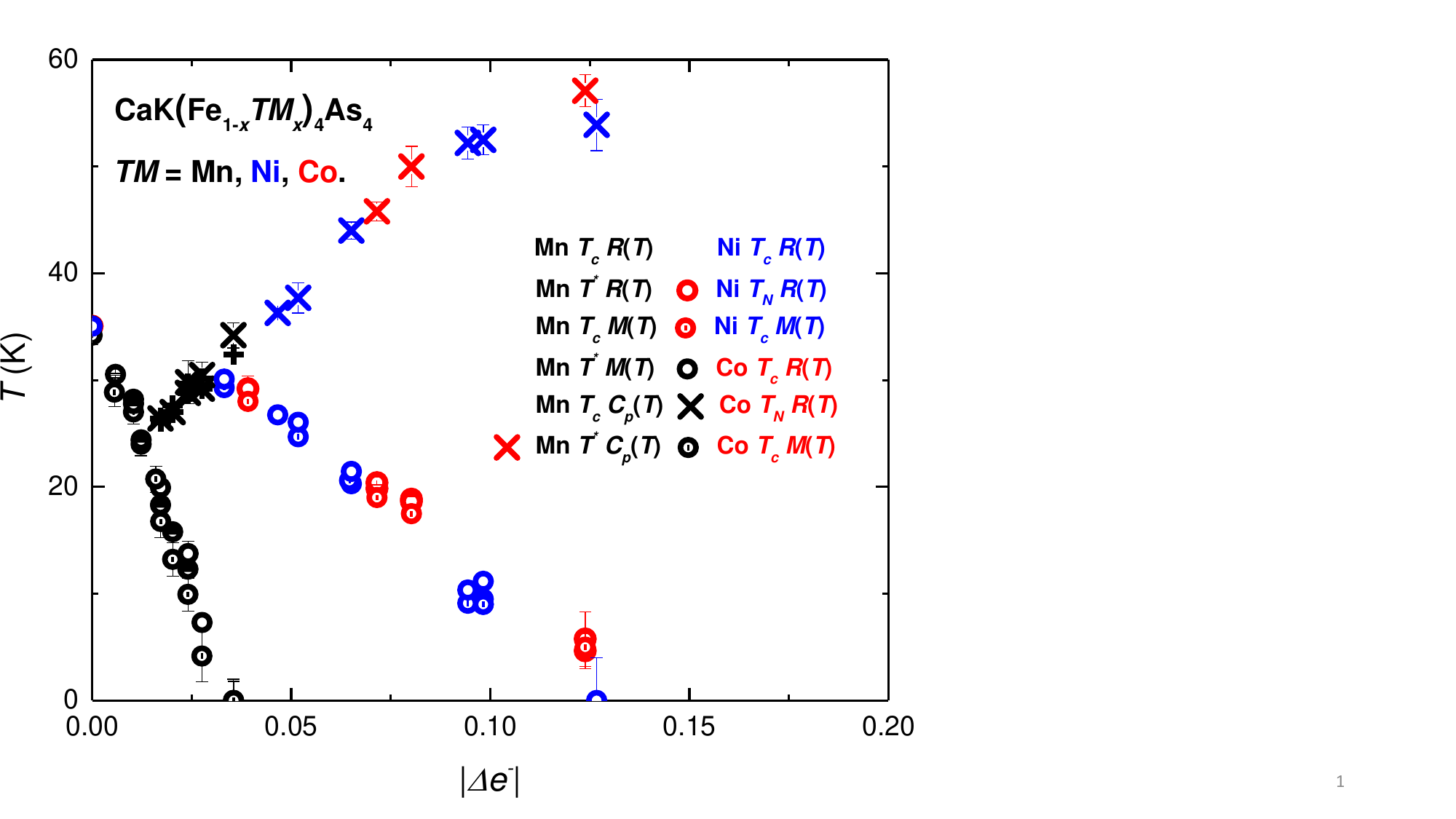}
	\caption{Temperature v.s. |$\Delta e^-$| change of electrons phase diagram of CaK(Fe$_{1-x}TM_{x}$)$_{4}$As$_{4}$ single crystals. $TM$ = Mn, Ni and Co.  The circular symbols denote the $T_{c}$ phase transitions and the cross-like symbols denote the $T^{*}$ and $T_N$ phase transitions, which are obtained from resistance, magnetic moment and specific heat measurements, denoted as "$R$($T$)", "$M$($T$)" and "$C_p$($T$)".\label{figure31}}
\end{figure}

For $TM$ = Co and Ni substitutions, the phase diagrams of CaK(Fe$_{1-x}TM_{x}$)$_{4}$As$_{4}$ scaled almost exactly when transition temperatures were plotted not as a function of $x$ but as a function of band filling change (i.e. when each Ni atom brings two extra electrons and each Co atom brings only one extra electron) {\color{blue}\cite{Meier2018,Meier2019T}}. This led to the conclusion that for electron doping of CaKFe$_{4}$As$_{4}$ the number of electrons added was the controlling factor for both the stabilization of magnetic ordering as well as for the suppression of superconductivity. In figure \ref{figure31} we show that as Mn is substituted into CaKFe$_{4}$As$_{4}$ there is a qualitatively similar suppression of superconductivity as well as the stabilization of magnetic order. This is at odds with the simple analogy that was developed between Co and Ni-substituted CaKFe$_{4}$As$_{4}$ and and (Ba$_{1-x}$K$_{x}$)Fe$_{2}$As$_{2}$. In that analogy CaKFe$_{4}$As$_{4}$ could be considered to have the same band filling as Ba$_{0.5}$K$_{0.5}$Fe$_{2}$As$_{2}$, giving rise to the expectations that if $TM$ substitution lead to rigid band changes in band filling, then (i) Co/Ni doping would suppress superconductivity and stabilize antiferromagnetic ordering (which does happen) and (ii) Mn doping would lead to a suppression of superconductivity, but not lead to any further magnetic ordering. So, the appearance of clear magnetic ordering in very slightly Mn substituted CaKFe$_{4}$As$_{4}$ is noteworthy.

More quantitatively, in figure \ref{figure31} the CaK(Fe$_{1-x}TM_{x}$)$_{4}$As$_{4}$ phase diagrams for $TM$ = Mn, Co and Ni are plotted on the same $T$ and |$\Delta$e$^{-}$| axes. This figure shows that whereas the $T^*$ phase lines do overlap, suggesting that the change electron count (positive or negative) can be correlated to the stabilization of magnetic order, the suppression of \textit{$T_{c}$} is much more rapid for Mn substitution than for Ni or Co. Indeed an unrealistic factor of four would be needed to bring the $T_{c}$ line into agreement with the Co and Ni data.  

Two possible explanations for the different suppression of $T_c$ readily come to mind. First is the simple acknowledgment that for CaKFe$_{4}$As$_{4}$ electron and hole doping via 3d-transition metal substitution for Fe may simply have different effects, most likely associated with asymmetries of the density of states of CaKFe$_{4}$As$_{4}$ on either side of E$_{F}$. The second explanation is given by Abrikosov–Gor'kov mechanism {\color{blue}\cite{A.A.AbrikosovandL.P.Gorkov1961}} which invokes the large, local moment behavior of the Mn ions and try to understand the enhanced suppression of $T_{c}$ as being associated with additional breaking of Cooper pairs by spin-flip scattering of electrons off of the local moment Mn impurity (such enhanced Mn pair breaking was already suggested in {\color{blue}\cite{Li2012a}} for Mn substitution into Ba$_{0.5}$K$_{0.5}$Fe$_{2}$As$_{2}$). This latter case is intriguing, but would require a rather curious type of high-$T_{c}$ superconductivity that is both built out of transition metal antiferromagnetic fluctuations and yet is aggressively suppressed by transition metal based magnetic impurities. This apparent contradiction could be reconciled by the fact that the Fe-moments that are providing the fluctuations are of order 0.5 $\mu_{B}$ whereas the Mn moments are estimated to be an order of magnitude larger, $\sim$ 5 $\mu_{B}$. In this sense Mn would be acting more like a rare earth impurity akin to Gd. It should be noted, though, that neither of these possible explanations for the different suppression of $T_c$ help explain or rationalize this fact the $T^*$ lines for Mn-, Co- and Ni-substitutions do seem to overlap.

In summary, we have been able to grow and study the CaK(Fe$_{1-x}$Mn$_x$)$_4$As$_4$ system. We have been able to assemble a $T$-$x$ phase diagram that clearly shows the suppression of the superconducting $T_{c}$ with the addition of Mn, with $T_{c}$ dropping from 35 K for $x$ = 0  to zero for $x$ > 0.036, as well as the stabilization of magnetic order for $x$ > 0.015, with 26 K $\leq$ $T^{*}$ $\leq$ 32 K. The similarity between the $T$-$x$ phase diagrams for Mn, Co and Ni substitution as well as the elastoresistivity data showing no strong indications of a nematic phase transition or fluctuations, strongly suggest that the magnetic ordering associated with $T^*$ is a hedgehog vortex crystal. As $x$ becomes greater than 0.015 and $T_{c}$ becomes less than $T^{*}$, a clear change in the behavior of $H^\prime_{c2}$($T$)/$T_c$ and the associated superconducting coherence length, $\xi$, can be seen. These are associated with the probable changes in the Fermi surface that accompany the AFM ordering at $T^{*}$. Comparable features in $H_{c1}$ or the London penetration depth are not clearly resolvable.

The $T$-$x$ phase diagram for CaK(Fe$_{1-x}$Mn$_x$)$_4$As$_4$ is qualitatively similar to what was found for Ni- and Co-1144 including the appearance of Mn-stabilized antiferromagnetic ordering. Whereas the stabilization of antiferromagnetic ordering occurs at the same rate (per change in band filling) as it does for Ni- or Co-substitutions, the suppression of $T_{c}$ is much faster (by a factor of roughly four).  The $\Delta$$C_{p}$ versus $T_{c}$ plot for CaK(Fe$_{1-x}$Mn$_x$)$_4$As$_4$ also drops somewhat faster that would be expected when compared to the CaK(Fe$_{1-x}$Ni$_x$)$_4$As$_4$ data.  Both of these enhanced suppression may be associated with the large effective moment found for the substituted Mn ions, $\sim$ 5 $\mu_{B}$, as compared to the size of the ordered Fe moments ($\sim$ 0.5 $\mu_{B}$).

\begin{acknowledgements}
	We thank A. Kreyssig and B. Kuthanazhi for useful discussions. Work at the Ames Laboratory was supported by the U.S. Department of Energy, Office of Science, Basic Energy Sciences, Materials Sciences and Engineering Division. The Ames Laboratory is operated for the U.S. Department of Energy by Iowa State University under Contract No. DE-AC02-07CH11358. E.G. and W.R.M. were supported, in part, by the Gordon and Betty Moore Foundations EPiQS Initiative through Grant GBMF4411. L.X. was supported by the W. M. Keck Foundation.
\end{acknowledgements}
\renewcommand\refname{[References]}
\bibliographystyle{apsrev}
\bibliography{Mn0229}

\begin{thebibliography}{49}
\expandafter\ifx\csname natexlab\endcsname\relax\def\natexlab#1{#1}\fi
\expandafter\ifx\csname bibnamefont\endcsname\relax
  \def\bibnamefont#1{#1}\fi
\expandafter\ifx\csname bibfnamefont\endcsname\relax
  \def\bibfnamefont#1{#1}\fi
\expandafter\ifx\csname citenamefont\endcsname\relax
  \def\citenamefont#1{#1}\fi
\expandafter\ifx\csname url\endcsname\relax
  \def\url#1{\texttt{#1}}\fi
\expandafter\ifx\csname urlprefix\endcsname\relax\def\urlprefix{URL }\fi
\providecommand{\bibinfo}[2]{#2}
\providecommand{\eprint}[2][]{\url{#2}}

\bibitem[{\citenamefont{Kamihara et~al.}(2008)\citenamefont{Kamihara, Watanabe,
  Hirano, and Hosono}}]{Kamihara2008}
\bibinfo{author}{\bibfnamefont{Y.}~\bibnamefont{Kamihara}},
  \bibinfo{author}{\bibfnamefont{T.}~\bibnamefont{Watanabe}},
  \bibinfo{author}{\bibfnamefont{M.}~\bibnamefont{Hirano}}, \bibnamefont{and}
  \bibinfo{author}{\bibfnamefont{H.}~\bibnamefont{Hosono}},
  \bibinfo{journal}{Journal of the American Chemical Society}
  \textbf{\bibinfo{volume}{130}}, \bibinfo{pages}{3296} (\bibinfo{year}{2008}).

\bibitem[{\citenamefont{Johnston}(2010)}]{Johnston2010}
\bibinfo{author}{\bibfnamefont{D.~C.} \bibnamefont{Johnston}},
  \bibinfo{journal}{Advances in Physics} \textbf{\bibinfo{volume}{59}},
  \bibinfo{pages}{803} (\bibinfo{year}{2010}).

\bibitem[{\citenamefont{Paglione and Greene}(2010)}]{Paglione2010}
\bibinfo{author}{\bibfnamefont{J.}~\bibnamefont{Paglione}} \bibnamefont{and}
  \bibinfo{author}{\bibfnamefont{R.~L.} \bibnamefont{Greene}},
  \bibinfo{journal}{Nature Physics} \textbf{\bibinfo{volume}{6}},
  \bibinfo{pages}{645} (\bibinfo{year}{2010}).

\bibitem[{\citenamefont{Hosono and Kuroki}(2015)}]{Hosono2015}
\bibinfo{author}{\bibfnamefont{H.}~\bibnamefont{Hosono}} \bibnamefont{and}
  \bibinfo{author}{\bibfnamefont{K.}~\bibnamefont{Kuroki}},
  \bibinfo{journal}{Physica C: Superconductivity and its Applications}
  \textbf{\bibinfo{volume}{514}}, \bibinfo{pages}{399} (\bibinfo{year}{2015}).

\bibitem[{\citenamefont{Canfield et~al.}(2009)\citenamefont{Canfield, Bud'ko,
  Ni, Yan, and Kracher}}]{Canfield2009a}
\bibinfo{author}{\bibfnamefont{P.~C.} \bibnamefont{Canfield}},
  \bibinfo{author}{\bibfnamefont{S.~L.} \bibnamefont{Bud'ko}},
  \bibinfo{author}{\bibfnamefont{N.}~\bibnamefont{Ni}},
  \bibinfo{author}{\bibfnamefont{J.~Q.} \bibnamefont{Yan}}, \bibnamefont{and}
  \bibinfo{author}{\bibfnamefont{A.}~\bibnamefont{Kracher}},
  \bibinfo{journal}{Physical Review B} \textbf{\bibinfo{volume}{80}},
  \bibinfo{pages}{060501(R)} (\bibinfo{year}{2009}).

\bibitem[{\citenamefont{Ni et~al.}(2009)\citenamefont{Ni, Thaler, Kracher, Yan,
  Bud'ko, and Canfield}}]{Ni2009}
\bibinfo{author}{\bibfnamefont{N.}~\bibnamefont{Ni}},
  \bibinfo{author}{\bibfnamefont{A.}~\bibnamefont{Thaler}},
  \bibinfo{author}{\bibfnamefont{A.}~\bibnamefont{Kracher}},
  \bibinfo{author}{\bibfnamefont{J.~Q.} \bibnamefont{Yan}},
  \bibinfo{author}{\bibfnamefont{S.~L.} \bibnamefont{Bud'ko}},
  \bibnamefont{and} \bibinfo{author}{\bibfnamefont{P.~C.}
  \bibnamefont{Canfield}}, \bibinfo{journal}{Physical Review B}
  \textbf{\bibinfo{volume}{80}}, \bibinfo{pages}{024511}
  (\bibinfo{year}{2009}).

\bibitem[{\citenamefont{Ni et~al.}(2010)\citenamefont{Ni, Thaler, Yan, Kracher,
  Colombier, Bud'ko, Canfield, and Hannahs}}]{Ni20101}
\bibinfo{author}{\bibfnamefont{N.}~\bibnamefont{Ni}},
  \bibinfo{author}{\bibfnamefont{A.}~\bibnamefont{Thaler}},
  \bibinfo{author}{\bibfnamefont{J.~Q.} \bibnamefont{Yan}},
  \bibinfo{author}{\bibfnamefont{A.}~\bibnamefont{Kracher}},
  \bibinfo{author}{\bibfnamefont{E.}~\bibnamefont{Colombier}},
  \bibinfo{author}{\bibfnamefont{S.~L.} \bibnamefont{Bud'ko}},
  \bibinfo{author}{\bibfnamefont{P.~C.} \bibnamefont{Canfield}},
  \bibnamefont{and} \bibinfo{author}{\bibfnamefont{S.~T.}
  \bibnamefont{Hannahs}}, \bibinfo{journal}{Physical Review B}
  \textbf{\bibinfo{volume}{82}}, \bibinfo{pages}{024519}
  (\bibinfo{year}{2010}).

\bibitem[{\citenamefont{Canfield and Bud'ko}(2010)}]{Canfield2010f}
\bibinfo{author}{\bibfnamefont{P.~C.} \bibnamefont{Canfield}} \bibnamefont{and}
  \bibinfo{author}{\bibfnamefont{S.~L.} \bibnamefont{Bud'ko}},
  \bibinfo{journal}{Annual Review of Condensed Matter Physics}
  \textbf{\bibinfo{volume}{1}}, \bibinfo{pages}{27} (\bibinfo{year}{2010}).

\bibitem[{\citenamefont{Stewart}(2011)}]{Stewart2011}
\bibinfo{author}{\bibfnamefont{G.~R.} \bibnamefont{Stewart}},
  \bibinfo{journal}{Reviews of Modern Physics} \textbf{\bibinfo{volume}{83}},
  \bibinfo{pages}{1589} (\bibinfo{year}{2011}).

\bibitem[{\citenamefont{Iyo et~al.}(2016)\citenamefont{Iyo, Kawashima, Kinjo,
  Nishio, Ishida, Fujihisa, Gotoh, Kihou, Eisaki, and Yoshida}}]{Yoshida2016}
\bibinfo{author}{\bibfnamefont{A.}~\bibnamefont{Iyo}},
  \bibinfo{author}{\bibfnamefont{K.}~\bibnamefont{Kawashima}},
  \bibinfo{author}{\bibfnamefont{T.}~\bibnamefont{Kinjo}},
  \bibinfo{author}{\bibfnamefont{T.}~\bibnamefont{Nishio}},
  \bibinfo{author}{\bibfnamefont{S.}~\bibnamefont{Ishida}},
  \bibinfo{author}{\bibfnamefont{H.}~\bibnamefont{Fujihisa}},
  \bibinfo{author}{\bibfnamefont{Y.}~\bibnamefont{Gotoh}},
  \bibinfo{author}{\bibfnamefont{K.}~\bibnamefont{Kihou}},
  \bibinfo{author}{\bibfnamefont{H.}~\bibnamefont{Eisaki}}, \bibnamefont{and}
  \bibinfo{author}{\bibfnamefont{Y.}~\bibnamefont{Yoshida}},
  \bibinfo{journal}{Journal of the American Chemical Society}
  \textbf{\bibinfo{volume}{138}}, \bibinfo{pages}{3410} (\bibinfo{year}{2016}).

\bibitem[{\citenamefont{Meier et~al.}(2016)\citenamefont{Meier, Kong,
  Kaluarachchi, Taufour, Jo, Drachuck, B{\"{o}}hmer, Saunders, Sapkota,
  Kreyssig et~al.}}]{Meier2016}
\bibinfo{author}{\bibfnamefont{W.~R.} \bibnamefont{Meier}},
  \bibinfo{author}{\bibfnamefont{T.}~\bibnamefont{Kong}},
  \bibinfo{author}{\bibfnamefont{U.~S.} \bibnamefont{Kaluarachchi}},
  \bibinfo{author}{\bibfnamefont{V.}~\bibnamefont{Taufour}},
  \bibinfo{author}{\bibfnamefont{N.~H.} \bibnamefont{Jo}},
  \bibinfo{author}{\bibfnamefont{G.}~\bibnamefont{Drachuck}},
  \bibinfo{author}{\bibfnamefont{A.~E.} \bibnamefont{B{\"{o}}hmer}},
  \bibinfo{author}{\bibfnamefont{S.~M.} \bibnamefont{Saunders}},
  \bibinfo{author}{\bibfnamefont{A.}~\bibnamefont{Sapkota}},
  \bibinfo{author}{\bibfnamefont{A.}~\bibnamefont{Kreyssig}},
  \bibnamefont{et~al.}, \bibinfo{journal}{Physical Review B}
  \textbf{\bibinfo{volume}{94}}, \bibinfo{pages}{064501}
  (\bibinfo{year}{2016}).

\bibitem[{\citenamefont{Meier et~al.}(2017)\citenamefont{Meier, Kong, Bud'ko,
  and Canfield}}]{Meier2017}
\bibinfo{author}{\bibfnamefont{W.~R.} \bibnamefont{Meier}},
  \bibinfo{author}{\bibfnamefont{T.}~\bibnamefont{Kong}},
  \bibinfo{author}{\bibfnamefont{S.~L.} \bibnamefont{Bud'ko}},
  \bibnamefont{and} \bibinfo{author}{\bibfnamefont{P.~C.}
  \bibnamefont{Canfield}}, \bibinfo{journal}{Physical Review Materials}
  \textbf{\bibinfo{volume}{1}}, \bibinfo{pages}{013401} (\bibinfo{year}{2017}).

\bibitem[{\citenamefont{Meier et~al.}(2018)\citenamefont{Meier, Ding, Kreyssig,
  Bud'ko, Sapkota, Kothapalli, Borisov, Valent{\'{i}}, Batista, Orth
  et~al.}}]{Meier2018}
\bibinfo{author}{\bibfnamefont{W.~R.} \bibnamefont{Meier}},
  \bibinfo{author}{\bibfnamefont{Q.-P.} \bibnamefont{Ding}},
  \bibinfo{author}{\bibfnamefont{A.}~\bibnamefont{Kreyssig}},
  \bibinfo{author}{\bibfnamefont{S.~L.} \bibnamefont{Bud'ko}},
  \bibinfo{author}{\bibfnamefont{A.}~\bibnamefont{Sapkota}},
  \bibinfo{author}{\bibfnamefont{K.}~\bibnamefont{Kothapalli}},
  \bibinfo{author}{\bibfnamefont{V.}~\bibnamefont{Borisov}},
  \bibinfo{author}{\bibfnamefont{R.}~\bibnamefont{Valent{\'{i}}}},
  \bibinfo{author}{\bibfnamefont{C.~D.} \bibnamefont{Batista}},
  \bibinfo{author}{\bibfnamefont{P.~P.} \bibnamefont{Orth}},
  \bibnamefont{et~al.}, \bibinfo{journal}{npj Quantum Materials}
  \textbf{\bibinfo{volume}{3}}, \bibinfo{pages}{5} (\bibinfo{year}{2018}).

\bibitem[{\citenamefont{Hsu et~al.}(2008)\citenamefont{Hsu, Luo, Yeh, Chen,
  Huang, Wu, Lee, Huang, Chu, Yan et~al.}}]{Hsu14262}
\bibinfo{author}{\bibfnamefont{F.-C.} \bibnamefont{Hsu}},
  \bibinfo{author}{\bibfnamefont{J.-Y.} \bibnamefont{Luo}},
  \bibinfo{author}{\bibfnamefont{K.-W.} \bibnamefont{Yeh}},
  \bibinfo{author}{\bibfnamefont{T.-K.} \bibnamefont{Chen}},
  \bibinfo{author}{\bibfnamefont{T.-W.} \bibnamefont{Huang}},
  \bibinfo{author}{\bibfnamefont{P.~M.} \bibnamefont{Wu}},
  \bibinfo{author}{\bibfnamefont{Y.-C.} \bibnamefont{Lee}},
  \bibinfo{author}{\bibfnamefont{Y.-L.} \bibnamefont{Huang}},
  \bibinfo{author}{\bibfnamefont{Y.-Y.} \bibnamefont{Chu}},
  \bibinfo{author}{\bibfnamefont{D.-C.} \bibnamefont{Yan}},
  \bibnamefont{et~al.}, \bibinfo{journal}{Proceedings of the National Academy
  of Sciences} \textbf{\bibinfo{volume}{105}}, \bibinfo{pages}{14262}
  (\bibinfo{year}{2008}).

\bibitem[{\citenamefont{B{\"{o}}hmer and Kreisel}(2018)}]{Bohmer2017a}
\bibinfo{author}{\bibfnamefont{A.~E.} \bibnamefont{B{\"{o}}hmer}}
  \bibnamefont{and} \bibinfo{author}{\bibfnamefont{A.}~\bibnamefont{Kreisel}},
  \bibinfo{journal}{Journal of Physics: Condensed Matter}
  \textbf{\bibinfo{volume}{30}}, \bibinfo{pages}{023001}
  (\bibinfo{year}{2018}).

\bibitem[{\citenamefont{Kawashima et~al.}(2016)\citenamefont{Kawashima, Kinjo,
  Nishio, Ishida, Fujihisa, Gotoh, Kihou, Eisaki, Yoshida, and
  Iyo}}]{Kawashima2016}
\bibinfo{author}{\bibfnamefont{K.}~\bibnamefont{Kawashima}},
  \bibinfo{author}{\bibfnamefont{T.}~\bibnamefont{Kinjo}},
  \bibinfo{author}{\bibfnamefont{T.}~\bibnamefont{Nishio}},
  \bibinfo{author}{\bibfnamefont{S.}~\bibnamefont{Ishida}},
  \bibinfo{author}{\bibfnamefont{H.}~\bibnamefont{Fujihisa}},
  \bibinfo{author}{\bibfnamefont{Y.}~\bibnamefont{Gotoh}},
  \bibinfo{author}{\bibfnamefont{K.}~\bibnamefont{Kihou}},
  \bibinfo{author}{\bibfnamefont{H.}~\bibnamefont{Eisaki}},
  \bibinfo{author}{\bibfnamefont{Y.}~\bibnamefont{Yoshida}}, \bibnamefont{and}
  \bibinfo{author}{\bibfnamefont{A.}~\bibnamefont{Iyo}},
  \bibinfo{journal}{Journal of the Physical Society of Japan}
  \textbf{\bibinfo{volume}{85}}, \bibinfo{pages}{064710}
  (\bibinfo{year}{2016}).

\bibitem[{\citenamefont{Bao et~al.}(2018)\citenamefont{Bao, Willa, Smylie,
  Chen, Welp, Chung, and Kanatzidis}}]{Bao2018}
\bibinfo{author}{\bibfnamefont{J.-K.} \bibnamefont{Bao}},
  \bibinfo{author}{\bibfnamefont{K.}~\bibnamefont{Willa}},
  \bibinfo{author}{\bibfnamefont{M.~P.} \bibnamefont{Smylie}},
  \bibinfo{author}{\bibfnamefont{H.}~\bibnamefont{Chen}},
  \bibinfo{author}{\bibfnamefont{U.}~\bibnamefont{Welp}},
  \bibinfo{author}{\bibfnamefont{D.~Y.} \bibnamefont{Chung}}, \bibnamefont{and}
  \bibinfo{author}{\bibfnamefont{M.~G.} \bibnamefont{Kanatzidis}},
  \bibinfo{journal}{Crystal Growth {\&} Design} \textbf{\bibinfo{volume}{18}},
  \bibinfo{pages}{3517} (\bibinfo{year}{2018}).

\bibitem[{\citenamefont{Gati et~al.}(2020)\citenamefont{Gati, Xiang, Bud'ko,
  and Canfield}}]{Gati2020}
\bibinfo{author}{\bibfnamefont{E.}~\bibnamefont{Gati}},
  \bibinfo{author}{\bibfnamefont{L.}~\bibnamefont{Xiang}},
  \bibinfo{author}{\bibfnamefont{S.~L.} \bibnamefont{Bud'ko}},
  \bibnamefont{and} \bibinfo{author}{\bibfnamefont{P.~C.}
  \bibnamefont{Canfield}}, \bibinfo{journal}{Annalen der Physik}
  \textbf{\bibinfo{volume}{532}}, \bibinfo{pages}{2000248}
  (\bibinfo{year}{2020}).

\bibitem[{\citenamefont{Mazin}(2010)}]{Mazin2010}
\bibinfo{author}{\bibfnamefont{I.~I.} \bibnamefont{Mazin}},
  \bibinfo{journal}{Nature} \textbf{\bibinfo{volume}{464}},
  \bibinfo{pages}{183} (\bibinfo{year}{2010}).

\bibitem[{\citenamefont{Meier}(2018)}]{Meier2019T}
\bibinfo{author}{\bibfnamefont{W.}~\bibnamefont{Meier}},
  \bibinfo{journal}{Thesis}  (\bibinfo{year}{2018}),
  \urlprefix\url{https://lib.dr.iastate.edu/etd/16856/}.

\bibitem[{\citenamefont{Li et~al.}(2012)\citenamefont{Li, Guo, Zhang, Yuan,
  Tsujimoto, Wang, Sathish, Sun, Yu, Yi et~al.}}]{Li2012a}
\bibinfo{author}{\bibfnamefont{J.}~\bibnamefont{Li}},
  \bibinfo{author}{\bibfnamefont{Y.~F.} \bibnamefont{Guo}},
  \bibinfo{author}{\bibfnamefont{S.~B.} \bibnamefont{Zhang}},
  \bibinfo{author}{\bibfnamefont{J.}~\bibnamefont{Yuan}},
  \bibinfo{author}{\bibfnamefont{Y.}~\bibnamefont{Tsujimoto}},
  \bibinfo{author}{\bibfnamefont{X.}~\bibnamefont{Wang}},
  \bibinfo{author}{\bibfnamefont{C.~I.} \bibnamefont{Sathish}},
  \bibinfo{author}{\bibfnamefont{Y.}~\bibnamefont{Sun}},
  \bibinfo{author}{\bibfnamefont{S.}~\bibnamefont{Yu}},
  \bibinfo{author}{\bibfnamefont{W.}~\bibnamefont{Yi}}, \bibnamefont{et~al.},
  \bibinfo{journal}{Physical Review B} \textbf{\bibinfo{volume}{85}},
  \bibinfo{pages}{214509} (\bibinfo{year}{2012}).

\bibitem[{\citenamefont{LeBoeuf et~al.}(2014)\citenamefont{LeBoeuf, Texier,
  Boselli, Forget, Colson, and Bobroff}}]{Leboeuf2014}
\bibinfo{author}{\bibfnamefont{D.}~\bibnamefont{LeBoeuf}},
  \bibinfo{author}{\bibfnamefont{Y.}~\bibnamefont{Texier}},
  \bibinfo{author}{\bibfnamefont{M.}~\bibnamefont{Boselli}},
  \bibinfo{author}{\bibfnamefont{A.}~\bibnamefont{Forget}},
  \bibinfo{author}{\bibfnamefont{D.}~\bibnamefont{Colson}}, \bibnamefont{and}
  \bibinfo{author}{\bibfnamefont{J.}~\bibnamefont{Bobroff}},
  \bibinfo{journal}{Physical Review B} \textbf{\bibinfo{volume}{89}},
  \bibinfo{pages}{035114} (\bibinfo{year}{2014}).

\bibitem[{\citenamefont{Thaler et~al.}(2011)\citenamefont{Thaler, Hodovanets,
  Torikachvili, Ran, Kracher, Straszheim, Yan, Mun, and Canfield}}]{Thaler2011}
\bibinfo{author}{\bibfnamefont{A.}~\bibnamefont{Thaler}},
  \bibinfo{author}{\bibfnamefont{H.}~\bibnamefont{Hodovanets}},
  \bibinfo{author}{\bibfnamefont{M.~S.} \bibnamefont{Torikachvili}},
  \bibinfo{author}{\bibfnamefont{S.}~\bibnamefont{Ran}},
  \bibinfo{author}{\bibfnamefont{A.}~\bibnamefont{Kracher}},
  \bibinfo{author}{\bibfnamefont{W.}~\bibnamefont{Straszheim}},
  \bibinfo{author}{\bibfnamefont{J.~Q.} \bibnamefont{Yan}},
  \bibinfo{author}{\bibfnamefont{E.}~\bibnamefont{Mun}}, \bibnamefont{and}
  \bibinfo{author}{\bibfnamefont{P.~C.} \bibnamefont{Canfield}},
  \bibinfo{journal}{Physical Review B} \textbf{\bibinfo{volume}{84}},
  \bibinfo{pages}{144528} (\bibinfo{year}{2011}).

\bibitem[{\citenamefont{Pandey et~al.}(2011)\citenamefont{Pandey, Anand, and
  Johnston}}]{Pandey2011}
\bibinfo{author}{\bibfnamefont{A.}~\bibnamefont{Pandey}},
  \bibinfo{author}{\bibfnamefont{V.~K.} \bibnamefont{Anand}}, \bibnamefont{and}
  \bibinfo{author}{\bibfnamefont{D.~C.} \bibnamefont{Johnston}},
  \bibinfo{journal}{Physical Review B} \textbf{\bibinfo{volume}{84}},
  \bibinfo{pages}{014405} (\bibinfo{year}{2011}).

\bibitem[{\citenamefont{Canfield}(2020)}]{Canfield2020}
\bibinfo{author}{\bibfnamefont{P.~C.} \bibnamefont{Canfield}},
  \bibinfo{journal}{Reports on Progress in Physics}
  \textbf{\bibinfo{volume}{83}}, \bibinfo{pages}{016501}
  (\bibinfo{year}{2020}).

\bibitem[{\citenamefont{Canfield et~al.}(2016)\citenamefont{Canfield, Kong,
  Kaluarachchi, and Jo}}]{Canfield2016a}
\bibinfo{author}{\bibfnamefont{P.~C.} \bibnamefont{Canfield}},
  \bibinfo{author}{\bibfnamefont{T.}~\bibnamefont{Kong}},
  \bibinfo{author}{\bibfnamefont{U.~S.} \bibnamefont{Kaluarachchi}},
  \bibnamefont{and} \bibinfo{author}{\bibfnamefont{N.~H.} \bibnamefont{Jo}},
  \bibinfo{journal}{Philosophical Magazine} \textbf{\bibinfo{volume}{96}},
  \bibinfo{pages}{84} (\bibinfo{year}{2016}).

\bibitem[{\citenamefont{Jesche et~al.}(2016)\citenamefont{Jesche, Fix,
  Kreyssig, Meier, and Canfield}}]{Jesche2016}
\bibinfo{author}{\bibfnamefont{A.}~\bibnamefont{Jesche}},
  \bibinfo{author}{\bibfnamefont{M.}~\bibnamefont{Fix}},
  \bibinfo{author}{\bibfnamefont{A.}~\bibnamefont{Kreyssig}},
  \bibinfo{author}{\bibfnamefont{W.~R.} \bibnamefont{Meier}}, \bibnamefont{and}
  \bibinfo{author}{\bibfnamefont{P.~C.} \bibnamefont{Canfield}},
  \bibinfo{journal}{Philosophical Magazine} \textbf{\bibinfo{volume}{96}},
  \bibinfo{pages}{2115} (\bibinfo{year}{2016}).

\bibitem[{\citenamefont{Newbury and Ritchie}(2014)}]{Newbury2014}
\bibinfo{author}{\bibfnamefont{D.~E.} \bibnamefont{Newbury}} \bibnamefont{and}
  \bibinfo{author}{\bibfnamefont{N.~W.~M.} \bibnamefont{Ritchie}}, in
  \emph{\bibinfo{booktitle}{Scanning Microscopies 2014}}, edited by
  \bibinfo{editor}{\bibfnamefont{M.~T.} \bibnamefont{Postek}},
  \bibinfo{editor}{\bibfnamefont{D.~E.} \bibnamefont{Newbury}},
  \bibinfo{editor}{\bibfnamefont{S.~F.} \bibnamefont{Platek}},
  \bibnamefont{and} \bibinfo{editor}{\bibfnamefont{T.~K.} \bibnamefont{Maugel}}
  (\bibinfo{year}{2014}), vol. \bibinfo{volume}{9236}, p.
  \bibinfo{pages}{92360H}.

\bibitem[{\citenamefont{Prozorov and Kogan}(2018)}]{Prozorov2018}
\bibinfo{author}{\bibfnamefont{R.}~\bibnamefont{Prozorov}} \bibnamefont{and}
  \bibinfo{author}{\bibfnamefont{V.~G.} \bibnamefont{Kogan}},
  \bibinfo{journal}{Physical Review Applied} \textbf{\bibinfo{volume}{10}},
  \bibinfo{pages}{014030} (\bibinfo{year}{2018}).

\bibitem[{\citenamefont{Kuo et~al.}(2016)\citenamefont{Kuo, Chu, Palmstrom,
  Kivelson, and Fisher}}]{Kuo2016}
\bibinfo{author}{\bibfnamefont{H.-H.} \bibnamefont{Kuo}},
  \bibinfo{author}{\bibfnamefont{J.-H.} \bibnamefont{Chu}},
  \bibinfo{author}{\bibfnamefont{J.~C.} \bibnamefont{Palmstrom}},
  \bibinfo{author}{\bibfnamefont{S.~A.} \bibnamefont{Kivelson}},
  \bibnamefont{and} \bibinfo{author}{\bibfnamefont{I.~R.}
  \bibnamefont{Fisher}}, \bibinfo{journal}{Science}
  \textbf{\bibinfo{volume}{352}}, \bibinfo{pages}{958} (\bibinfo{year}{2016}).

\bibitem[{\citenamefont{Kuo et~al.}(2013)\citenamefont{Kuo, Shapiro, Riggs, and
  Fisher}}]{Kuo2013}
\bibinfo{author}{\bibfnamefont{H.-H.} \bibnamefont{Kuo}},
  \bibinfo{author}{\bibfnamefont{M.~C.} \bibnamefont{Shapiro}},
  \bibinfo{author}{\bibfnamefont{S.~C.} \bibnamefont{Riggs}}, \bibnamefont{and}
  \bibinfo{author}{\bibfnamefont{I.~R.} \bibnamefont{Fisher}},
  \bibinfo{journal}{Physical Review B} \textbf{\bibinfo{volume}{88}},
  \bibinfo{pages}{085113} (\bibinfo{year}{2013}).

\bibitem[{\citenamefont{Fisher}(1962)}]{MFisher1962}
\bibinfo{author}{\bibfnamefont{M.~E.} \bibnamefont{Fisher}},
  \bibinfo{journal}{Philosophical Magazine} \textbf{\bibinfo{volume}{7}},
  \bibinfo{pages}{1731} (\bibinfo{year}{1962}).

\bibitem[{\citenamefont{Xiang et~al.}(2018)\citenamefont{Xiang, Meier, Xu,
  Kaluarachchi, Bud'ko, and Canfield}}]{Xiang2018}
\bibinfo{author}{\bibfnamefont{L.}~\bibnamefont{Xiang}},
  \bibinfo{author}{\bibfnamefont{W.~R.} \bibnamefont{Meier}},
  \bibinfo{author}{\bibfnamefont{M.}~\bibnamefont{Xu}},
  \bibinfo{author}{\bibfnamefont{U.~S.} \bibnamefont{Kaluarachchi}},
  \bibinfo{author}{\bibfnamefont{S.~L.} \bibnamefont{Bud'ko}},
  \bibnamefont{and} \bibinfo{author}{\bibfnamefont{P.~C.}
  \bibnamefont{Canfield}}, \bibinfo{journal}{Physical Review B}
  \textbf{\bibinfo{volume}{97}}, \bibinfo{pages}{174517}
  (\bibinfo{year}{2018}).

\bibitem[{\citenamefont{Kaluarachchi et~al.}(2016)\citenamefont{Kaluarachchi,
  Taufour, B{\"{o}}hmer, Tanatar, Bud'ko, Kogan, Prozorov, and
  Canfield}}]{Kaluarachchi2016}
\bibinfo{author}{\bibfnamefont{U.~S.} \bibnamefont{Kaluarachchi}},
  \bibinfo{author}{\bibfnamefont{V.}~\bibnamefont{Taufour}},
  \bibinfo{author}{\bibfnamefont{A.~E.} \bibnamefont{B{\"{o}}hmer}},
  \bibinfo{author}{\bibfnamefont{M.~A.} \bibnamefont{Tanatar}},
  \bibinfo{author}{\bibfnamefont{S.~L.} \bibnamefont{Bud'ko}},
  \bibinfo{author}{\bibfnamefont{V.~G.} \bibnamefont{Kogan}},
  \bibinfo{author}{\bibfnamefont{R.}~\bibnamefont{Prozorov}}, \bibnamefont{and}
  \bibinfo{author}{\bibfnamefont{P.~C.} \bibnamefont{Canfield}},
  \bibinfo{journal}{Physical Review B} \textbf{\bibinfo{volume}{93}},
  \bibinfo{pages}{064503} (\bibinfo{year}{2016}).

\bibitem[{\citenamefont{Xiang et~al.}(2017)\citenamefont{Xiang, Kaluarachchi,
  B{\"{o}}hmer, Taufour, Tanatar, Prozorov, Bud'ko, and Canfield}}]{Xiang2017}
\bibinfo{author}{\bibfnamefont{L.}~\bibnamefont{Xiang}},
  \bibinfo{author}{\bibfnamefont{U.~S.} \bibnamefont{Kaluarachchi}},
  \bibinfo{author}{\bibfnamefont{A.~E.} \bibnamefont{B{\"{o}}hmer}},
  \bibinfo{author}{\bibfnamefont{V.}~\bibnamefont{Taufour}},
  \bibinfo{author}{\bibfnamefont{M.~A.} \bibnamefont{Tanatar}},
  \bibinfo{author}{\bibfnamefont{R.}~\bibnamefont{Prozorov}},
  \bibinfo{author}{\bibfnamefont{S.~L.} \bibnamefont{Bud'ko}},
  \bibnamefont{and} \bibinfo{author}{\bibfnamefont{P.~C.}
  \bibnamefont{Canfield}}, \bibinfo{journal}{Physical Review B}
  \textbf{\bibinfo{volume}{96}}, \bibinfo{pages}{024511}
  (\bibinfo{year}{2017}).

\bibitem[{\citenamefont{Taufour et~al.}(2014)\citenamefont{Taufour, Foroozani,
  Tanatar, Lim, Kaluarachchi, Kim, Liu, Lograsso, Kogan, Prozorov
  et~al.}}]{Taufour2014}
\bibinfo{author}{\bibfnamefont{V.}~\bibnamefont{Taufour}},
  \bibinfo{author}{\bibfnamefont{N.}~\bibnamefont{Foroozani}},
  \bibinfo{author}{\bibfnamefont{M.~A.} \bibnamefont{Tanatar}},
  \bibinfo{author}{\bibfnamefont{J.}~\bibnamefont{Lim}},
  \bibinfo{author}{\bibfnamefont{U.}~\bibnamefont{Kaluarachchi}},
  \bibinfo{author}{\bibfnamefont{S.~K.} \bibnamefont{Kim}},
  \bibinfo{author}{\bibfnamefont{Y.}~\bibnamefont{Liu}},
  \bibinfo{author}{\bibfnamefont{T.~A.} \bibnamefont{Lograsso}},
  \bibinfo{author}{\bibfnamefont{V.~G.} \bibnamefont{Kogan}},
  \bibinfo{author}{\bibfnamefont{R.}~\bibnamefont{Prozorov}},
  \bibnamefont{et~al.}, \bibinfo{journal}{Physical Review B}
  \textbf{\bibinfo{volume}{89}}, \bibinfo{pages}{220509(R)}
  (\bibinfo{year}{2014}).

\bibitem[{\citenamefont{Chu et~al.}(2012)\citenamefont{Chu, Kuo, Analytis, and
  Fisher}}]{Chu2012}
\bibinfo{author}{\bibfnamefont{J.-H.} \bibnamefont{Chu}},
  \bibinfo{author}{\bibfnamefont{H.-H.} \bibnamefont{Kuo}},
  \bibinfo{author}{\bibfnamefont{J.~G.} \bibnamefont{Analytis}},
  \bibnamefont{and} \bibinfo{author}{\bibfnamefont{I.~R.}
  \bibnamefont{Fisher}}, \bibinfo{journal}{Science}
  \textbf{\bibinfo{volume}{338}}, \bibinfo{pages}{469} (\bibinfo{year}{2012}).

\bibitem[{\citenamefont{B{\"{o}}hmer et~al.}(2020)\citenamefont{B{\"{o}}hmer,
  Chen, Meier, Xu, Drachuck, Merz, Wiecki, Bud'ko, Borisov, Valent{\'{i}}
  et~al.}}]{Bohmer2020}
\bibinfo{author}{\bibfnamefont{A.~E.} \bibnamefont{B{\"{o}}hmer}},
  \bibinfo{author}{\bibfnamefont{F.}~\bibnamefont{Chen}},
  \bibinfo{author}{\bibfnamefont{W.~R.} \bibnamefont{Meier}},
  \bibinfo{author}{\bibfnamefont{M.}~\bibnamefont{Xu}},
  \bibinfo{author}{\bibfnamefont{G.}~\bibnamefont{Drachuck}},
  \bibinfo{author}{\bibfnamefont{M.}~\bibnamefont{Merz}},
  \bibinfo{author}{\bibfnamefont{P.~W.} \bibnamefont{Wiecki}},
  \bibinfo{author}{\bibfnamefont{S.~L.} \bibnamefont{Bud'ko}},
  \bibinfo{author}{\bibfnamefont{V.}~\bibnamefont{Borisov}},
  \bibinfo{author}{\bibfnamefont{R.}~\bibnamefont{Valent{\'{i}}}},
  \bibnamefont{et~al.} (\bibinfo{year}{2020}), \eprint{2011.13207}.

\bibitem[{\citenamefont{Cho et~al.}(2017)\citenamefont{Cho, Fente, Teknowijoyo,
  Tanatar, Joshi, Nusran, Kong, Meier, Kaluarachchi, Guillam{\'{o}}n
  et~al.}}]{Cho2017}
\bibinfo{author}{\bibfnamefont{K.}~\bibnamefont{Cho}},
  \bibinfo{author}{\bibfnamefont{A.}~\bibnamefont{Fente}},
  \bibinfo{author}{\bibfnamefont{S.}~\bibnamefont{Teknowijoyo}},
  \bibinfo{author}{\bibfnamefont{M.~A.} \bibnamefont{Tanatar}},
  \bibinfo{author}{\bibfnamefont{K.~R.} \bibnamefont{Joshi}},
  \bibinfo{author}{\bibfnamefont{N.~M.} \bibnamefont{Nusran}},
  \bibinfo{author}{\bibfnamefont{T.}~\bibnamefont{Kong}},
  \bibinfo{author}{\bibfnamefont{W.~R.} \bibnamefont{Meier}},
  \bibinfo{author}{\bibfnamefont{U.}~\bibnamefont{Kaluarachchi}},
  \bibinfo{author}{\bibfnamefont{I.}~\bibnamefont{Guillam{\'{o}}n}},
  \bibnamefont{et~al.}, \bibinfo{journal}{Physical Review B}
  \textbf{\bibinfo{volume}{95}}, \bibinfo{pages}{100502(R)}
  (\bibinfo{year}{2017}).

\bibitem[{\citenamefont{Dordevic et~al.}(2013)\citenamefont{Dordevic, Basov,
  and Homes}}]{Dordevic2013}
\bibinfo{author}{\bibfnamefont{S.~V.} \bibnamefont{Dordevic}},
  \bibinfo{author}{\bibfnamefont{D.~N.} \bibnamefont{Basov}}, \bibnamefont{and}
  \bibinfo{author}{\bibfnamefont{C.~C.} \bibnamefont{Homes}},
  \bibinfo{journal}{Scientific Reports} \textbf{\bibinfo{volume}{3}},
  \bibinfo{pages}{1713} (\bibinfo{year}{2013}).

\bibitem[{\citenamefont{Kogan}(2013)}]{Kogan2013a}
\bibinfo{author}{\bibfnamefont{V.~G.} \bibnamefont{Kogan}},
  \bibinfo{journal}{Physical Review B} \textbf{\bibinfo{volume}{87}},
  \bibinfo{pages}{220507(R)} (\bibinfo{year}{2013}).

\bibitem[{\citenamefont{Kogan et~al.}(2013)\citenamefont{Kogan, Prozorov, and
  Mishra}}]{Kogan2013}
\bibinfo{author}{\bibfnamefont{V.~G.} \bibnamefont{Kogan}},
  \bibinfo{author}{\bibfnamefont{R.}~\bibnamefont{Prozorov}}, \bibnamefont{and}
  \bibinfo{author}{\bibfnamefont{V.}~\bibnamefont{Mishra}},
  \bibinfo{journal}{Physical Review B} \textbf{\bibinfo{volume}{88}},
  \bibinfo{pages}{224508} (\bibinfo{year}{2013}), ISSN
  \bibinfo{issn}{1098-0121},
  \urlprefix\url{https://link.aps.org/doi/10.1103/PhysRevB.88.224508}.

\bibitem[{\citenamefont{Bud'ko et~al.}(2018)\citenamefont{Bud'ko, Kogan,
  Prozorov, Meier, Xu, and Canfield}}]{BudKo2018}
\bibinfo{author}{\bibfnamefont{S.~L.} \bibnamefont{Bud'ko}},
  \bibinfo{author}{\bibfnamefont{V.~G.} \bibnamefont{Kogan}},
  \bibinfo{author}{\bibfnamefont{R.}~\bibnamefont{Prozorov}},
  \bibinfo{author}{\bibfnamefont{W.~R.} \bibnamefont{Meier}},
  \bibinfo{author}{\bibfnamefont{M.}~\bibnamefont{Xu}}, \bibnamefont{and}
  \bibinfo{author}{\bibfnamefont{P.~C.} \bibnamefont{Canfield}},
  \bibinfo{journal}{Physical Review B} \textbf{\bibinfo{volume}{98}},
  \bibinfo{pages}{144520} (\bibinfo{year}{2018}).

\bibitem[{\citenamefont{Bud'ko et~al.}(2009)\citenamefont{Bud'ko, Ni, and
  Canfield}}]{BudKo2009}
\bibinfo{author}{\bibfnamefont{S.~L.} \bibnamefont{Bud'ko}},
  \bibinfo{author}{\bibfnamefont{N.}~\bibnamefont{Ni}}, \bibnamefont{and}
  \bibinfo{author}{\bibfnamefont{P.~C.} \bibnamefont{Canfield}},
  \bibinfo{journal}{Physical Review B} \textbf{\bibinfo{volume}{79}},
  \bibinfo{pages}{220516(R)} (\bibinfo{year}{2009}).

\bibitem[{\citenamefont{Skalski et~al.}(1964)\citenamefont{Skalski,
  Betbeder-Matibet, and Weiss}}]{Skalski1964}
\bibinfo{author}{\bibfnamefont{S.}~\bibnamefont{Skalski}},
  \bibinfo{author}{\bibfnamefont{O.}~\bibnamefont{Betbeder-Matibet}},
  \bibnamefont{and} \bibinfo{author}{\bibfnamefont{P.~R.} \bibnamefont{Weiss}},
  \bibinfo{journal}{Physical Review} \textbf{\bibinfo{volume}{136}},
  \bibinfo{pages}{A1500} (\bibinfo{year}{1964}).

\bibitem[{\citenamefont{Bud'ko et~al.}(2010)\citenamefont{Bud'ko, Kogan,
  Hodovanets, Ran, Moser, Lampe, and Canfield}}]{BudKo2010}
\bibinfo{author}{\bibfnamefont{S.~L.} \bibnamefont{Bud'ko}},
  \bibinfo{author}{\bibfnamefont{V.~G.} \bibnamefont{Kogan}},
  \bibinfo{author}{\bibfnamefont{H.}~\bibnamefont{Hodovanets}},
  \bibinfo{author}{\bibfnamefont{S.}~\bibnamefont{Ran}},
  \bibinfo{author}{\bibfnamefont{S.~A.} \bibnamefont{Moser}},
  \bibinfo{author}{\bibfnamefont{M.~J.} \bibnamefont{Lampe}}, \bibnamefont{and}
  \bibinfo{author}{\bibfnamefont{P.~C.} \bibnamefont{Canfield}},
  \bibinfo{journal}{Physical Review B} \textbf{\bibinfo{volume}{82}},
  \bibinfo{pages}{174513} (\bibinfo{year}{2010}).

\bibitem[{\citenamefont{{A. A. Abrikosov and L. P.
  Gor'kov}}(1961)}]{A.A.AbrikosovandL.P.Gorkov1961}
\bibinfo{author}{\bibnamefont{{A. A. Abrikosov and L. P. Gor'kov}}},
  \bibinfo{journal}{Sov. Phys. JETP 12, 1243 (1961)}
  \textbf{\bibinfo{volume}{12}}, \bibinfo{pages}{1243} (\bibinfo{year}{1961}).

\bibitem[{\citenamefont{OriginLab}()}]{OriginH}
\bibinfo{author}{\bibnamefont{OriginLab}}, \emph{\bibinfo{title}{{Origin
  Help}}},
  \urlprefix\url{https://www.originlab.com/doc/Origin-Help/Interpret-Regression-Result{\#}R-Square{\_}.28COD.29}.

\bibitem[{\citenamefont{Tomioka et~al.}(1993)\citenamefont{Tomioka, Naito, and
  Kitazawa}}]{Tomioka1993}
\bibinfo{author}{\bibfnamefont{Y.}~\bibnamefont{Tomioka}},
  \bibinfo{author}{\bibfnamefont{M.}~\bibnamefont{Naito}}, \bibnamefont{and}
  \bibinfo{author}{\bibfnamefont{K.}~\bibnamefont{Kitazawa}},
  \bibinfo{journal}{Physica C: Superconductivity}
  \textbf{\bibinfo{volume}{215}}, \bibinfo{pages}{297} (\bibinfo{year}{1993}).

\end{thebibliography}

\clearpage
\section{Appendix}

\begin{figure}[h]
	\includegraphics[width=1.5\columnwidth]{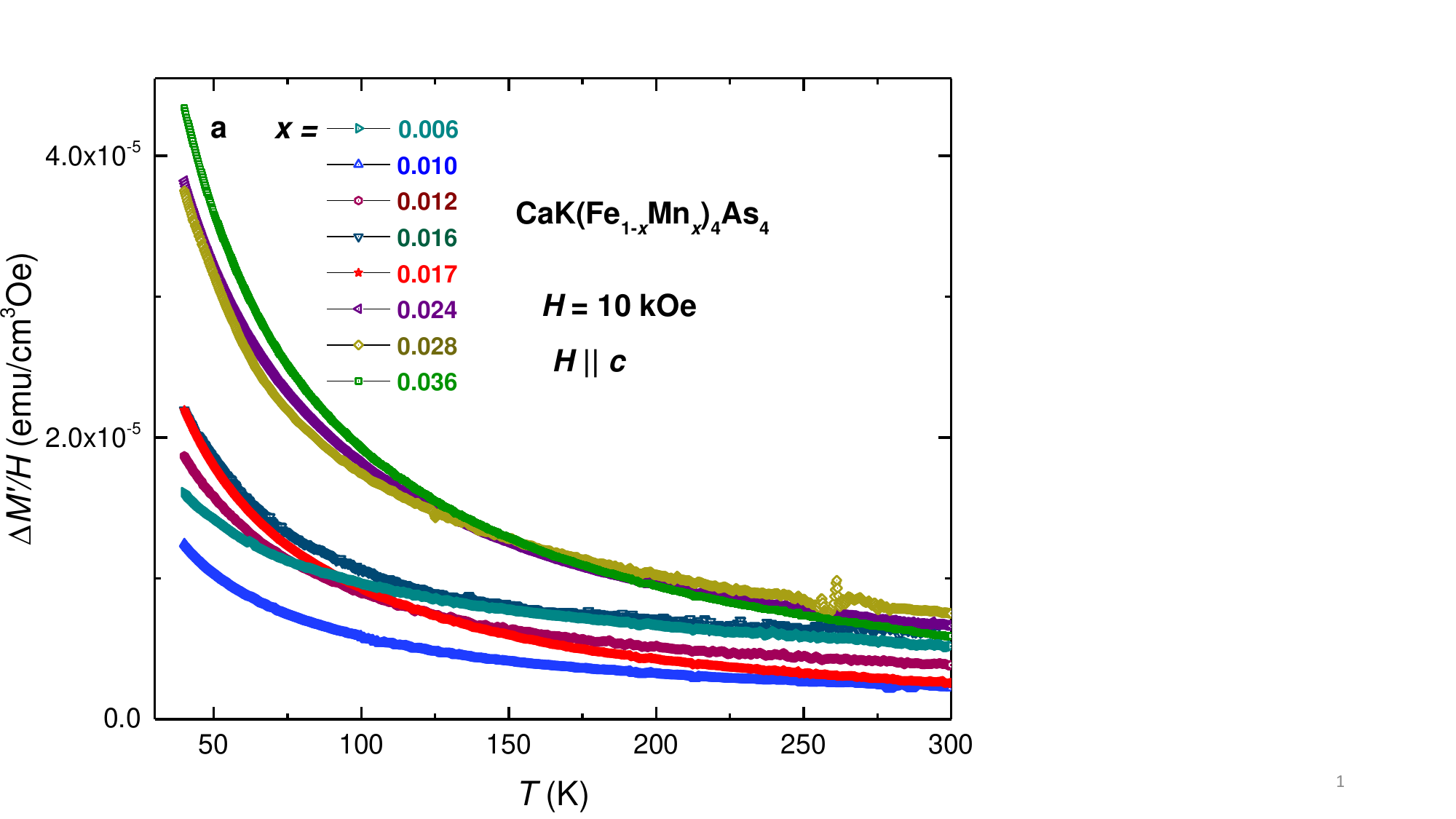}
	\caption{Difference in the magnetization ($\Delta M^{\prime}$) between CaKFe$_{4}$As$_{4}$ and CaK(Fe$_{1-x}$Mn$_{x}$)$_{4}$As$_{4}$ single crystals as a function of temperature from 40 K to 300 K for with a field of 10 kOe applied parallel to the crystallographic \textit{ab} plane shows appearance of Curie-Weiss tail.}
	\label{figure 13}
\end{figure}

Figure \ref{figure 13} shows the difference in the magnetization ($\Delta M^{\prime}$) between CaK(Fe$_{1-x}$Mn$_{x}$)$_{4}$As$_{4}$ and CaKFe$_{4}$As$_{4}$ single crystals as a function of temperature from 40~K to 300~K with a field of 10~kOe applied parallel to the crystallographic \textit{ab} plane. The magnetization plots have the appearance of Curie-Weiss tails suggesting that the Mn substituted system has local moment behavior. In order to determine if these are indeed Curie-Weiss tails, fitting was done by assuming that local moment behavior is only due to Mn and fixing $\theta$ values to be determined by $T^*$  value or its extrapolation by linear fitting $T^*$ to $x$ = 0.006 for low $x$, where $T^*$ is absent to $x$ = 0.006 for low $x$, where $T^*$ is absent. Due to the small signals in the high temperature range, we performed fitting from 40~K to 200~K. Table \ref{Table1} shows the result of fitting. The value of $\mu_{eff}$ is around 5 $\mu_{B}$ and $R^2$ (COD) values are larger than 0.99. R (COD) is also known as coefficient of determination which is a statistical measure to qualify the linear regression. If $R^2$ is 1, it indicates that the fitted line explains all the variability of the response data around its mean {\color{blue}\cite{OriginH}}. The fitting results strongly suggest that Mn, unlike Ni or Co in CaK(Fe$_{1-x}$$TM_{x}$)$_4$As$_4$, {\color{blue}\cite{Meier2019T}} is more local-like moment bearing.

\begin{table}[htbp]

	\begin{tabular}{rrrrrr}
		\midrule
		\multicolumn{1}{l}{$x_{EDS}$} & \multicolumn{1}{l}{ \ $\mu_{eff}$, $\mu_{B}$} & \multicolumn{1}{l}{\ \ $\theta$, K} & \multicolumn{1}{l}{\ $\chi_0$, $\dfrac{emu}{mol\ Oe}$} & \multicolumn{1}{l}{ \ \ $R^2 (COD)$}  \\
		\midrule
		0.006 &  \ 5.47 (0.18) &\ \  22.0 &\ \ -0.0132 &\  0.9977  \\	
		0.010 & 4.26 (0.22) & 23.9 & -0.0021 & 0.9985  \\
		0.012 & 4.82 (0.14) & 24.6 & -0.0021 & 0.9985  \\
		0.016 & 4.55 (0.09) & 25.5 & -0.0010 & 0.9948  \\
		0.017 & 4.69 (0.08) & 26.5 & -0.0060 & 0.9975  \\
		0.024 & 5.18 (0.07) & 28.8 & -0.0042 & 0.9986  \\
		0.028 & 4.63 (0.08) & 30.0 & -0.0046 & 0.9911  \\
		0.036 & 4.93 (0.05) & 35.5 & -0.0065 & 0.9954  \\
		\midrule
	\end{tabular}%
	\caption{Table shows effective, $\mu_{eff}$, Curie-Weiss temperature, $\theta$, temperature-independent susceptibility, $\chi_0$ obtained from Curie-Weiss fit to the difference magnetization ($\Delta M^{\prime}$) between CaKFe$_{4}$As$_{4}$ and CaK(Fe$_{1-x}$Mn$_{x}$)$_{4}$As$_{4}$ single crystals as a function of temperature from 40~K to 150~K for with a field of 10 kOe applied parallel to the crystallographic \textit{ab} plane. The $\theta$ values were fixed to the inferred values of $T^{*}$ or their extrapolated values (see text for details)}
	\label{Table1}%
\end{table}%

\begin{figure}[H]
	\includegraphics[width=1.5\columnwidth]{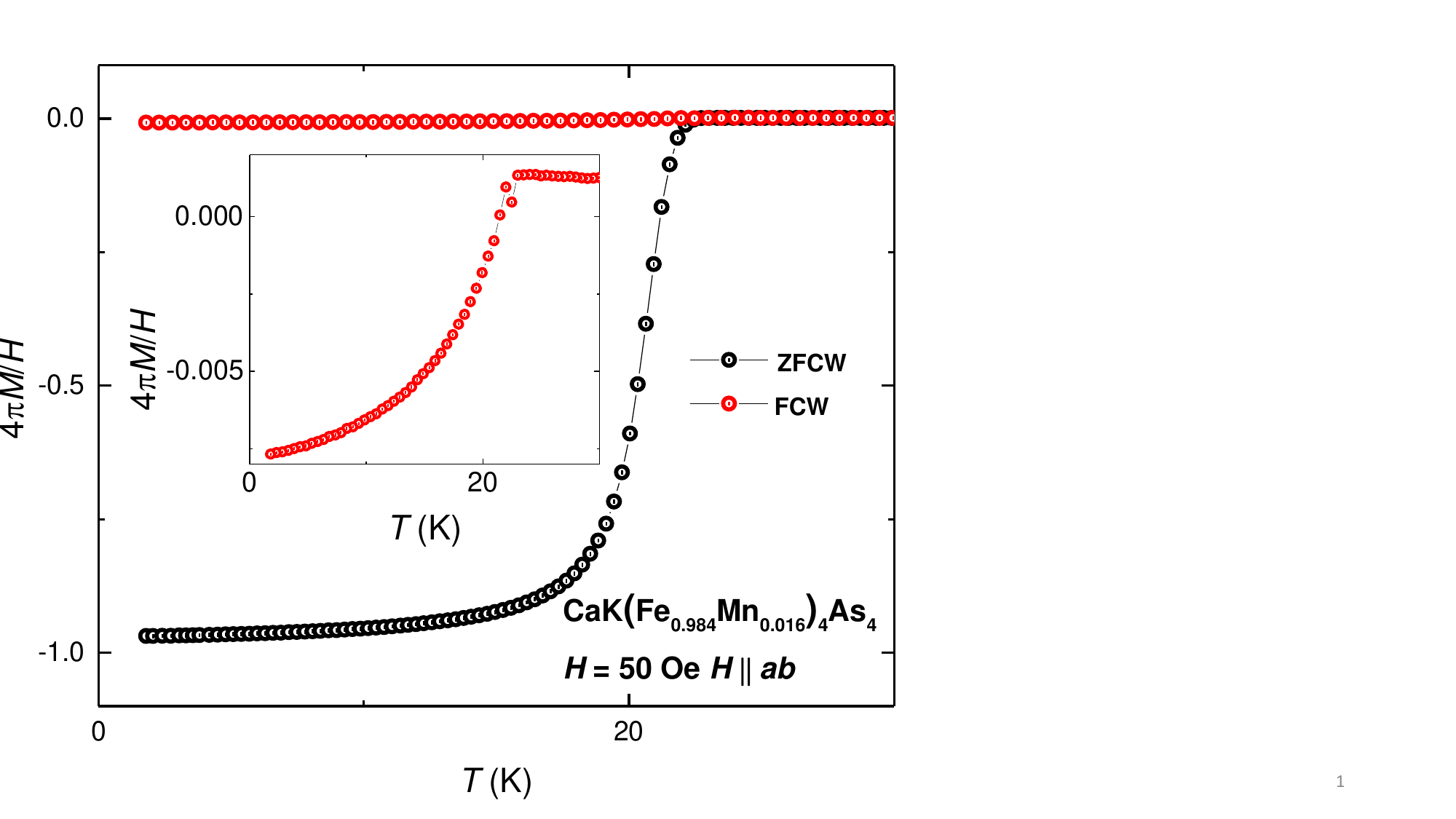}
	\caption{Zero-field-cooled-warming (ZFCW) and Field-cooled (FCW) low temperature magnetization as a function of temperature for CaK(Fe$_{0.984}$Mn$_{0.016}$)$_{4}$As$_{4}$ single crystals with a field of 50~Oe applied parallel to the crystallographic \textit{ab} plane}. $M$ is the volumetric magnetic moment with cgs unit emu cm$^{-3}$ or Oe.  \label{50OeFCZFCz}
\end{figure}

Figure \ref{50OeFCZFCz} shows Zero-field-cooled-warming (ZFCW) and Field-cooled-warming (FCW) low temperature magnetization as a function of temperature for CaK(Fe$_{0.984}$Mn$_{0.016}$)$_{4}$As$_{4}$ single crystals with a field of 50 Oe applied parallel to \textit{ab} plane. The large difference between ZFCW and FCW is consistent with the large pinning found even in CaKFe$_4$As$_4$\cite{Tomioka1993}.

  \begin{figure}[H]
	\centering
	\begin{minipage}{0.44\textwidth}
		\centering
		\includegraphics[width=1.5\columnwidth]{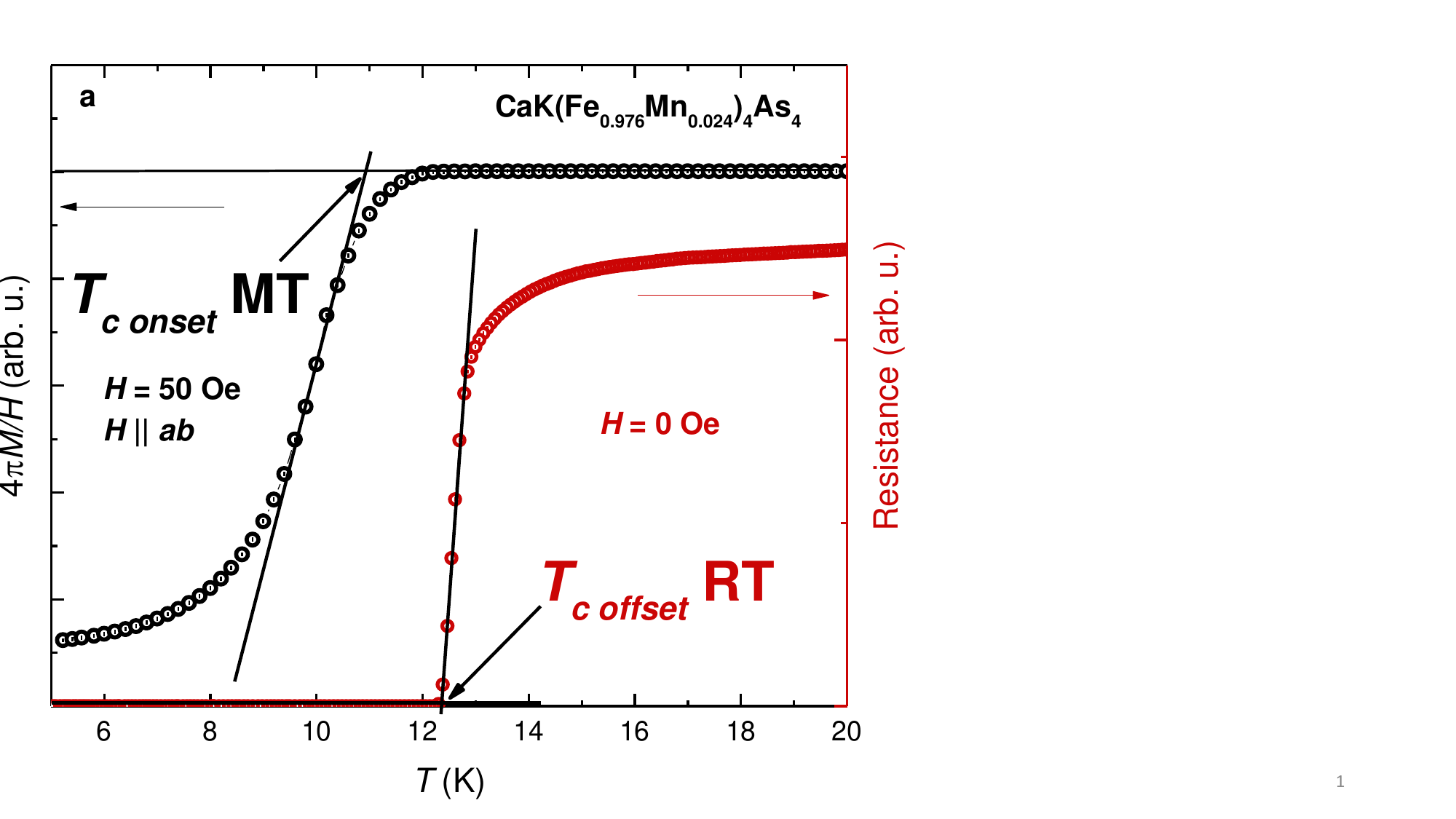}		
	\end{minipage}\hfill
	\begin{minipage}{0.42\textwidth}
		\centering
		\includegraphics[width=1.5\columnwidth]{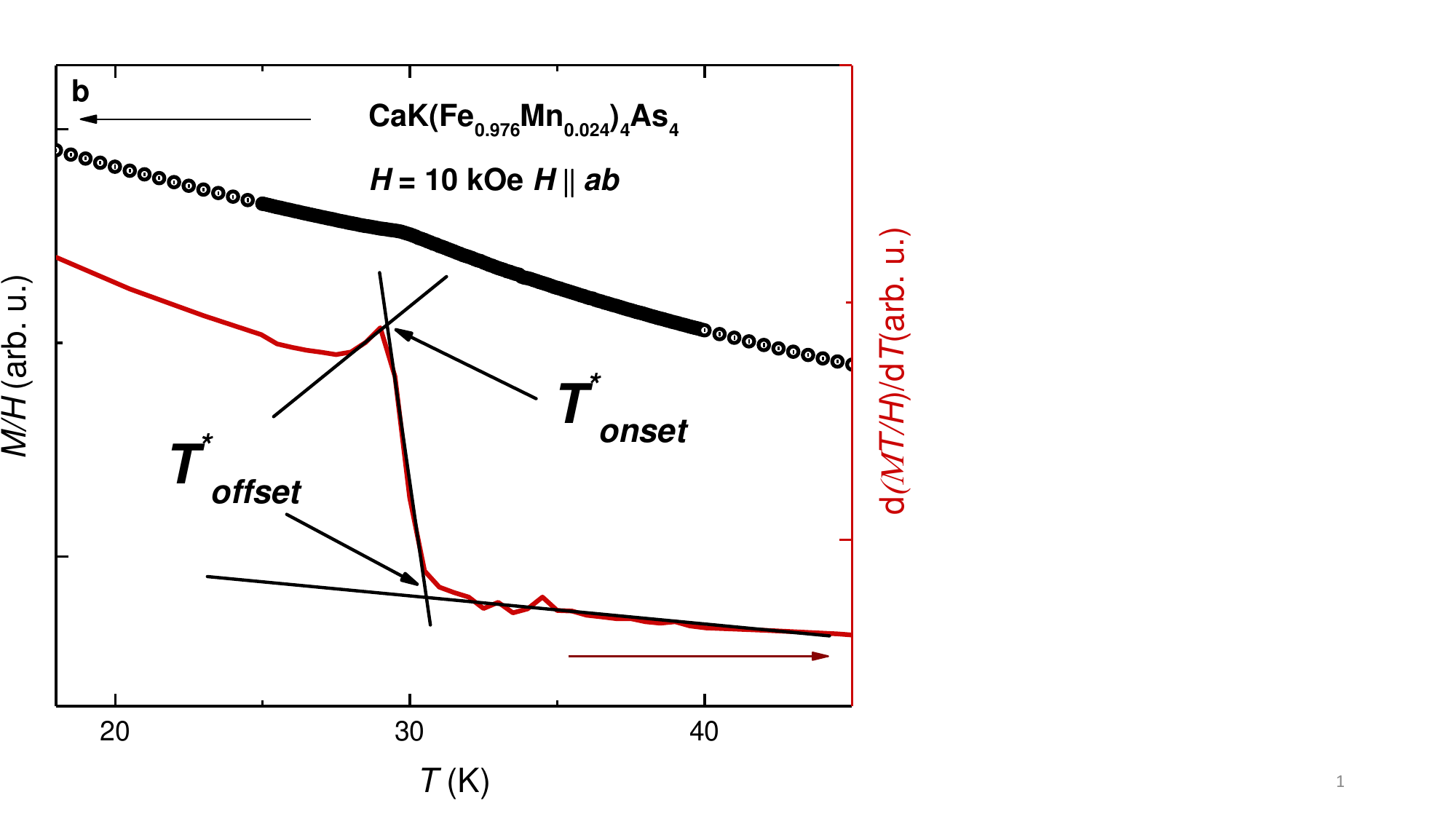}
		
	\end{minipage}\hfill
	\begin{minipage}{0.44\textwidth}
		\centering
		\includegraphics[width=1.5\columnwidth]{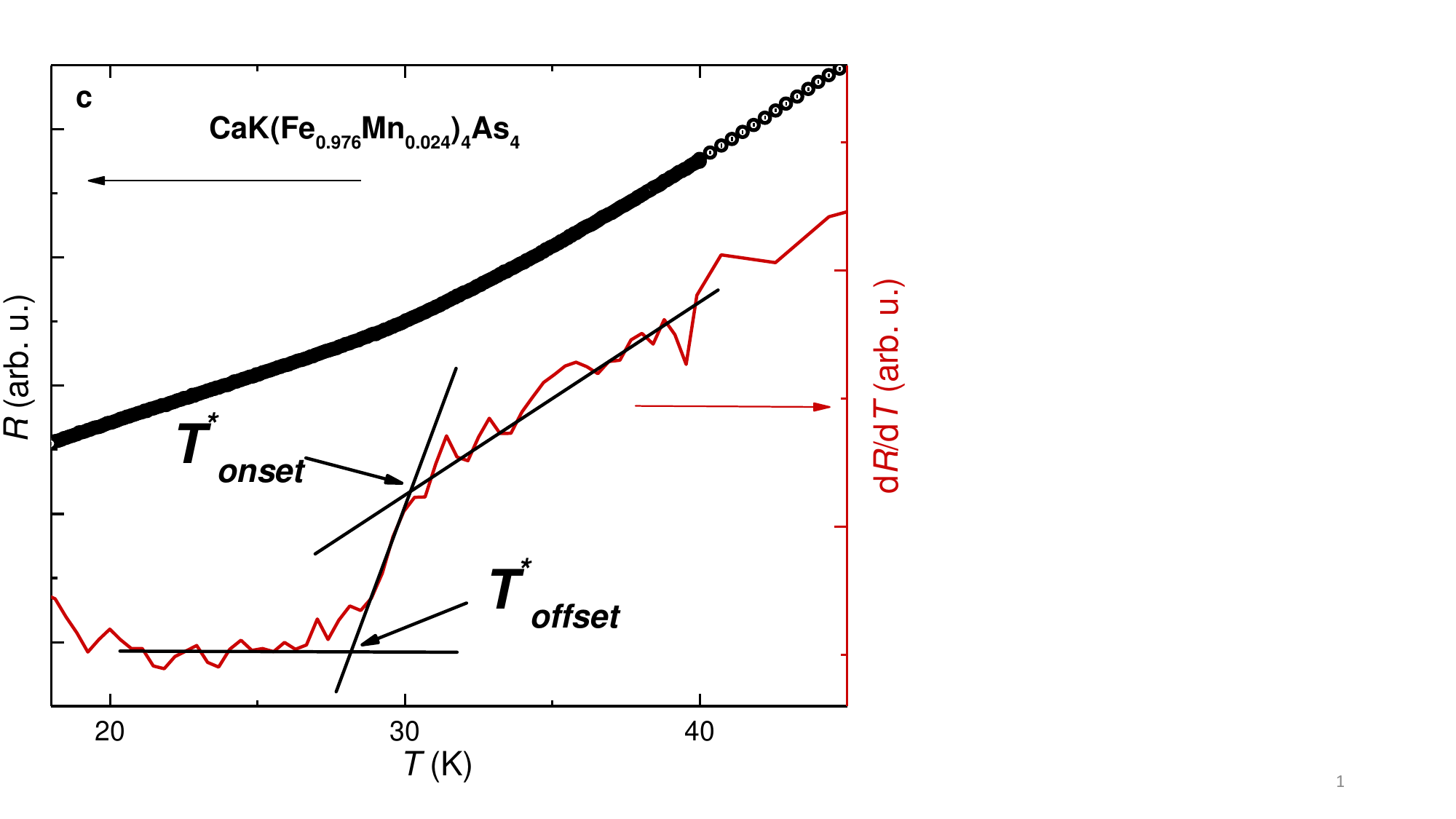}
		\caption{Onset and offset criteria for $T_c$ and $T^*$ based on magnetization and resistance measurement.   \label{figure92}}
	\end{minipage}
\end{figure}

Criteria for inferring $T_{c}$ and $T^{*}$ are shown in Fig. \ref{figure92}.  For $T_{c}$ (Fig. \ref{figure92}a) we use an onset criterion for our \textit{M}(\textit{T}) data and an offset criterion for our \textit{R}(\textit{T}) data. As is often the case, these criteria agree well, especially in the low field limit. For $T^{*}$, although the feature is much clearer for Mn substitution than it was for Ni or Co substitution {\color{blue}\cite{Meier2019T}}, the features in \textit{M}(\textit{T}) and \textit{R}(\textit{T}) are still somewhat subtle. In order to infer $T^{*}$ we take the average of onset and offset value and use the difference as error.

 \begin{figure}[H]
	\centering
		\includegraphics[width=2.6\columnwidth]{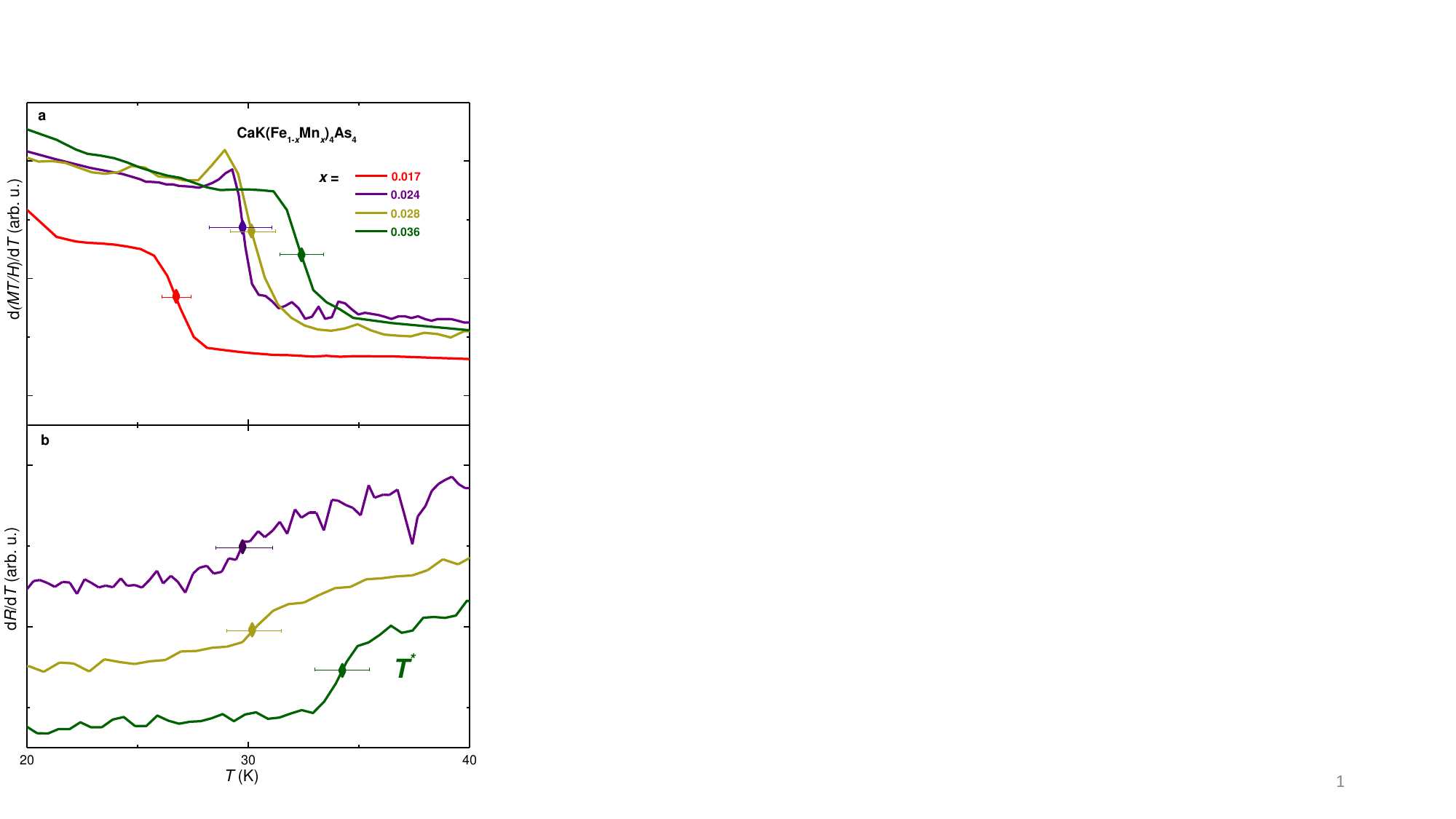}
		\caption{The $T^*$ anomaly appears clearly as a step in both plot d($MT/H$)/d$T$ and the derivative of resistance, d\textit{R}/d\textit{T}. Only the data above $T_c$ are plotted. \label{figure 14}}
	\end{figure}
d\textit{R}(\textit{T})/d\textit{T} and d\textit{M}(T)/d\textit{T} data for several different \textit{x}-values are shown in figure \ref{figure 14}, showing good agreement between the position of the $T^{*}$ features.

\begin{figure}
	\includegraphics[width=1.5\columnwidth]{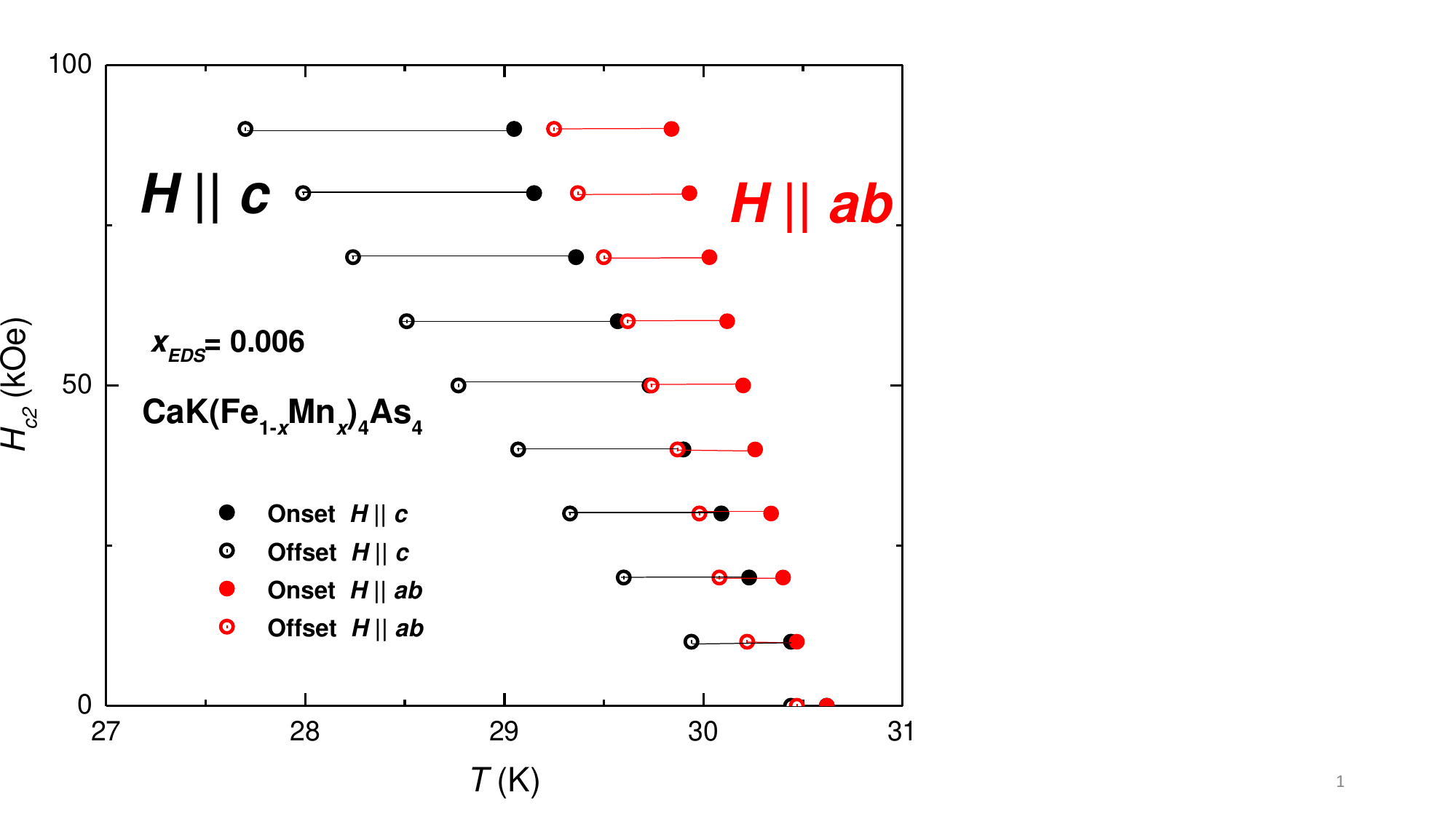}
	\caption{Anisotropic $H_{c2}(T)$ data determined for two single crystalline samples of $x$$_{EDS}$=0.006 CaK(Fe$_{1-x}$Mn$_{x}$)$_{4}$As$_{4}$ using onset criterion (solid) and offset criterion (hollow) inferred from the temperature-dependent electrical resistance data.   \label{figure81}}
\end{figure}

\begin{figure}[H]
	\includegraphics[width=1.5\columnwidth]{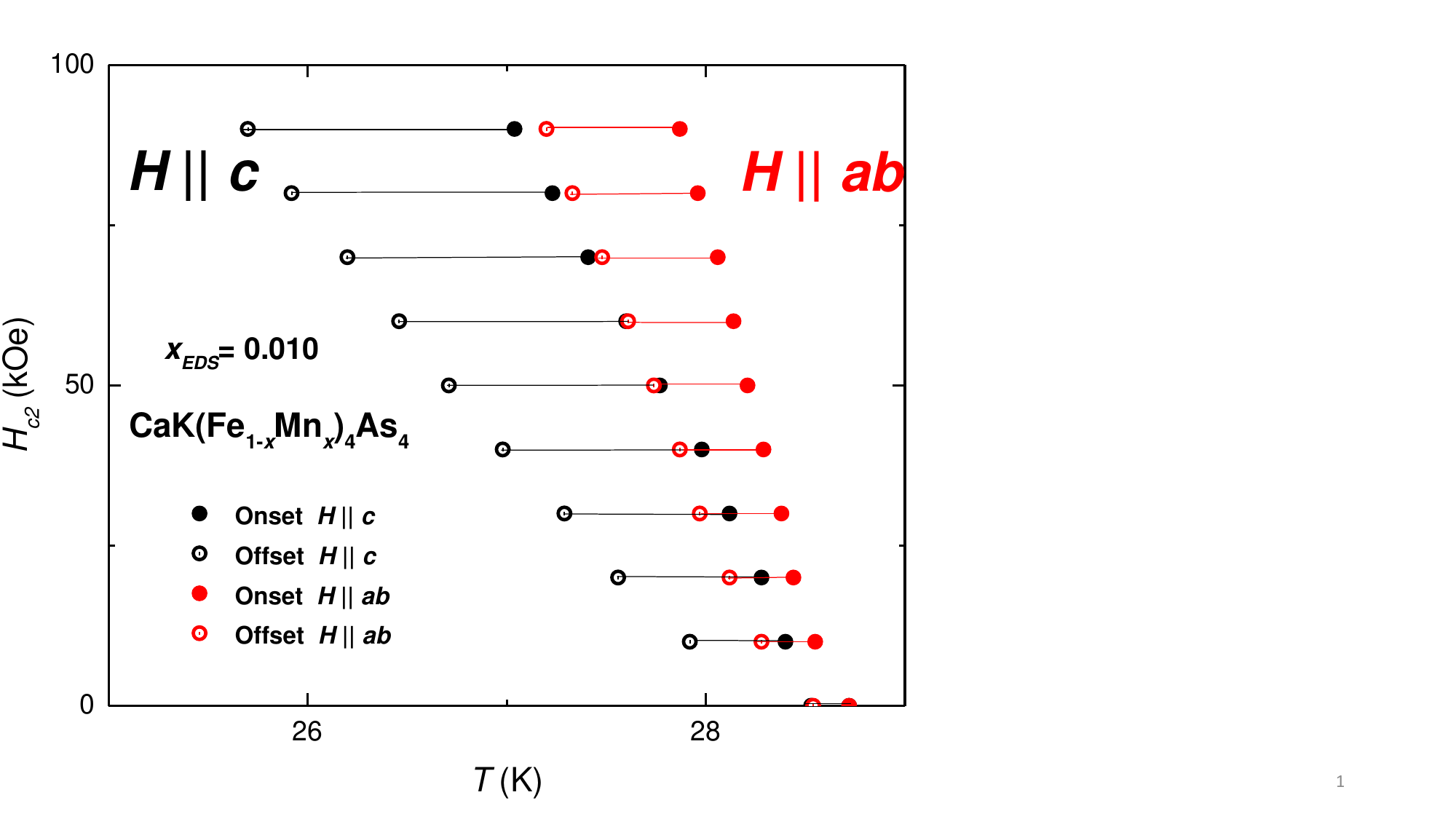}	
	\caption{Anisotropic $H_{c2}(T)$ data determined for two single crystalline samples of $x$$_{EDS}$=0.010 CaK(Fe$_{1-x}$Mn$_{x}$)$_{4}$As$_{4}$ using onset criterion (solid) and offset criterion (hollow) inferred from the temperature-dependent electrical resistance data.   \label{figure82}}
\end{figure}

\begin{figure}[H]
	\includegraphics[width=1.5\columnwidth]{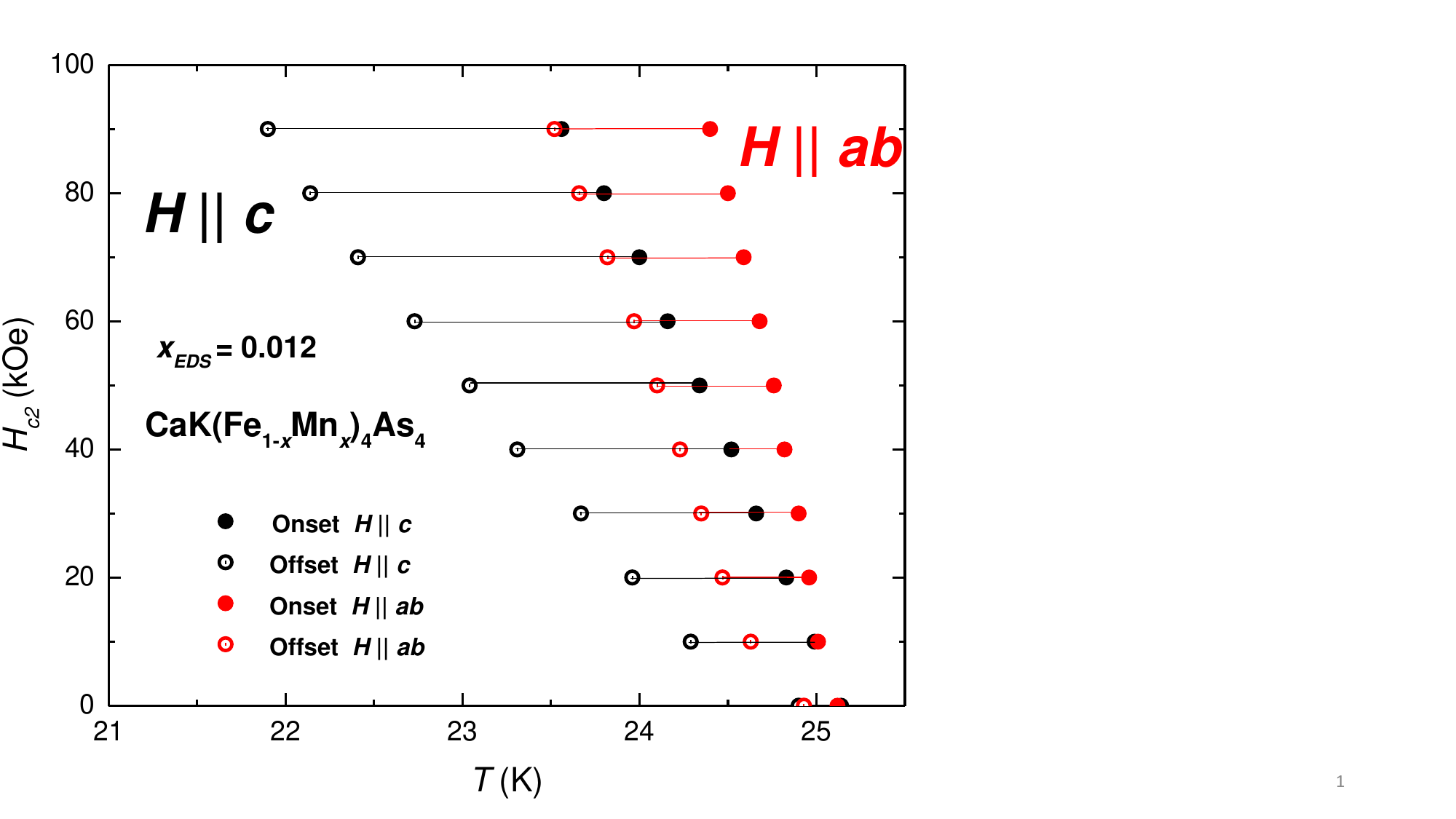}	
	\caption{Anisotropic $H_{c2}(T)$ data determined for two single crystalline samples of $x$$_{EDS}$=0.012 CaK(Fe$_{1-x}$Mn$_{x}$)$_{4}$As$_{4}$ using onset criterion (solid) and offset criterion (hollow) inferred from the temperature-dependent electrical resistance data.   \label{figure83}}
\end{figure}

\begin{figure}[H]
	\includegraphics[width=1.5\columnwidth]{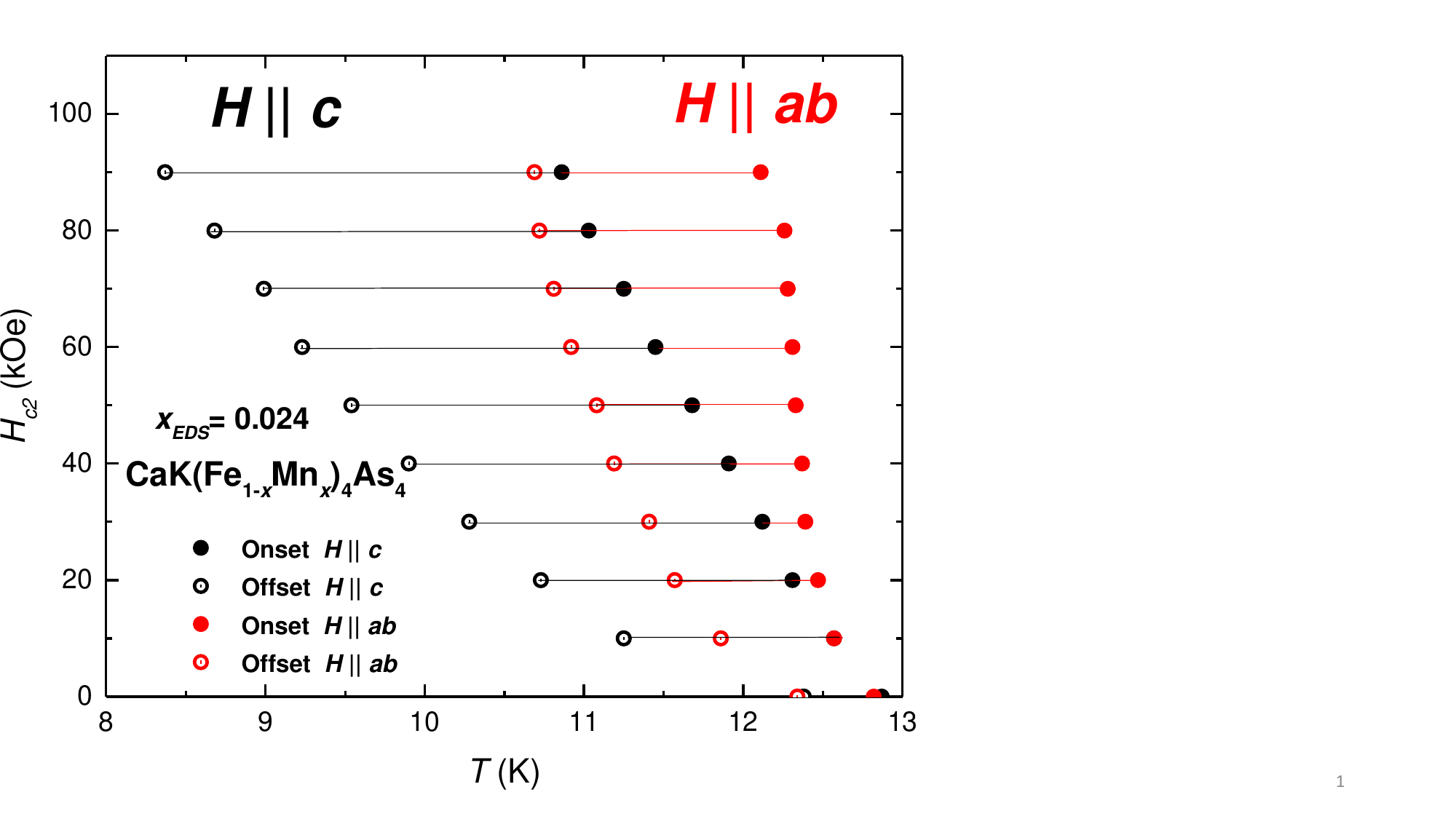}	
	\caption{Anisotropic $H_{c2}(T)$ data determined for two single crystalline samples of $x$$_{EDS}$=0.024 CaK(Fe$_{1-x}$Mn$_{x}$)$_{4}$As$_{4}$ using onset criterion (solid) and offset criterion (hollow) inferred from the temperature-dependent electrical resistance data.   \label{figure84}}
\end{figure}

\begin{figure}[H]
	\includegraphics[width=1.5\columnwidth]{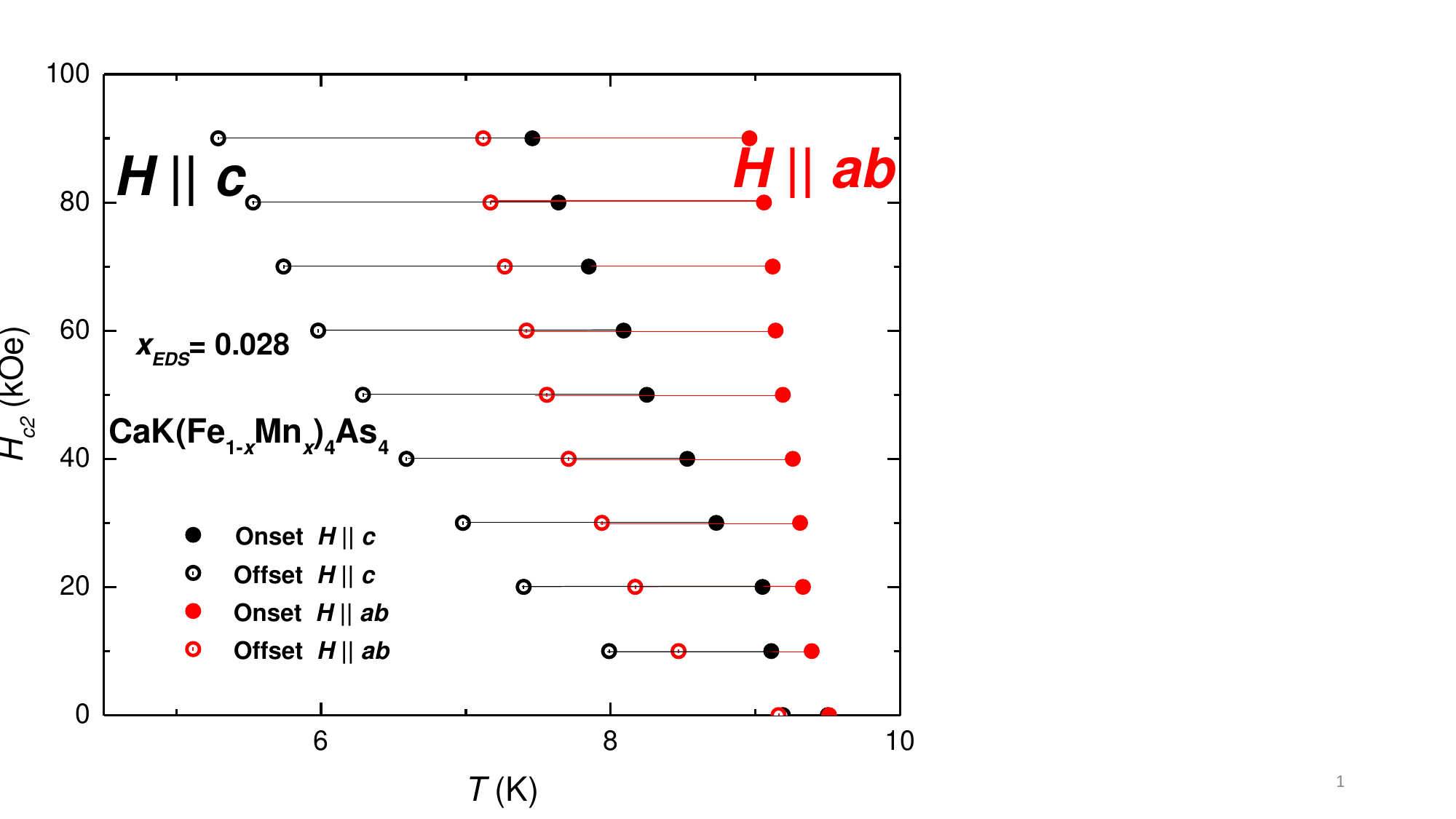}	
	\caption{Anisotropic $H_{c2}(T)$ data determined for two single crystalline samples of $x$$_{EDS}$=0.028 CaK(Fe$_{1-x}$Mn$_{x}$)$_{4}$As$_{4}$ using onset criterion (solid) and offset criterion (hollow) inferred from the temperature-dependent electrical resistance data.   \label{figure86}}
\end{figure}

Figures \ref{figure81} - \ref{figure86} present $H_{c2}(T)$ curves which is obtained from $R$($T$) data for fixed applied fields and the criteria shown in Figs \ref{figure7}a of CaK(Fe$_{1-x}$Mn$_{x}$)$_{4}$As$_{4}$ single crystals for $x$ = 0.006, 0.010, 0.012, 0.024 and 0.028.

\end{document}